# Continuum Limits and Exact Finite-Size-Scaling Functions for One-Dimensional $O(N)$-Invariant Spin Models

Attilio Cucchieri
Tereza Mendes
*Department of Physics*
*New York University*
*4 Washington Place*
*New York, NY 10003 USA*
Internet: `ATTILIOC@ACF2.NYU.EDU, MENDES@MAFALDA.PHYSICS.NYU.EDU`

Andrea Pelissetto
*Dipartimento di Fisica*
*Università degli Studi di Pisa*
*Pisa 56100, ITALIA*
Internet: `PELISSET@SUNTHPI1.DIFI.UNIPI.IT`

Alan D. Sokal
*Department of Physics*
*New York University*
*4 Washington Place*
*New York, NY 10003 USA*
Internet: `SOKAL@ACF4.NYU.EDU`

September 11, 1995

**Abstract**

We solve exactly the general one-dimensional $O(N)$-invariant spin model taking values in the sphere $S^{N-1}$, with nearest-neighbor interactions, in finite volume with periodic boundary conditions, by an expansion in hyperspherical harmonics. The possible continuum limits are discussed for a general one-parameter family of interactions, and an infinite number of universality classes is found. For these classes we compute the finite-size-scaling functions and the leading corrections to finite-size scaling. A special two-parameter family of interactions (which includes the mixed isovector/isotensor model) is also treated, and no additional universality classes appear. In the appendices we give new formulae for the Clebsch-Gordan coefficients and 6–$j$ symbols of the $O(N)$ group, and some new generalizations of the Poisson summation formula; these may be of independent interest.

# Contents

# 1  Introduction

The purpose of this paper is to study the continuum limits and finite-size-scaling functions in a general class of one-dimensional $O(N)$-invariant spin models (also called nonlinear $\sigma$-models). Despite the relatively trivial nature of physics in one dimension, this exercise is interesting for several reasons:

1) *Two*-dimensional nonlinear $\sigma$-models are of direct interest in condensed-matter physics, and they are of indirect interest in elementary-particle physics because they share with four-dimensional gauge theories the property of perturbative asymptotic freedom [1,2,3,4]. In particular, recent work [5,6,7], combining Monte Carlo simulations and heuristic analytic arguments, has given evidence for the existence of new universality classes for the two-dimensional $O(N)$-invariant lattice $\sigma$-model with mixed isovector/isotensor action. The present work was motivated by the idea of investigating the occurrence of analogous universality classes in the *one*-dimensional case, where an exact analytic treatment is possible.[1]

2) A second motivation was to perform the computation of an exact finite-size-scaling function (as well as the leading correction to it) for a non-trivial spin model. Finite-size scaling has become increasingly important in the analysis of Monte Carlo data [8,9]. (For example, the functions derived in this paper can be used for comparison in the multigrid Monte Carlo study of the one-dimensional $O(4)$-symmetric non-linear $\sigma$-model [10].) Moreover, finite-size scaling is the basis of an important new method for extrapolation of finite-volume Monte Carlo data to infinite volume [11,12,13]. It is also useful to know something about the corrections to finite-size scaling. In particular, in the new methods for extrapolation to infinite volume, it is crucial to understand the corrections to finite-size scaling because they induce systematic errors in the extrapolation.

3) Finally, our solution method makes use of the functions defined by the generalization of the usual spherical harmonics to the $N$-dimensional unit sphere $S^{N-1}$, which we call *hyperspherical harmonics*. Although these functions are well known [14,15,16,17,18,19], we were unable to find any convenient list of their properties in the literature, and therefore we thought that it would be useful to make a compendium of the relevant properties and formulae. In particular, we were unable to find the Clebsch-Gordan coefficients anywhere in the literature (although they too are probably known). Using the representation of hyperspherical harmonics as completely symmetric and traceless tensors[2], the computation of the Clebsch-Gordan coefficients is a straightforward combinatoric exercise. Indeed, we can go further and compute many of the 6–$j$ symbols. We believe that hyperspherical harmonics constitute the most efficient approach to the derivation of high-temperature expansions for $O(N)$-invariant spin models taking values in $S^{N-1}$. Indeed, they have been used for this purpose by the King's College group [20,21,22,23,24] and others [25,26,27]; but the methods were cumbersome, in part due to the lack of convenient expressions for the Clebsch-Gordan coefficients. In addition to the work reported here, we are now using these methods to extend various high-temperature expansions for *two*- and *three*-dimensional $O(N)$-invariant and $U(N)$-invariant spin models [28].

---

[1] We thank Erhard Seiler for the suggestion to do this.

[2] This representation is of course well known, but it is not (so far as we know) employed in any of the standard treatises on hyperspherical harmonics. As we shall show here, this representation is an extremely convenient one; one of the purposes of this paper is to make some advertising on its behalf.

This paper is organized as follows: The hyperspherical harmonics are introduced in Section 2, where we also explain how they are used in the expansion of the Gibbs weight $\exp(-\mathcal{H})$. In Section 3 we give the exact solution for the general one-dimensional $S^{N-1}$ $\sigma$-model in finite volume, as well as its infinite-volume limit. All expressions are written in terms of the normalized expansion coefficients $v_{N,l}$ (which generalize the well-known $v = \tanh J$ for the Ising case $N = 1$). In Section 4.1 we discuss in detail the possible continuum limits for one-parameter Hamiltonians by performing the large-$J$ (i.e. low-temperature) expansion of $v_{N,l}(J)$, and we show the appearance of infinite families of universality classes. The finite-size-scaling functions and the corresponding corrections to finite-size scaling are given in Sections 4.2 and 4.3, respectively. Finally, in Section 5, we analyze a class of two-parameter Hamiltonians — which includes, among others, the mixed isovector/isotensor model studied in [5,6,7] — and we show that no additional universality classes appear beyond the ones already found in Section 4.1. In Appendix A we provide proofs of various properties of the hyperspherical harmonics, including the Clebsch-Gordan coefficients and some of the 6–$j$ symbols. In Appendix B we analyze the finite-size-scaling functions for a one-parameter family of universality classes that includes those of the mixed isovector/isotensor model; this analysis is based on generalized Poisson summation formulae applied to some generalized theta functions. (We think these formulae may be of independent interest; as far as we know they are new.) In Appendix C we study the limit $N \to \infty$ of the finite-size-scaling functions for the standard $N$-vector universality class.

## 2   Hyperspherical Harmonics

The purpose of this section is to introduce the *hyperspherical harmonics* that will give the basis for expanding the Gibbs weight $e^{-\mathcal{H}}$ for our spin models. From the mathematical point of view this is connected with doing *harmonic analysis* on the unit sphere $S^{N-1} \subset \mathbb{R}^N$ acted on transitively by the compact connected Lie group $SO(N)$ [30]. More precisely, let us consider:

- $\boldsymbol{\sigma} \in S^{N-1}$

- $R \in SO(N)$

- the normalized rotation-invariant measure $d\Omega(\boldsymbol{\sigma})$ on $S^{N-1}$

- the (complex-valued) square-integrable functions $f \in L^2(S^{N-1})$

- the unitary representation $T(R)$ of $SO(N)$ on $L^2(S^{N-1})$ defined by $(T(R)f)(\boldsymbol{\sigma}) = f(R^{-1}\boldsymbol{\sigma})$

Then, we want to find an orthogonal Hilbert-space decomposition of $L^2(S^{N-1})$ into subspaces such that the representation $T(R)$ restricted to each subspace is irreducible. The needed decomposition turns out to be precisely the decomposition of $L^2(S^{N-1})$ into eigenspaces of the *Laplace-Beltrami operator* $\mathcal{L} = \mathcal{L}_{S^{N-1}}$.[3,4] In fact, it can be proved[5] that:

(a) The eigenvalues[6] of $\mathcal{L}$ are

$$\lambda_{N,k} = k(N + k - 2) \geq 0, \tag{2.1}$$

where $k = 0, 1, 2, \ldots$. The corresponding eigenspace $E_{N,k}$ has dimension[7]

$$\mathcal{N}_{N,k} \equiv \dim E_{N,k} = \frac{N + 2k - 2}{k!} \frac{\Gamma(N + k - 2)}{\Gamma(N - 1)} \tag{2.2}$$

and can be given several equivalent descriptions:

---

[3] The Laplace-Beltrami operator on $S^{N-1}$ can be defined as follows: Define on $\mathbb{R}^N$ the vector fields ("angular-momentum operators")

$$L^{\alpha\beta} = i\left(x^\alpha \frac{\partial}{\partial x^\beta} - x^\beta \frac{\partial}{\partial x^\alpha}\right).$$

Then the restriction to $S^{N-1}$ of each $L^{\alpha\beta}$ is a vector field on $S^{N-1}$, and

$$\mathcal{L} \equiv \sum_{1 \leq \alpha < \beta \leq N} L^{\alpha\beta} L^{\alpha\beta}$$

is the Laplace-Beltrami operator on $S^{N-1}$.

[4] We remark that, for $N \geq 3$, $\mathcal{L}$ generates the algebra $\mathbf{D}(S^{N-1})$ of $SO(N)$-invariant differential operators on $S^{N-1}$. (For $N = 2$ this is not the case, because $\partial/\partial\theta$ is an $SO(2)$-invariant differential operator not belonging to the algebra generated by $\mathcal{L}$. But if we consider differential operators invariant under $O(N)$ instead of $SO(N)$, then the assertion is true also for $N = 2$.)

[5] See [30], Theorem 3.1 (pp. 17–19).

[6] Note that our $\mathcal{L}$ is the *negative* of the usual Laplacian, i.e. it is a *positive*-semidefinite operator.

[7] For a proof see [30], Exercise A.5(i) (pp. 74, 552) and [15], Lemma 3 (p. 4). See also Appendix A.1 below. Usually we are interested in the case $N \geq 3$, for which formula (2.2) is unambiguous. But (2.2) is also valid for $N = 1, 2$, if it is interpreted as an analytic (in fact polynomial) function of $N$ for each *fixed* integer $k \geq 0$. Thus, for $N = 1$ and $N = 2$ we have

$$\mathcal{N}_{1,k} \equiv \dim E_{1,k} = \lim_{N \to 1} \frac{N + 2k - 2}{k!} \frac{\Gamma(N + k - 2)}{\Gamma(N - 1)} = \begin{cases} 1 & \text{for } k = 0, 1 \\ 0 & \text{for } k \geq 2 \end{cases}$$

and

$$\mathcal{N}_{2,k} \equiv \dim E_{2,k} = \lim_{N \to 2} \frac{N + 2k - 2}{k!} \frac{\Gamma(N + k - 2)}{\Gamma(N - 1)} = \begin{cases} 1 & \text{for } k = 0 \\ 2 & \text{for } k \geq 1 \end{cases}.$$

Note also that $\mathcal{N}_{N,0} = 1$ and $\mathcal{N}_{N,1} = N$ for all $N$.

(i) $E_{N,k}$ consists of the restrictions to $S^{N-1}$ of the *harmonic polynomials* of degree $k$ on $\mathbb{R}^N$ (namely, the homogeneous polynomials of degree $k$ that satisfy Laplace's equation on $\mathbb{R}^N$).

(ii) $E_{N,k}$ is spanned by the functions $f(\boldsymbol{\sigma}) = (\boldsymbol{a} \cdot \boldsymbol{\sigma})^k$ with $\boldsymbol{a} \in \mathbb{C}^N$ and $\sum_{i=1}^N a_i^2 = 0$.

(iii) $E_{N,k}$ is spanned by the *completely symmetric and traceless tensors* $Y_{N,k}^{\alpha_1 \ldots \alpha_k}(\boldsymbol{\sigma})$ of rank $k$, as the indices $\alpha_1, \alpha_2, \ldots, \alpha_k$ range over the $N^k$ allowable values.[8] (These tensors are described in more detail below.) Of course, since in general $N^k > \dim E_{N,k}$, the $Y_{N,k}^{\alpha_1 \ldots \alpha_k}(\boldsymbol{\sigma})$ form an *overcomplete* set.

(b) Each eigenspace $E_{N,k}$ is left invariant by $T(R)$. Moreover, for $N \geq 3$ the representation $T(R){\upharpoonright} E_{N,k}$ of $SO(N)$ is irreducible.[9]

(c) $L^2(S^{N-1}) = \bigoplus_{k=0}^{\infty} E_{N,k}$ (orthogonal Hilbert space decomposition).

To make all this concrete, we can write:
$$Y_{N,k}^{\alpha_1 \ldots \alpha_k}(\boldsymbol{\sigma}) \equiv \mu_{N,k} \left( \sigma^{\alpha_1} \cdots \sigma^{\alpha_k} - \text{Traces} \right) \tag{2.3}$$
where $\boldsymbol{\sigma} \in S^{N-1}$, "Traces" is such that $Y_{N,k}^{\alpha_1 \ldots \alpha_k}(\boldsymbol{\sigma})$ is completely symmetric and traceless (namely[10] $\delta_{\alpha_i \alpha_j} Y_{N,k}^{\alpha_1 \ldots \alpha_k}(\boldsymbol{\sigma}) = 0$ for any $i \neq j$), and
$$\mu_{N,k} = \left[ \frac{2^k \, \Gamma\left(\frac{N}{2} + k\right)}{k! \, \Gamma\left(\frac{N}{2}\right)} \right]^{1/2} \tag{2.4}$$

Explicit examples are:
$$Y_{N,0}(\boldsymbol{\sigma}) = 1 \tag{2.5}$$
$$Y_{N,1}^{\alpha}(\boldsymbol{\sigma}) = \sqrt{N} \, \sigma^{\alpha} \tag{2.6}$$
$$Y_{N,2}^{\alpha\beta}(\boldsymbol{\sigma}) = \sqrt{\frac{N(N+2)}{2}} \left( \sigma^{\alpha} \sigma^{\beta} - \frac{1}{N} \delta^{\alpha\beta} \right) \tag{2.7}$$
$$Y_{N,3}^{\alpha\beta\gamma}(\boldsymbol{\sigma}) = \sqrt{\frac{N(N+2)(N+4)}{6}}$$

---

[8] The functions $(\boldsymbol{a} \cdot \boldsymbol{\sigma})^k$ used in description (ii) above are linear combinations of the $Y$'s (see eq. (2.3)), namely $(\boldsymbol{a} \cdot \boldsymbol{\sigma})^k = \mu_{N,k}^{-1} \sum_{\{\alpha\}} a_{\alpha_1} \ldots a_{\alpha_k} Y_{N,k}^{\alpha_1 \ldots \alpha_k}(\boldsymbol{\sigma})$. The condition $\sum_{i=1}^N a_i^2 = 0$ ensures that the "Traces" in (2.3) make no contribution.

[9] For $N = 2$ the group is abelian, and the spin-$k$ representation for $k \geq 1$ decomposes into the *two* irreducible representations $e^{\pm i k \theta}$. However, if we consider $O(N)$ rather than $SO(N)$, then the representation is irreducible also for $N = 2$.

[10] The usual summation convention will be used in this paper from now on.

$$\times \left[\sigma^\alpha \sigma^\beta \sigma^\gamma - \frac{1}{N+2} \left(\delta^{\alpha\beta}\sigma^\gamma + \delta^{\alpha\gamma}\sigma^\beta + \delta^{\beta\gamma}\sigma^\alpha\right)\right] \quad (2.8)$$

$$\begin{aligned}Y_{N,4}^{\alpha\beta\gamma\delta}(\boldsymbol{\sigma}) &= \sqrt{\frac{N(N+2)(N+4)(N+6)}{24}} \\ &\times \left[\sigma^\alpha \sigma^\beta \sigma^\gamma \sigma^\delta - \frac{1}{N+4}\left(\delta^{\alpha\beta}\sigma^\gamma\sigma^\delta + \text{5 permutations}\right)\right. \\ &\left. + \frac{1}{(N+2)(N+4)}\left(\delta^{\alpha\beta}\delta^{\gamma\delta} + \delta^{\alpha\gamma}\delta^{\beta\delta} + \delta^{\alpha\delta}\delta^{\beta\gamma}\right)\right] \quad (2.9)\end{aligned}$$

[The general formula is given in equation (A.17).] We note that for $N = 3$ the $Y$'s are linear combinations of the usual spherical harmonics, and $\dim E_{3,k} = 2k + 1$. Similarly, for $N = 2$ the $Y$'s are linear combinations of $\cos k\theta$ and $\sin k\theta$ (or equivalently of $e^{\pm ik\theta}$), and $\dim E_{2,k} = 2$ for $k \geq 1$. For $N = 1$, $Y_{N,k}$ vanishes for $k \geq 2$, while $Y_{1,0} = 1$ and $Y_{1,1} = \sigma$.

The normalization $\mu_{N,k}$ is chosen so that the following orthogonality relation holds (see Appendix A.2):

$$\int d\Omega(\boldsymbol{\sigma}) \; Y_{N,k}^{\alpha_1\ldots\alpha_k}(\boldsymbol{\sigma}) \; Y_{N,l}^{\beta_1\ldots\beta_l}(\boldsymbol{\sigma}) = \delta_{kl} \; I_{N,k}^{\alpha_1\ldots\alpha_k;\beta_1\ldots\beta_k} , \quad (2.10)$$

where $I_{N,k}^{\alpha_1\ldots\alpha_k;\beta_1\ldots\beta_k}$ is the unique orthogonal projector onto the space of completely symmetric and traceless tensors of rank $k$, defined by the following properties (see Appendix A.3):

1. complete symmetry in the indices $\alpha$, and in the indices $\beta$

2. symmetry under the total exchange $\alpha_i \leftrightarrow \beta_i$ for all $i$

3. $\delta_{\alpha_i \alpha_j} \; I_{N,k}^{\alpha_1\ldots\alpha_k;\beta_1\ldots\beta_k} = 0$ for any $i \neq j$

4. $I_{N,k}^{\alpha_1\ldots\alpha_k;\beta_1\ldots\beta_k} \; T_{N,k}^{\beta_1\ldots\beta_k} = T_{N,k}^{\alpha_1\ldots\alpha_k}$ for any completely symmetric and traceless tensor $T_{N,k}$

As special cases of condition 4 we have

$$I_{N,k}^2 = I_{N,k} \quad (2.11)$$

and

$$I_{N,k}^{\alpha_1\ldots\alpha_k;\beta_1\ldots\beta_k} \; Y_{N,k}^{\beta_1\ldots\beta_k}(\boldsymbol{\sigma}) = Y_{N,k}^{\alpha_1\ldots\alpha_k}(\boldsymbol{\sigma}) . \quad (2.12)$$

For example we have:

$$I_{N,1}^{\alpha;\beta} = \delta^{\alpha\beta} \quad (2.13)$$

$$I_{N,2}^{\alpha_1\alpha_2;\beta_1\beta_2} = \frac{1}{2}\left(\delta^{\alpha_1\beta_1}\delta^{\alpha_2\beta_2} + \delta^{\alpha_1\beta_2}\delta^{\alpha_2\beta_1}\right) - \frac{1}{N}\delta^{\alpha_1\alpha_2}\delta^{\beta_1\beta_2} \quad (2.14)$$

$$I_{N,3}^{\alpha_1\alpha_2\alpha_3;\beta_1\beta_2\beta_3} = \frac{1}{6}\left[\delta^{\alpha_1\beta_1}\delta^{\alpha_2\beta_2}\delta^{\alpha_3\beta_3} + \delta^{\alpha_1\beta_2}\delta^{\alpha_2\beta_3}\delta^{\alpha_3\beta_1} + \delta^{\alpha_1\beta_3}\delta^{\alpha_2\beta_1}\delta^{\alpha_3\beta_2}\right.$$

$$+ \delta^{\alpha_1\beta_1}\delta^{\alpha_2\beta_3}\delta^{\alpha_3\beta_2} + \delta^{\alpha_1\beta_3}\delta^{\alpha_2\beta_2}\delta^{\alpha_3\beta_1} + \delta^{\alpha_1\beta_2}\delta^{\alpha_2\beta_1}\delta^{\alpha_3\beta_3}\Big]$$

$$- \frac{1}{3(N+2)} \Big[ \delta^{\alpha_1\alpha_2} \left( \delta^{\beta_1\beta_2}\delta^{\alpha_3\beta_3} + \delta^{\beta_1\beta_3}\delta^{\alpha_3\beta_2} + \delta^{\beta_2\beta_3}\delta^{\alpha_3\beta_1} \right)$$

$$+ \delta^{\alpha_1\alpha_3} \left( \delta^{\beta_1\beta_2}\delta^{\alpha_2\beta_3} + \delta^{\beta_1\beta_3}\delta^{\alpha_2\beta_2} + \delta^{\beta_2\beta_3}\delta^{\alpha_2\beta_1} \right)$$

$$+ \delta^{\alpha_2\alpha_3} \left( \delta^{\beta_1\beta_2}\delta^{\alpha_1\beta_3} + \delta^{\beta_1\beta_3}\delta^{\alpha_1\beta_2} + \delta^{\beta_2\beta_3}\delta^{\alpha_1\beta_1} \right) \Big]$$

(2.15)

[The general formula is given in equation (A.27).] The trace of this operator is given by [see (A.35)/(A.36)]

$$I_{N,k}^{\alpha_1\ldots\alpha_k;\alpha_1\ldots\alpha_k} = \mathcal{N}_{N,k} \equiv \dim E_{N,k}, \qquad (2.16)$$

as of course it must be. We remark that $Y_{N,k}(\boldsymbol{\sigma}) \cdot Y_{N,k}(\boldsymbol{\sigma}) \equiv Y_{N,k}^{\alpha_1\ldots\alpha_k}(\boldsymbol{\sigma})Y_{N,k}^{\alpha_1\ldots\alpha_k}(\boldsymbol{\sigma})$ is independent of $\boldsymbol{\sigma}$ [by $O(N)$ invariance], and hence

$$Y_{N,k}(\boldsymbol{\sigma}) \cdot Y_{N,k}(\boldsymbol{\sigma}) = \mathcal{N}_{N,k} \qquad (2.17)$$

by (2.10) and (2.16).

As stated in the theorem given at the beginning of this section, the hyperspherical harmonics are a complete set of functions on $L^2(S^{N-1})$. Thus any function $f(\boldsymbol{\sigma})$ can be expanded as

$$f(\boldsymbol{\sigma}) = \sum_{k=0}^{\infty} \widetilde{f}_k^{\alpha_1\ldots\alpha_k} Y_{N,k}^{\alpha_1\ldots\alpha_k}(\boldsymbol{\sigma}) \qquad (2.18)$$

where

$$\widetilde{f}_k^{\alpha_1\ldots\alpha_k} = \int d\Omega(\boldsymbol{\tau}) f(\boldsymbol{\tau}) Y_{N,k}^{\alpha_1\ldots\alpha_k}(\boldsymbol{\tau}). \qquad (2.19)$$

For smooth functions this expansion converges very fast. Indeed, if $f(\boldsymbol{\sigma})$ is infinitely differentiable, then, for $k \to \infty$, the coefficients of the expansion go to zero faster than any inverse power of $k$ (see Appendix A.4).

The completeness of the hyperspherical harmonics can be expressed through the relation[11]

$$\sum_{k=0}^{\infty} Y_{N,k}^{\alpha_1\ldots\alpha_k}(\boldsymbol{\sigma}) Y_{N,k}^{\alpha_1\ldots\alpha_k}(\boldsymbol{\tau}) = \delta(\boldsymbol{\sigma}, \boldsymbol{\tau}) \qquad (2.20)$$

where the $\delta$-function is defined with respect to the measure $d\Omega(\boldsymbol{\sigma})$.

Finally, let us consider an invariant function of *two* "spins" $\boldsymbol{\sigma}, \boldsymbol{\tau} \in S^{N-1}$, i.e. a function of $\boldsymbol{\sigma}\cdot\boldsymbol{\tau}$.[12] We want now to compute its expansion in terms of hyperspherical harmonics. Using Schur's lemma (see Appendix A.4) we can write

$$f(\boldsymbol{\sigma}\cdot\boldsymbol{\tau}) = \sum_{k=0}^{\infty} F_{N,k} Y_{N,k}(\boldsymbol{\sigma}) \cdot Y_{N,k}(\boldsymbol{\tau}). \qquad (2.21)$$

---

[11] Note that the normalization here follows directly from the one defined for (2.10).

[12] For $N = 2$ there are functions of $\boldsymbol{\sigma}, \boldsymbol{\tau}$ which are $SO(2)$-invariant [but not $O(2)$-invariant] and are *not* functions of $\boldsymbol{\sigma}\cdot\boldsymbol{\tau}$: namely, they can depend also on $\boldsymbol{\sigma}\times\boldsymbol{\tau}$. We are not interested in such functions.

We can drop the "Traces" terms of either one of the $Y$'s in the scalar product above, since the other $Y$ is traceless. Also, since the scalar product is rotationally invariant, we can rotate $\boldsymbol{\sigma}$ to $\mathbf{w} \equiv (1, 0, \ldots, 0)$ and correspondingly rotate $\boldsymbol{\tau}$ to some $\boldsymbol{\rho}$ with $\boldsymbol{\sigma} \cdot \boldsymbol{\tau} = \mathbf{w} \cdot \boldsymbol{\rho} = \rho^1$. In this way we obtain

$$\begin{aligned} Y_{N,k}(\boldsymbol{\sigma}) \cdot Y_{N,k}(\boldsymbol{\tau}) &= Y_{N,k}(\mathbf{w}) \cdot Y_{N,k}(\boldsymbol{\rho}) \\ &= \mu_{N,k}\, w^{\alpha_1} \ldots w^{\alpha_k}\, Y_{N,k}^{\alpha_1 \ldots \alpha_k}(\boldsymbol{\rho}) = \mu_{N,k}\, Y_{N,k}^{1\ldots 1}(\boldsymbol{\rho}) \,. \end{aligned} \qquad (2.22)$$

Now $Y_{N,k}^{1\ldots 1}(\boldsymbol{\rho})$ can be expressed in terms of *Gegenbauer polynomials*[13] (this corresponds to the relation between $Y_{l0}$ and Legendre polynomials for the usual spherical harmonics) as[14] (see Appendix A.2)

$$Y_{N,k}^{1\ldots 1}(\boldsymbol{\rho}) = \frac{\mathcal{N}_{N,k}}{\mu_{N,k}} \frac{C_k^{N/2-1}(\rho^1)}{C_k^{N/2-1}(1)} \qquad (2.23)$$

and therefore

$$Y_{N,k}(\boldsymbol{\sigma}) \cdot Y_{N,k}(\boldsymbol{\tau}) = \mathcal{N}_{N,k} \frac{C_k^{N/2-1}(\boldsymbol{\sigma} \cdot \boldsymbol{\tau})}{C_k^{N/2-1}(1)} \,. \qquad (2.24)$$

In particular, for $\mathbf{w} \equiv (1, 0, \ldots, 0)$, we have

$$Y_{N,k}^{1\ldots 1}(\mathbf{w}) = \frac{\mathcal{N}_{N,k}}{\mu_{N,k}} \,. \qquad (2.25)$$

From equation (2.21), using the orthogonality relations, the rotational invariance of the measure, equation (2.17) and (2.24), we get

$$F_{N,k} = \int d\Omega(\boldsymbol{\rho})\, f(\rho^1)\, \frac{C_k^{N/2-1}(\rho^1)}{C_k^{N/2-1}(1)} \,. \qquad (2.26)$$

Now the integrand depends only on $\rho^1$ and we can integrate out the other coordinates. We finally get

$$F_{N,k} = \frac{\mathcal{S}_{N-1}}{\mathcal{S}_N} \int_{-1}^{1} dt\, (1-t^2)^{(N-3)/2}\, f(t)\, \frac{C_k^{N/2-1}(t)}{C_k^{N/2-1}(1)} \,, \qquad (2.27)$$

---

[13] See [31], pp. 1029–1031.

[14] For $N = 2$ this relation is singular, since $C_k^0(x) = 0$. This singularity is due simply to the normalization convention of the Gegenbauer polynomials, and indeed the limit $N \to 2$ is well-defined. The result is simply

$$\lim_{N \to 2} \frac{C_k^{N/2-1}(\cos\theta)}{C_k^{N/2-1}(1)} = \cos k\theta = \frac{T_k(\cos\theta)}{T_k(1)} \,.$$

where $T_k(\theta)$ are the Chebyshev polynomials of the first kind (see [31], formulae 8.934.4 (p. 1030) and 8.940.1 (p. 1032)).

where $\mathcal{S}_N$ is the surface area of the $N$-dimensional unit sphere:

$$\mathcal{S}_N = \frac{2\pi^{N/2}}{\Gamma(N/2)} \ . \tag{2.28}$$

From the general properties of the hyperspherical harmonics we can derive the following properties of the coefficients $F_{N,k}$ (for the proofs of properties 1 and 2, see Appendix A.4):

1. If $f(t)$ is positive[15] for $t \in [-1, 1]$, then $|F_{N,k}| < F_{N,0}$ for all $k \neq 0$.

2. If $f(t)$ is smooth (i.e. $C^\infty$), then $\lim_{k \to \infty} k^n F_{N,k} = 0$ for every $n$.

3. If $f(t) = t^l$, then the integral in (2.27) can be performed explicitly[16] and the coefficients $F_{N,k}$ are given by

$$F_{N,k}^{(l)} = \begin{cases} \dfrac{\Gamma\left(\frac{N}{2}\right) \Gamma(l+1)}{2^l \, \Gamma\left(\frac{N+k+l}{2}\right) \Gamma\left(\frac{l-k}{2} + 1\right)} & \text{if } k+l \text{ is even and } k \leq l \\ 0 & \text{otherwise} \end{cases} \tag{2.29}$$

and are, in particular, always nonnegative. It immediately follows that for a generic function

$$f(t) = \sum_{l=0}^{\infty} f_l \, t^l \ , \tag{2.30}$$

the coefficients $F_{N,k}$ are given by

$$F_{N,k} = \sum_{l=k}^{\infty} f_l \, F_{N,k}^{(l)} \ . \tag{2.31}$$

Therefore, if all the coefficients $f_l$ are nonnegative, then so are the $F_{N,k}$.

In particular, using (2.27) or (2.31) it is possible to compute the coefficients $F_{N,k}$ for the functions $\exp[J(\boldsymbol{\sigma} \cdot \boldsymbol{\tau})]$ and $\exp\left[\frac{J}{2}(\boldsymbol{\sigma} \cdot \boldsymbol{\tau})^2\right]$. In the first case we obtain

$$F_{N,k} = \Gamma\left(\frac{N}{2}\right) \left(\frac{J}{2}\right)^{1-\frac{N}{2}} I_{\frac{N}{2}+k-1}(J) \tag{2.32}$$

where $I_\nu$ is the *modified Bessel function*[17]; in the second case the integration gives

$$F_{N,k} = \begin{cases} \dfrac{\Gamma\left(\frac{N}{2}\right) \Gamma\left(\frac{k+1}{2}\right)}{\sqrt{\pi} \, \Gamma\left(\frac{N}{2}+k\right)} \left(\frac{J}{2}\right)^{k/2} {}_1F_1\left(\frac{k+1}{2} \, ; \, k+\frac{N}{2} \, ; \, \frac{J}{2}\right) & \text{for even } k \\ 0 & \text{for odd } k \end{cases} \tag{2.33}$$

---

[15] More precisely, it suffices that $f$ be *nonnegative* and *not almost-everywhere-vanishing*.

[16] See [31], formula 7.311.2, p. 826.

[17] In particular, for $N = 1$ (the Ising model) we get $F_{1,0} = \cosh J$ and $F_{1,1} = \sinh J$, and therefore the formulae in the following sections will be written in terms of the usual high-temperature expansion parameter $v_{1,1} \equiv F_{1,1}/F_{1,0} = \tanh J$.

where $_1F_1$ is the *confluent (degenerate) hypergeometric function*.[18] These two expansions will be used in the next section.

Let us now compute the Clebsch-Gordan coefficients. In general we can write

$$Y_{N,k}^{\alpha_1...\alpha_k}(\boldsymbol{\sigma}) \, Y_{N,l}^{\beta_1...\beta_l}(\boldsymbol{\sigma}) \;=\; \sum_m \mathcal{C}_{N;\,k,l,m}^{\alpha_1...\alpha_k;\beta_1...\beta_l;\gamma_1...\gamma_m} \; Y_{N,m}^{\gamma_1...\gamma_m}(\boldsymbol{\sigma}) \qquad (2.34)$$

Using the orthogonality relations (2.10) we obtain

$$\mathcal{C}_{N;\,k,l,m}^{\alpha_1...\alpha_k;\beta_1...\beta_l;\gamma_1...\gamma_m} \;=\; \int d\Omega(\boldsymbol{\sigma}) \; Y_{N,k}^{\alpha_1...\alpha_k}(\boldsymbol{\sigma}) Y_{N,l}^{\beta_1...\beta_l}(\boldsymbol{\sigma}) Y_{N,m}^{\gamma_1...\gamma_m}(\boldsymbol{\sigma}) \;. \qquad (2.35)$$

This integral can be computed explicitly. We get (see Appendix A.5)

$$\begin{aligned}\mathcal{C}_{N;\,k,l,m}^{\alpha_1...\alpha_k;\beta_1...\beta_l;\gamma_1...\gamma_m} \;&=\; \frac{\mu_{N,k}\,\mu_{N,l}\,\mu_{N,m}}{\mu_{N,k+j}^2\,(k+j)!}\,\frac{k!\,l!\,m!}{i!\,j!\,h!} \\ &\quad\times\; I_{N,k}^{\alpha_1...\alpha_k;a_1...a_i b_1...b_h} \; I_{N,l}^{\beta_1...\beta_l;b_1...b_h c_1...c_j} \; I_{N,m}^{\gamma_1...\gamma_m;c_1...c_j a_1...a_i} \end{aligned} \qquad (2.36)$$

if $|l-k| \leq m \leq l+k$ and $k+l+m$ is even, with $i=(m+k-l)/2$, $j=(m+l-k)/2$, $h=(l+k-m)/2$, and vanishes otherwise. (Of course we are considering $k,l,m \geq 0$.)

In the following we will be interested in the scalar quantity

$$\mathcal{C}_{N;\,k,l,m}^2 \;=\; \mathcal{C}_{N;\,k,l,m} \cdot \mathcal{C}_{N;\,k,l,m} \;. \qquad (2.37)$$

The general formula is reported in Appendix A.5 [see (A.63)]. A particularly simple case is $m=l+k$:

$$\mathcal{C}_{N;\,k,l,l+k}^2 \;=\; \mathcal{N}_{N,l+k}\,\frac{\mu_{N,k}^2\,\mu_{N,l}^2}{\mu_{N,l+k}^2}\;, \qquad (2.38)$$

which can be obtained directly from (2.36), using the properties of the $I_{N,k}$ tensor and (2.16). If $k=1$ this gives

$$\mathcal{C}_{N;\,1,l,l+1}^2 \;=\; N\binom{N+l-2}{l}\;. \qquad (2.39)$$

It follows immediately from (2.35)–(2.37) that $\mathcal{C}_{N;\,k,l,m}^2$ is symmetric in the variables $k$, $l$ and $m$. This implies, for example, that

$$\mathcal{C}_{N;\,k,l,l-k}^2 \;=\; \mathcal{C}_{N;\,k,l-k,l}^2 \;=\; \mathcal{N}_{N,l}\,\frac{\mu_{N,k}^2\,\mu_{N,l-k}^2}{\mu_{N,l}^2} \qquad (2.40)$$

[from (2.38)]. It also implies that, for $k$ fixed, it suffices to find $\mathcal{C}_{N;\,k,l,m}^2$ for $l \leq m \leq l+k$. Thus, the two coefficients needed (for each $l$) for the case $k=1$ are obtained from (2.39). For the case $k=2$, which will be used later on, we have from (2.38) that

$$\mathcal{C}_{N;\,2,l,l+2}^2 \;=\; \frac{N(N+2)(N+l-1)}{2(N+2l)}\binom{N+l-2}{l}\;, \qquad (2.41)$$

---
[18]See [31], pp. 1058–1059.

and from (A.63) we obtain[19]

$$\mathcal{C}^2_{N;\,2,l,l} = \mathcal{N}_{N,l}\,\frac{l(N+2)(N-2)(N+l-2)}{(N+2l)(N+2l-4)}\,. \tag{2.42}$$

Using the completeness relation (2.20), formula (2.35) and (2.17) it is easy to verify the identity

$$\sum_{k=0}^{\infty} \mathcal{C}^2_{N;\,k,l,m} = \mathcal{N}_{N,l}\,\mathcal{N}_{N,m}\,. \tag{2.43}$$

# 3 Exact Solution for a Generic $h$

## 3.1 Finite Volume

In this section we want to discuss the most general $O(N)$-invariant $\sigma$-model taking values in $S^{N-1}$, with nearest-neighbor interactions, defined on a one-dimensional lattice with $L$ sites and periodic boundary conditions. We consider a Hamiltonian of the form

$$\mathcal{H}(\{\boldsymbol{\sigma}\}) = -\sum_{x=0}^{L-1} h(\boldsymbol{\sigma}_x \cdot \boldsymbol{\sigma}_{x+1}) \tag{3.1}$$

with $\boldsymbol{\sigma}_L \equiv \boldsymbol{\sigma}_0$. Interesting special cases are the *N-vector model*

$$h(\boldsymbol{\sigma}_x \cdot \boldsymbol{\sigma}_y) = J\,\boldsymbol{\sigma}_x \cdot \boldsymbol{\sigma}_y \tag{3.2}$$

and the $RP^{N-1}$ *model*

$$h(\boldsymbol{\sigma}_x \cdot \boldsymbol{\sigma}_y) = \frac{J}{2}(\boldsymbol{\sigma}_x \cdot \boldsymbol{\sigma}_y)^2\,. \tag{3.3}$$

The coefficients $F_{N,k}$ have already been evaluated for both of these models [see (2.32) and (2.33)].

We want to evaluate the following quantities:

- Partition function:

$$Z_N(h;L) = \int \mathcal{D}\boldsymbol{\sigma}\,\prod_{x=0}^{L-1} e^{h(\boldsymbol{\sigma}_x \cdot \boldsymbol{\sigma}_{x+1})} \tag{3.4}$$

- Spin-$k$ two-point function ($k = 1, 2, \ldots$):

$$G_{N,k}(x,h;L) = \frac{1}{\mathcal{N}_{N,k}}\,\langle Y_{N,k}(\boldsymbol{\sigma}_0)\cdot Y_{N,k}(\boldsymbol{\sigma}_x)\rangle_L \tag{3.5}$$

$$\widetilde{G}_{N,k}(p,h;L) = \sum_{x=0}^{L-1} e^{ipx}\,G_{N,k}(x,h;L) \tag{3.6}$$

---

[19]Formula (2.42) is potentially ambiguous if $N+2l-4 = 0$, which can happen for $(N=2, l=1)$ and $(N=4, l=0)$. In fact $\mathcal{C}^2_{N;\,2,1,1} = N(N-1)$ and $\mathcal{C}^2_{N;\,4,0,0} = 0$; these results can be obtained by interpreting (2.42) as an analytic function of $N$ for each fixed $l$.

where $0 \leq x < L$, and $p$ is an integer multiple of $2\pi/L$. Note that the normalization $\mathcal{N}_{N,k}$ [defined in (2.16)] ensures that $G_{N,k}(0, h; L) = 1$.[20]

- Susceptibility (= two-point function at zero momentum):

$$\chi_{N,k}(h; L) = \widetilde{G}_{N,k}(0, h; L) \tag{3.7}$$

- Two-point function at the smallest nonzero momentum:

$$\mathcal{F}_{N,k}(h; L) = \widetilde{G}_{N,k}\left(\pm\frac{2\pi}{L}, h; L\right) \tag{3.8}$$

- Second-moment correlation length:

$$\xi_{N,k}^{(2nd)}(h; L) = \begin{cases} \dfrac{([\chi_{N,k}(h; L)/\mathcal{F}_{N,k}(h; L)] - 1)^{1/2}}{2\sin(\pi/L)} & \text{if } \chi_{N,k} \geq \mathcal{F}_{N,k} \\ \text{undefined} & \text{otherwise} \end{cases} \tag{3.9}$$

In all of these formulae, we have used the abbreviations

$$\mathcal{D}\boldsymbol{\sigma} \equiv \prod_{x=0}^{L-1} d\Omega(\boldsymbol{\sigma}_x) \tag{3.10}$$

$$\langle f(\{\boldsymbol{\sigma}\})\rangle_L \equiv \frac{1}{Z_N(h; L)} \int \mathcal{D}\boldsymbol{\sigma}\ f(\{\boldsymbol{\sigma}\})\ e^{-\mathcal{H}(\{\boldsymbol{\sigma}\})} \tag{3.11}$$

To compute all these quantities, we expand $e^{-\mathcal{H}}$ in terms of the hyperspherical harmonics $Y_{N,k}$, as described in the previous section:

$$\exp[h(\boldsymbol{\sigma}_x \cdot \boldsymbol{\sigma}_y)] = \sum_{k=0}^{\infty} F_{N,k}(h)\ Y_{N,k}(\boldsymbol{\sigma}_x) \cdot Y_{N,k}(\boldsymbol{\sigma}_y)\ . \tag{3.12}$$

The integration over $\mathcal{D}\boldsymbol{\sigma}$ is then immediate using the orthogonality relations (2.10) and the integral (2.35). In this way (using also the symmetry of $\mathcal{C}^2_{N;\,k,l,m}$ in the indices $l$ and $m$) we obtain:

$$Z_N(h; L) = F_{N,0}(h)^L \sum_{l=0}^{\infty} \mathcal{N}_{N,l}\ v_{N,l}(h)^L \tag{3.13}$$

---

[20]Since we are using periodic boundary conditions, $G_{N,k}(x, h; L) = G_{N,k}(L - x, h; L)$ and therefore $\widetilde{G}_{N,k}(p, h; L)$ is real. Indeed, $\widetilde{G}_{N,k}(p, h; L) \geq 0$ for all $p$ because equation (3.6) can be written as

$$\widetilde{G}_{N,k}(p, h; L) = \frac{1}{L}\left\langle \left|\sum_{x=0}^{L-1} e^{ipx} Y_{N,k}(\boldsymbol{\sigma}_x)\right|^2 \right\rangle$$

by using translational invariance. On the other hand, $G_{N,k}(x, h; L)$ may in some cases be negative ("antiferromagnetism").

$$G_{N,k}(x, h; L) = \frac{1}{\mathcal{N}_{N,k}} \frac{F_{N,0}(h)^L}{Z_N(h;L)} \sum_{l,m=0}^{\infty} \mathcal{C}_{N;\,k,l,m}^2 \, v_{N,m}(h)^x \, v_{N,l}(h)^{L-x} \quad (3.14)$$

$$\widetilde{G}_{N,k}(p, h; L) = \frac{1}{\mathcal{N}_{N,k}} \frac{F_{N,0}(h)^L}{Z_N(h;L)} \sum_{l,m=0}^{\infty} \left\{ \mathcal{C}_{N;\,k,l,m}^2 \, v_{N,l}(h)^L \right.$$
$$\left. \times \frac{v_{N,l}(h)^2 - v_{N,m}(h)^2}{v_{N,l}(h)^2 - 2(\cos p)\, v_{N,l}(h)v_{N,m}(h) + v_{N,m}(h)^2} \right\} \quad (3.15)$$

where $0 \leq x \leq L-1$; here we have defined the normalized expansion coefficients

$$v_{N,k}(h) = \frac{F_{N,k}(h)}{F_{N,0}(h)}, \quad (3.16)$$

which will play a central role in the subsequent analysis.[21] Notice that, because of the properties of the coefficients $F_{N,k}$ discussed in Section 2, we have $|v_{N,k}(h)| < 1$ for all $k \neq 0$.[22] Moreover, all these series converge very fast (at least if $h$ is smooth): this is because, for $k \to \infty$, $v_{N,k}(h)$ goes to zero faster than any power of $k$, while $\mathcal{N}_{N,k} \sim k^{N-2}$ and [see (2.43)]

$$\mathcal{C}_{N;\,k,l,m}^2 \leq \min(\mathcal{N}_{N,k}\mathcal{N}_{N,l},\ \mathcal{N}_{N,k}\mathcal{N}_{N,m},\ \mathcal{N}_{N,m}\mathcal{N}_{N,l}). \quad (3.17)$$

Finally, using (2.43), it is trivial to check that $G_{N,k}(0, h; L) = 1$ in (3.14).

## 3.2 Infinite Volume

We want now to consider the infinite-volume limit $L \to \infty$ in the expressions from Section 3.1, keeping the parameters of $h$ fixed. Since $|v_{N,k}(h)| < 1$ for all $k \neq 0$, $v_{N,k}(h)^L$ goes to zero for $L \to \infty$ unless $k = 0$. Thus in (3.13), (3.14) and (3.15) only the term with $l = 0$ survives in the infinite-volume limit. Since

---

[21]The summand in braces in (3.15) is potentially ambiguous in two cases:

(i) $p = 0$ and $v_{N,l} = v_{N,m}$;

(ii) $p = \pi$ and $v_{N,l} = -v_{N,m}$.

In these cases the correct summand is $\mathcal{C}_{N;\,k,l,m}^2 L\, v_{N,l}(h)^L$, as can be seen by going back to (3.14) and performing the sum over $x$. [The same result can be obtained formally by symmetrizing the summand in $l$ and $m$ (using $\mathcal{C}_{N;\,k,l,m}^2 = \mathcal{C}_{N;\,k,m,l}^2$), i.e. replacing $v_{N,l}^L$ by $(v_{N,l}^L - v_{N,m}^L)/2$, and then treating $v_{N,l}$ and $v_{N,m}$ as independent variables for which one can take the limit $v_{N,m} \to \pm v_{N,l}$.]

[22]For the $RP^{N-1}$ model, or more generally if $h(\boldsymbol{\sigma}_x \cdot \boldsymbol{\sigma}_y)$ is an even function, all the coefficients $F_{N,l}(h)$ [and the corresponding $v_{N,l}(h)$] with $l$ *odd* are equal to zero (by symmetry). Therefore, in the above formulae, only even values of $l$ and $m$ can appear in the sums (except, of course, in $G_{N,k}(x, h; L)$ for $x = 0$). From this and the properties of the quantities $\mathcal{C}_{N;\,k,l,m}^2$, i.e. that they are nonzero only if $k + l + m$ is even, it follows that for $k$ *odd* the spin-$k$ two-point function vanishes for all $x \neq 0$. Of course, this follows equivalently from the $Z_2$-gauge-invariance of the model when $h$ is an even function.

$C^2_{N;\,k,0,m} = \delta_{km}\,\mathcal{N}_{N,k}$, we get the well-known results[23] [29,32]

$$Z_N(h;L) = F_{N,0}(h)^L \left[1 + O(e^{-\mu L})\right] \tag{3.18}$$

$$G_{N,k}(x,h;L) = v_{N,k}(h)^{|x|} \left[1 + O(e^{-\mu L})\right] \tag{3.19}$$

$$\widetilde{G}_{N,k}(p,h;L) = \frac{1 - v_{N,k}(h)^2}{1 - 2(\cos p)v_{N,k}(h) + v_{N,k}(h)^2} \left[1 + O(e^{-\mu L})\right] \tag{3.20}$$

where

$$\mu = -\min_{k \neq 0} \log |v_{N,k}(h)| \;. \tag{3.21}$$

In particular we obtain

$$\chi_{N,k}(h;\infty) = \frac{1 + v_{N,k}(h)}{1 - v_{N,k}(h)} \tag{3.22}$$

and

$$\xi^{(2nd)}_{N,k}(h;\infty) = \begin{cases} \dfrac{v_{N,k}(h)^{1/2}}{1 - v_{N,k}(h)} & \text{if } v_{N,k} \geq 0 \\ \text{undefined} & \text{if } v_{N,k} < 0 \end{cases} \tag{3.23}$$

Let us notice that in infinite volume the correlation functions are simple exponentials. In fact, if we define the *masses* $m_{N,k}(h)$ for $k = 1, 2, \ldots$ by

$$m_{N,k}(h) = \begin{cases} -\log v_{N,k}(h) & \text{for } 0 \leq v_{N,k} < 1 \\ \text{undefined} & \text{for } -1 < v_{N,k} < 0 \end{cases} \tag{3.24}$$

then, in the usual case[24] in which $v_{N,k} > 0$, the correlation functions are

$$G_{N,k}(x,h;\infty) = e^{-m_{N,k}|x|} \;. \tag{3.25}$$

We can also define the *exponential correlation length* by

$$\xi^{(exp)}_{N,k}(h;\infty) = \lim_{x \to \pm\infty} \frac{-|x|}{\log G_{N,k}(x,h;\infty)} = \frac{1}{m_{N,k}(h)} \;. \tag{3.26}$$

## 4 A One-Parameter Family of Hamiltonians

In this section we want to study the continuum limits and finite-size-scaling functions in a one-parameter family of interactions of the form

$$h(\boldsymbol{\sigma} \cdot \boldsymbol{\tau}) = J\widetilde{h}(\boldsymbol{\sigma} \cdot \boldsymbol{\tau}) \;, \tag{4.1}$$

---

[23] The formulae in Section 3.1 are written for $x \geq 0$. By translation invariance, we obviously have $G_{N,k}(x) = G_{N,k}(-x)$. Therefore, we can obtain formulae valid for all $x$ by systematically replacing $x$ by $|x|$; we have done that here.

[24] As we will see in Section 4.1, the case of negative $v_{N,k}$ does not give rise to a valid continuum limit.

where $\widetilde{h}$ is some fixed function. Therefore, $F_{N,k}$, $v_{N,k}$ and all the quantities introduced in the previous sections are now functions of $J$. As $\widetilde{h}$ is arbitrary it suffices to consider the case $J > 0$ only. Since we are in one dimension, there are no critical points at finite $J$; the only way of obtaining a continuum limit is to take $J \to +\infty$. We will do this by obtaining an asymptotic expansion of the coefficients $F_{N,k}(J)$ for large $J$. Using the general formula (2.27) with $f(t) = \exp[J\widetilde{h}(t)]$, the problem reduces to expanding the integrand around the absolute maxima of $\widetilde{h}(t)$ in the interval $[-1, 1]$.

In Section 4.1 we will study the continuum limit in infinite volume. In Sections 4.2 and 4.3 we will study the finite-size-scaling limit and the corrections to it.

The discussion in Section 4.1 of the possible continuum limits will be restricted to the case $N \geq 3$, since $N = 2$ displays different properties (related to the different topological structures of the sphere for $N \geq 3$ and $N = 2$, and to the fact that the only nontrivial normal subgroup of $O(N)$ for $N \geq 3$ is $\{\pm I\}$, while for $N = 2$ there are many others).[25] Although for $N = 2$ the analysis of possible continuum limits is not *complete*, it is nevertheless valid for the limits included, and so are the finite-size-scaling functions and their corrections.[26]

## 4.1 Continuum Limits and Universality Classes for $N \geq 3$

### 4.1.1 Generalities on Continuum Limits

Consider a sequence $\langle \cdot \rangle^{(n)}$ of infinite-volume lattice models. A continuum limit is defined by choosing length rescaling factors $\Xi^{(n)} \to \infty$ and field-strength rescaling factors $\zeta_{N,k}^{(n)}$ such that the limits[27]

$$G_{N,k}^{(cont)}(\overline{\mathbf{x}}) = \lim_{n \to \infty} \zeta_{N,k}^{(n)} G_{N,k}^{(n)}\left(\Xi^{(n)} \overline{\mathbf{x}}\right) \tag{4.2}$$

$$\widetilde{G}_{N,k}^{(cont)}(\overline{\mathbf{p}}) = \lim_{n \to \infty} \zeta_{N,k}^{(n)} \Xi^{(n)-d} \widetilde{G}_{N,k}^{(n)}\left(\Xi^{(n)-1} \overline{\mathbf{p}}\right) \tag{4.3}$$

exist (in the sense of distributions), where $d$ is the spatial dimension. (For simplicity we are considering only the two-point correlation functions.) In other words, a continuum distance of $\overline{x}$ centimeters corresponds to $x \equiv \Xi^{(n)}\overline{x}$ lattice spacings; and conversely, one lattice spacing corresponds to $\Xi^{(n)-1}$ centimeters, which tends to zero in the limit.

In our case of a $d = 1$ nearest-neighbor model, the correlation functions are pure exponentials [see (3.19)]; the only parameter is the mass parameter $v_{N,k}^{(n)}$. It

---

[25] In particular, the discussion following (4.36) does not apply for $N = 2$.

[26] The case $N = 1$ is even more trivial, as the *only* possible function $\widetilde{h}$ is $\widetilde{h}(t) = t$.

[27] We use the Fourier-transform convention

$$\widetilde{G}_{N,k}^{(cont)}(\overline{\mathbf{p}}) \equiv \int d^d\overline{\mathbf{x}} \; e^{i\overline{\mathbf{p}} \cdot \overline{\mathbf{x}}} \; G_{N,k}^{(cont)}(\overline{\mathbf{x}}) \; .$$

is easiest to work in $p$-space: for any fixed continuum momentum $\overline{p}$, the lattice momentum $p \equiv \Xi^{(n)^{-1}} \overline{p}$ tends to zero as $n \to \infty$, so we can approximate

$$\cos p \approx 1 - p^2/2 = 1 - \Xi^{(n)^{-2}} \overline{p}^2/2 \,. \tag{4.4}$$

Thus, the denominator in (3.20) is

$$(1 - v_{N,k}^{(n)})^2 + \Xi^{(n)^{-2}} \overline{p}^2 \, v_{N,k}^{(n)} \,. \tag{4.5}$$

(Note that $\Xi^{(n)^{-2}} \to 0$.)

Now consider a ratio of the correlation function for two different values $\overline{p}$, $\overline{p}'$. If $1 - v_{N,k}^{(n)}$ does *not* go to zero as $n \to \infty$ at least as fast as $\Xi^{(n)^{-1}}$, then the ratio $\widetilde{G}_{N,k}^{(cont)}(\overline{p}')/\widetilde{G}_{N,k}^{(cont)}(\overline{p})$ is 1, i.e. $\widetilde{G}_{N,k}^{(cont)}(\overline{p})$ is independent of $\overline{p}$. This is a physically trivial theory (white noise). On the other hand, if $1 - v_{N,k}^{(n)}$ goes to zero *faster* than $\Xi^{(n)^{-1}}$, then the limit (if any) will be $\text{const}/\overline{p}^2$, i.e. a massless free field, which is ill-defined in dimension $d = 1$. Therefore, a sensible continuum limit can be obtained only when the product $(1 - v_{N,k}^{(n)}) \Xi^{(n)}$ tends to a nonzero finite constant (which is of course $k$-dependent); and this limiting constant is in fact the mass $m_{N,k}^{(cont)}$ of the theory. Moreover, in dimension $d = 1$ it easily follows from (3.19) that $\zeta_{N,k}^{(n)}$ should likewise tend to a nonzero finite ($k$-dependent) constant $\zeta_{N,k}^{(cont)}$; the continuum correlation function is then a massive free field

$$\widetilde{G}_{N,k}^{(cont)}(\overline{p}) = \frac{\mathcal{Z}_{N,k}^{(cont)}}{\overline{p}^2 + m_{N,k}^{(cont)^2}} \tag{4.6}$$

with mass

$$m_{N,k}^{(cont)} \equiv \lim_{n \to \infty} \Xi^{(n)} m_{N,k}^{(n)} = \lim_{n \to \infty} \Xi^{(n)} (1 - v_{N,k}^{(n)}) \tag{4.7}$$

and field-strength normalization

$$\mathcal{Z}_{N,k}^{(cont)} \equiv 2 \, \zeta_{N,k}^{(cont)} \, m_{N,k}^{(cont)} \,. \tag{4.8}$$

Going back to $x$-space, we have

$$G_{N,k}^{(cont)}(\overline{x}) = \zeta_{N,k}^{(cont)} \, \exp\left[ -m_{N,k}^{(cont)} \, |\overline{x}| \right] \,. \tag{4.9}$$

In summary, continuum limits can be obtained from sequences of lattice theories in which $v_{N,k}^{(n)} \to 1$ (i.e. $m_{N,k}^{(n)} \to 0$), and only from such sequences. In particular, continuum limits in this sense *cannot* be obtained from sequences of theories in which $v_{N,k}^{(n)} \to -1$, i.e. antiferromagnetic models with slow decay of correlations. As can be seen from (4.9), a continuum limit is uniquely defined by the limiting masses $m_{N,k}^{(cont)}$ and the limiting normalizations $\zeta_{N,k}^{(cont)}$. Moreover, we shall consider two continuum theories which differ only by rescalings of $\overline{x}$ and the field strengths to be essentially identical. We therefore label the different universality classes by the *limiting mass ratios*, defined as[28]

$$\mathcal{R}_{N,k} \equiv \frac{m_{N,2}^{(cont)}}{m_{N,k}^{(cont)}} = \lim_{n\to\infty} \frac{m_{N,2}^{(n)}}{m_{N,k}^{(n)}} \,. \tag{4.10}$$

In the case at hand [Hamiltonians of the family (4.1)], we are considering a sequence of theories $h = J\widetilde{h}$ parametrized by $J$ (which plays the role of $n$). As already mentioned, the only possibility for having $v_{N,k}(J) \to 1$ is to let $J \to +\infty$. In the next subsection we will perform an asymptotic expansion of $v_{N,k}(J)$ for large $J$, and we will typically find a behavior of the form[29]

$$v_{N,k}(J) = 1 - \widetilde{a}_{N,k}\, \Lambda(J) + o(\Lambda(J))\,, \tag{4.11}$$

where the mass scale $\Lambda(J)$ and the coefficients $\widetilde{a}_{N,k}$ will be computed in each case.[30] In this situation, $\Xi(J)$ should clearly be taken to be proportional to $\Lambda(J)^{-1}$, and the continuum masses will be

$$m_{N,k}^{(cont)} = \widetilde{a}_{N,k}\, \lim_{J\to\infty} \Lambda(J)\, \Xi(J)\,. \tag{4.12}$$

**Remark**: For some choices of $\widetilde{h}$ we will find that the mass parameters $v_{N,k}(J)$ behave *differently* according to whether $k$ is even or odd. In such a case we shall take $\Lambda(J)$ to be of the order of the *smallest* mass in the theory — which, it turns out, is always in the even sector — and we shall write (4.11) only for $k$ even. We shall then take $\Xi(J)$ proportional to $\Lambda(J)^{-1}$, and obtain a good continuum limit *in the even sector*. Of course, in the odd sector we have simply white noise ($m_{N,k}^{(cont)} = +\infty$).

### 4.1.2 Two Simple Cases; $N$-Vector and $RP^{N-1}$ Universality Classes

Before considering the general case of one-parameter Hamiltonians, let us discuss two simple cases of Hamiltonians which generalize, respectively, the $N$-vector model and the $RP^{N-1}$ model:

**First simple case**: $t = +1$ is the *only* absolute maximum of $\widetilde{h}(t)$, and $\widetilde{h}'(1) > 0$. [This is a subset of what will later be called the Hamiltonians of *Type I*.]

---

[28] We choose $m_{N,2}$ in the numerator for reasons that will become clear later.

[29] In Section 4.3 we will assume an expansion to the next order [see (4.90)], which will be used to compute the corrections to finite-size scaling. In Section 4.1.2 we will explicitly compute such an expansion for two simple Hamiltonians (the *first case* and those belonging to the *second case* with $\widetilde{h}$ even): see (4.14) ff.

[30] Obviously there is some arbitrariness in the definition of $\Lambda(J)$: if $\overline{\Lambda}(J)$ is a function satisfying $C \equiv \lim_{J\to\infty} \Lambda(J)/\overline{\Lambda}(J)$ with $0 < C < \infty$, then the pair $\overline{\Lambda}(J)$, $\widetilde{\overline{a}}_{N,k} \equiv C\widetilde{a}_{N,k}$ is just as good as the pair $\Lambda(J)$, $\widetilde{a}_{N,k}$.

Starting from (2.27) we first expand the integrand around $t = 1$ using the relation

$$\frac{C_k^{N/2-1}(t)}{C_k^{N/2-1}(1)} = {}_2F_1\left(N + k - 2, -k; \frac{N-1}{2}; \frac{1-t}{2}\right) \qquad (4.13)$$

where ${}_2F_1(a, b; c; z)$ is the *hypergeometric function*.[31] Then, extending the integration in $t$ from $[-1, 1]$ to $[-\infty, 1]$, we obtain the asymptotic expansion

$$F_{N,k}(J) = f_N(J)\left[1 - \frac{a_{N,k}}{J} + \frac{b_{N,k}}{J^2} + O(J^{-3})\right] \qquad (4.14)$$

with

$$f_N(J) = \frac{e^{J\tilde{h}(1)}}{\left[2\pi J\tilde{h}'(1)\right]^{1/2}} \Gamma\left(\frac{N}{2}\right)\left(\frac{J\tilde{h}'(1)}{2}\right)^{1-N/2} \qquad (4.15)$$

$$a_{N,k} = \frac{1}{2\tilde{h}'(1)}\left[\lambda_{N,k} + \frac{1}{4}(N-1)(N-3) - \frac{N^2-1}{4}r\right] \qquad (4.16)$$

$$b_{N,k} = \frac{1}{8\tilde{h}'(1)^2}\left\{\frac{(N+2k+1)(N+2k-1)(N+2k-3)(N+2k-5)}{16}\right.$$
$$- \frac{(N+2k-1)(N+2k-3)(N+3)(N+1)}{8}r$$
$$\left.+ \frac{(N+5)(N+3)(N+1)(N-1)}{16}r^2 - \frac{(N+3)(N+1)(N-1)}{6}s\right\} \qquad (4.17)$$

where $\lambda_{N,k} (\geq 0)$ are the eigenvalues of the Laplace-Beltrami operator on the sphere [given in (2.1)], and we have defined

$$r \equiv \tilde{h}''(1)/\tilde{h}'(1) \qquad (4.18\text{a})$$
$$s \equiv \tilde{h}'''(1)/\tilde{h}'(1) \qquad (4.18\text{b})$$

For the normalized expansion coefficients $v_{N,k}(J)$, we therefore have

$$v_{N,k}(J) = 1 - \frac{\tilde{a}_{N,k}}{2\tilde{h}'(1)J} + \frac{\tilde{b}_{N,k}}{4\tilde{h}'(1)^2 J^2} + O(J^{-3}) \qquad (4.19)$$

where

$$\tilde{a}_{N,k} = \lambda_{N,k} \qquad (4.20)$$

$$\tilde{b}_{N,k} = \tilde{a}_{N,k}\left[\frac{\tilde{a}_{N,k}}{2} - (N+1)r - 1\right] \qquad (4.21)$$

---

[31]See [31], formulae 9.100 and 9.14.2.

The $N$-vector model corresponds to $\widetilde{h}(1) = \widetilde{h}'(1) = 1$, $r = s = 0$. Notice that in this case formulae (4.14)–(4.17) could alternatively have been gotten through a direct expansion of the Bessel functions in (2.32).

Thus, for $J \to +\infty$ all masses [see (3.24)] go to zero as

$$m_{N,k}(J) \approx \lambda_{N,k}\, \Lambda(J)\,, \qquad (4.22)$$

where $\Lambda(J) \equiv 1/[2J\widetilde{h}'(1)]$ is a non-universal scale factor that goes to zero for $J \to +\infty$. (Here $\approx$ means that the ratio of the left and right sides tends to 1 as $J \to +\infty$.) If we consider the *mass ratios* defined by[32]

$$\mathcal{R}_{N,k}(J) \equiv \frac{m_{N,2}(J)}{m_{N,k}(J)}\,, \qquad (4.23)$$

we obtain, in the continuum limit,

$$\mathcal{R}_{N,k} = \frac{\lambda_{N,2}}{\lambda_{N,k}}\,. \qquad (4.24)$$

Therefore, all these Hamiltonians give rise to the same continuum limit and belong to what we will call the *$N$-vector universality class*.

**Second simple case:** $t = \pm 1$ are the only absolute maxima of $\widetilde{h}(t)$ [hence $\widetilde{h}(1) = \widetilde{h}(-1)$], and $\widetilde{h}'(1) > 0$, $\widetilde{h}'(-1) < 0$. [This is a subset of what will later be called the Hamiltonians of *Type II*.]

In this case one must sum the contributions of the two maxima, that is $F_{N,k}(J) = F_{N,k}^+(J) + F_{N,k}^-(J)$. The contribution $F_{N,k}^+$ coming from $t = 1$ has already been computed. Using the fact that $C_k^{N/2-1}(-1) = (-1)^k C_k^{N/2-1}(1)$, we see that the contribution $F_{N,k}^-$ coming from $t = -1$ can be obtained from $F_{N,k}^+$ by replacing the derivatives $\widetilde{h}^{(n)}(1)$ with $(-1)^n \widetilde{h}^{(n)}(-1)$ and then multiplying the whole thing by $(-1)^k$. Thus, keeping only the leading terms, we get

$$v_{N,k}(J) = 1 - \frac{\lambda_{N,k}}{2J} \frac{\widetilde{h}'(1)^{-(N+1)/2} + |\widetilde{h}'(-1)|^{-(N+1)/2}}{\widetilde{h}'(1)^{(1-N)/2} + |\widetilde{h}'(-1)|^{(1-N)/2}} + O(J^{-2}) \qquad (4.25)$$

for $k$ even, and

$$v_{N,k}(J) = \frac{\widetilde{h}'(1)^{(1-N)/2} - |\widetilde{h}'(-1)|^{(1-N)/2}}{\widetilde{h}'(1)^{(1-N)/2} + |\widetilde{h}'(-1)|^{(1-N)/2}} + O(J^{-1}) \qquad (4.26)$$

for $k$ odd.

From these formulae we immediately see that $\lim\limits_{J \to +\infty} |v_k(J)| < 1$ for $k$ odd, so that the odd-spin sector of the theory remains non-critical even at $J = +\infty$. [In the special case where the function $\widetilde{h}$ is *even*[33] (as in e.g. the $RP^{N-1}$ model), we have in fact $v_{N,k}(J) = 0$ for $k$ odd, for all $J$; while for $k$ even we get the same $v_{N,k}(J)$ as in the first simple case, given by formulae (4.19)–(4.21).] On the other hand, the even-spin masses go to zero as

$$m_{N,k}(J) \approx \lambda_{N,k}\,\Lambda(J) \qquad \text{for } k \text{ even}, \qquad (4.27)$$

where again

$$\Lambda(J) \equiv \frac{1}{2J}\frac{\widetilde{h}'(1)^{-(N+1)/2} + |\widetilde{h}'(-1)|^{-(N+1)/2}}{\widetilde{h}'(1)^{(1-N)/2} + |\widetilde{h}'(-1)|^{(1-N)/2}} \qquad (4.28)$$

is a non-universal scale factor that goes to zero for $J \to +\infty$. Thus the limiting mass ratio $\mathcal{R}_{N,k}$ in this case is the same as in the $N$-vector universality class for even $k$ and is zero for odd $k$. That is,

$$\mathcal{R}_{N,k} = \begin{cases} \lambda_{N,2}/\lambda_{N,k} & \text{for } k \text{ even} \\ 0 & \text{for } k \text{ odd} \end{cases} \qquad (4.29)$$

---

[32] We use $m_{N,2}$ rather than $m_{N,1}$ in the numerator in order to facilitate comparison with the second simple case below, in which $m_{N,k} = +\infty$ for all odd $k$.

All these Hamiltonians belong to the same universality class, which we will call the $RP^{N-1}$ *universality class*. Notice that the exact $Z_2$ gauge symmetry, which holds for the usual $RP^{N-1}$ Hamiltonian (and more generally whenever $h$ is even), plays here no role. Provided that the Hamiltonian has a two-maxima structure at $t = \pm 1$ with $\widetilde{h}'(\pm 1) \neq 0$, the *continuum limit* will be $Z_2$-gauge-symmetric. For instance, a Hamiltonian with $\widetilde{h}(t) = t^2 + \alpha(t - t^3)$, with $|\alpha| < 1$ belongs to this universality class.

In summary, we have thus far defined two universality classes:

(i) the $N$-vector universality class, where all the masses go to zero as $J \to +\infty$ at the *same rate* and the limiting mass ratio $\mathcal{R}_{N,k}$ is given by (4.24) for all $k$; and

(ii) the $RP^{N-1}$ universality class, where as $J \to +\infty$ the even sector displays the same behavior as for the $N$-vector universality class [i.e. the masses go to zero at the same rate with $\mathcal{R}_{N,k}$ given by (4.24) for all even values of $k$] while in the odd sector the masses either

   (a) do not go to zero [as in the second simple case above] or else

   (b) go to zero at a rate slower than for the even sector [as will occur in some examples below],

and therefore $\mathcal{R}_{N,k}$ is zero for all odd values of $k$.

---

[33] In this case we have $F_{N,k}(J) = 0$ for $k$ odd. For $k$ even we have $F_{N,k}(J) = 2F^+_{N,k}(J)$, and therefore the coefficients $a_{N,k}$ and $b_{N,k}$ are given by (4.16) and (4.17), and $f_N(J)$ has twice the value in (4.15). For $r = 1$ and $s = 0$ we obtain the expansion for the $RP^{N-1}$ model, which can also be obtained by direct expansion of the coefficients (2.33).

Formulae (4.22) and (4.27) had to be expected on general grounds. Indeed, the continuum limit of the $N$-vector model (or more generally of any model belonging to the first simple case above) is simply Brownian motion on $S^{N-1}$, and the generator of Brownian motion is the Laplace-Beltrami operator.[34] Thus we expect $m_{N,k}(J) \approx \Lambda(J) \lambda_{N,k}$ where $\Lambda(J)$ is a non-universal scale factor depending on the chosen sequence of lattice Hamiltonians. An analogous discussion applies to the $RP^{N-1}$ case: here the continuum limit is Brownian motion on $RP^{N-1}$, and thus the corresponding masses are related to the eigenvalues of the Laplace-Beltrami operator on $RP^{N-1}$ (which are simply the even-spin eigenvalues of the Laplace-Beltrami operator on $S^{N-1}$).

### 4.1.3 General One-Parameter Family

We want now to address the general problem of studying the limit $J \to +\infty$ for an *arbitrary* interaction $\widetilde{h}$; in particular, we want to know whether the two universality classes we have just discussed are the only ones which can appear as a critical limit of interactions of the form (4.1). As we shall see, the situation is much more complicated than this, and in fact an infinite number of universality classes appears.

Let us assume henceforth that $\widetilde{h}$ is smooth, and that it has finitely many absolute maxima, all of finite order. In particular, suppose it has $M$ absolute maxima on the interval $[-1,1]$ at points $t_1, \ldots t_M$ with $\widetilde{h}(t_1) = \ldots = \widetilde{h}(t_M) = \widetilde{h}_{max}$. Let $n_i$ be the order of the maximum at $t_i$, i.e. the smallest (nonzero) integer such that $\widetilde{h}^{(n_i)}(t_i) \neq 0$. (When $t_i \neq \pm 1$ the order $n_i$ is of course even and $\geq 2$, and $\widetilde{h}^{(n_i)}(t_i) < 0$. When $t_i = -1$ we have $\widetilde{h}^{(n_i)}(t_i) < 0$, and when $t_i = +1$ we have $(-1)^{n_i} \widetilde{h}^{(n_i)}(t_i) < 0$.) For $J \to +\infty$ we have

$$F_{N,k} \approx \sum_{i=1}^{M} F_{N,k}^{(i)}, \quad (4.30)$$

where $F_{N,k}^{(i)}$ is the contribution of the $i$-th maximum; to leading order in $J$ it is given by

$$F_{N,k}^{(i)} \approx e^{J\widetilde{h}_{max}} A_i \frac{C_k^{N/2-1}(t_i)}{C_k^{N/2-1}(1)} J^{-\alpha_i} \quad (4.31)$$

where

$$\alpha_i = \begin{cases} (N-1)/(2n_i) & \text{if } t_i = \pm 1 \\ 1/n_i & \text{if } t_i \neq \pm 1 \end{cases} \quad (4.32)$$

and $A_i$ is a positive constant, independent of $J$ and $k$, given explicitly by

$$A_i = \left(\frac{|\widetilde{h}^{(n_i)}(t_i)|}{n_i!}\right)^{-\alpha_i} \frac{\Gamma\left(\frac{N}{2}\right)}{\Gamma\left(\frac{N-1}{2}\right)} \frac{\Gamma(\alpha_i)}{n_i} \widetilde{A}_i \quad (4.33)$$

---

[34]An arbitrary second-order elliptic differential operator on a manifold $M$ generates a diffusion process on $M$; Brownian motion is the special case in which the generator is the Laplace-Beltrami operator. For the general theory of diffusions on a manifold, see e.g. [33, sections 4.1–4.3].

where

$$\widetilde{A}_i \equiv \begin{cases} 2^{(N-3)/2} & \text{if } t_i = \pm 1 \\ 2\left(1 - t_i^2\right)^{(N-3)/2} & \text{if } t_i \neq \pm 1 \end{cases} \qquad (4.34)$$

For $J \to +\infty$ the leading contribution comes from those terms with the smallest $\alpha_i$; we call these maxima the *principal maxima*. Setting $\alpha = \min_i \alpha_i$, we thus have

$$F_{N,k} \approx e^{J\widetilde{h}_{max}} J^{-\alpha} \sum_{i:\, \alpha_i = \alpha} A_i \, \frac{C_k^{N/2-1}(t_i)}{C_k^{N/2-1}(1)} \,. \qquad (4.35)$$

We want now to know under what conditions the mass $m_{N,k}(J)$ tends to zero as $J \to +\infty$. For this analysis it is sufficient to use the leading-order expansion (4.35). Equivalently we want to see under what conditions $v_{N,k}(J) \to 1$, i.e. when (notice that $C_0^{N/2-1}(t) = 1$ and $A_i > 0$)

$$\sum_{i:\, \alpha_i = \alpha} A_i \, \frac{C_k^{N/2-1}(t_i)}{C_k^{N/2-1}(1)} = \sum_{i:\, \alpha_i = \alpha} A_i \,. \qquad (4.36)$$

Since[35], for $N \geq 3$, $|C_k^{N/2-1}(t)| < C_k^{N/2-1}(1)$ when $t \neq \pm 1$, this condition cannot be satisfied for any $k$ if there is in the sum an $i$ such that $t_i \neq \pm 1$. Thus the principal maxima can only be at $1$ or $-1$. Moreover, if $t = -1$ appears in the sum, the condition can be satisfied only for even $k$, since $C_k^{N/2-1}(-1) = (-1)^k C_k^{N/2-1}(1)$. We end up with the following result:

1. If $t_i = 1$ is the only principal maximum of $\widetilde{h}(t)$, then

$$\lim_{J \to +\infty} v_{N,k}(J) = 1 \qquad (4.37)$$

   for all $k \geq 1$. In this case all correlations become critical.

2. If $t_i = \pm 1$ are the only principal maxima of $\widetilde{h}(t)$, then

$$\lim_{J \to +\infty} v_{N,k}(J) = \begin{cases} 1 & \text{for } k \text{ even} \\ c_{N,k} & \text{for } k \text{ odd} \end{cases} \qquad (4.38)$$

   with $-1 < c_{N,k} < 1$. In this case only the even-spin sector becomes critical. In detail, we have

$$c_{N,k} = \frac{|\widetilde{h}^{(n)}(1)|^{-\alpha} - |\widetilde{h}^{(n)}(-1)|^{-\alpha}}{|\widetilde{h}^{(n)}(1)|^{-\alpha} + |\widetilde{h}^{(n)}(-1)|^{-\alpha}} \qquad (4.39)$$

   where $n = (N-1)/2\alpha$.

3. If $t_i = -1$ is the only principal maximum of $\widetilde{h}(t)$, then

$$\lim_{J \to +\infty} v_{N,k}(J) = \begin{cases} 1 & \text{for } k \text{ even} \\ -1 & \text{for } k \text{ odd} \end{cases} \qquad (4.40)$$

   As in the preceding case, only the even-spin sector becomes critical.

---

[35]See Appendix A.4.

4. If there exists at least one $t_i \neq \pm 1$ such that $\alpha_i = \alpha$ (i.e. there are principal maxima other than $\pm 1$), then

$$\lim_{J \to +\infty} v_{N,k}(J) = c'_{N,k} \qquad (4.41)$$

with $-1 < c'_{N,k} < 1$ for all $k$. In this case there is no continuum limit for any $k$. In particular, if there is *exactly one* principal maximum, and this is a point $t_i \neq \pm 1$, then

$$c'_{N,k} = \frac{C_k^{N/2-1}(t_i)}{C_k^{N/2-1}(1)} . \qquad (4.42)$$

These results can be understood heuristically: If some $t_i \neq \pm 1$ contributes at leading order to the asymptotic expansion of $F_{N,k}$ (case 4), then for large $J$ the typical configurations have $\boldsymbol{\sigma}_x \cdot \boldsymbol{\sigma}_{x+1} \approx t_i$ on a significant fraction of the bonds. For $N \geq 3$ there are *many* configurations on each bond with this property (since the azimuthal angles are undetermined), and they keep the system disordered even at $J = +\infty$. In case 1, by contrast, the system orders and thus for $J = +\infty$ the correlation length becomes infinite. In case 2 the system orders modulo a sign; the even-spin correlations are insensitive to the sign and thus display critical behavior, while the odd-spin ones remain disordered even at $J = +\infty$. In case 3, the system develops antiferromagnetic order as $J = +\infty$; the even-spin correlations are insensitive to the sign and thus display critical behavior, while the odd-spin correlations have no continuum limit.[36] In the following we will disregard the theories belonging to case 3 [since for the odd-spin sector they do not have a continuum limit, and for the even-spin sector they are identical to theories of case 1 with $\tilde{h}(t) \to \tilde{h}(-t)$] and to case 4 (since we have proven that they do not exhibit any non-trivial critical behavior). The Hamiltonians described in case 1 (respectively case 2) will be called Hamiltonians of *Type I* (respectively *Type II*).

To characterize the different universality classes we want now to derive the behavior of the masses $m_{N,k}$ in the limit $J \to +\infty$. In order to do this, we must carry the asymptotic expansion of $F_{N,k}(J)$ to the first subleading order for the principal maxima, and also consider the leading contributions from the non-principal maxima. Let us first consider **theories of Type I**. The relevant expansion for $F_{N,k}$ is (we set $t_+ \equiv t_1 = 1$)

$$F_{N,k} = e^{J\tilde{h}_{max}} \left\{ A_+ J^{-\alpha} \left[ 1 - \frac{c_+ d_{k,+}}{J^{1/n_+}} + o(J^{-1/n_+}) \right] \right.$$

---

[36] The antiferromagnetic case 3 can be transformed into the ferromagnetic case 1 by the change of variables $\widehat{\boldsymbol{\sigma}_x} = (-1)^x \boldsymbol{\sigma}_x$ together with $\widehat{\tilde{h}}(t) = \tilde{h}(-t)$. The correlation functions then transform as $G_{N,k}(x, \widehat{\tilde{h}}; \infty) = (-1)^{kx} G_{N,k}(x, \tilde{h}; \infty)$ and

$$\widetilde{G}_{N,k}(p, \widehat{\tilde{h}}; \infty) = \begin{cases} \widetilde{G}_{N,k}(p, \tilde{h}; \infty) & \text{for } k \text{ even} \\ \widetilde{G}_{N,k}(p + \pi, \tilde{h}; \infty) & \text{for } k \text{ odd} \end{cases}$$

Thus, case 3 is *identical* to case 1 for the even-spin correlation functions; and it has *no* continuum limit for the odd-spin correlation functions (since there is no divergence at $p = 0$). This mapping also works in finite volume, *provided that $L$ is even*.

$$+ \sum_{i=2}^{M} A_i \frac{C_k^{N/2-1}(t_i)}{C_k^{N/2-1}(1)} \left[ J^{-\alpha_i} + o(J^{-\alpha_i}) \right] \right\} \qquad (4.43)$$

where

$$c_+ = \frac{1}{N-1} \frac{\Gamma\left(\frac{N+1}{2n_+}\right)}{\Gamma\left(\frac{N-1}{2n_+}\right)} \left[ \frac{n_+!}{|\tilde{h}^{(n_+)}(1)|} \right]^{1/n_+} \qquad (4.44)$$

$$d_{k,+} = \lambda_{N,k} + \frac{1}{4}(N-3)(N-1) - \frac{N^2-1}{2\, n_+\,(n_++1)} \frac{\tilde{h}^{(n_++1)}(1)}{\tilde{h}^{(n_+)}(1)} \qquad (4.45)$$

(Note that $c_+$ does not depend on $k$, while $d_{k,+}$ does.) The first correction to the leading term depends now on the relation between $\alpha$ and $\beta \equiv \min_{2 \leq i \leq M} \alpha_i$. We have

$$m_{N,k}(J) \approx \Lambda(J)\, \lambda_{N,k} + o(J^{-1/n_+})$$
$$+ \frac{1}{A_+ J^{\beta - \alpha}} \sum_{i:\, \alpha_i = \beta} A_i \left( 1 - \frac{C_k^{N/2-1}(t_i)}{C_k^{N/2-1}(1)} \right) + o(J^{\alpha - \beta}) \qquad (4.46)$$

with

$$\Lambda(J) = c_+\, J^{-1/n_+} . \qquad (4.47)$$

(We call the maxima with $\alpha_i = \beta$ the *next-to-principal maxima*.) Here it should be understood that only the dominant term is to be kept:

(a) If $\beta > \alpha + 1/n_+$, the first term is dominant and the model belongs to the $N$-vector universality class (4.22).

(b) If $\beta < \alpha + 1/n_+$, then the third term (the term of order $1/J^{\beta - \alpha}$) dominates *provided that its coefficient is not zero*. The coefficient is zero if $k$ is even and the only next-to-principal maximum is $t_i = -1$; otherwise the coefficient is nonzero. Thus, for $k$ odd the mass is

$$m_{N,k}(J) \approx \frac{1}{A_+ J^{\beta - \alpha}} \sum_{i:\, \alpha_i = \beta} A_i \left( 1 - \frac{C_k^{N/2-1}(t_i)}{C_k^{N/2-1}(1)} \right) . \qquad (4.48)$$

(b1) If the coefficient (for all $k$) is nonzero, the limiting mass ratio (for any $k$) is

$$\mathcal{R}_{N,k} = \frac{\sum_{i:\, \alpha_i = \beta} A_i \left( 1 - \frac{C_2^{N/2-1}(t_i)}{C_2^{N/2-1}(1)} \right)}{\sum_{i:\, \alpha_i = \beta} A_i \left( 1 - \frac{C_k^{N/2-1}(t_i)}{C_k^{N/2-1}(1)} \right)} . \qquad (4.49)$$

Clearly there is a multi-parameter family of new universality classes, obtainable by varying the $\{t_i, A_i\}_{i=1}^M$ appropriately.

(b2) If the coefficient (for $k$ even) is zero, the behavior depends on whether the correction $o(J^{-(\beta-\alpha)})$ is larger or smaller than $J^{-1/n_+}$. If it is smaller, then (for $k$ even) the term $\Lambda(J)\,\lambda_{N,k}$ will dominate and therefore, the ratios $\mathcal{R}_{N,k}$ will be those of the $RP^{N-1}$ universality class.[37] If the $o(J^{-(\beta-\alpha)})$ correction is larger than or equal to $J^{-1/n_+}$, then a more detailed investigation is needed. (We note that, also in this last case, the limiting mass ratio is zero for $k$ odd just as for the $RP^{N-1}$ universality class.)

(c) Finally, if $\beta = \alpha + 1/n_+$, then both terms are of the same order. Again we obtain new (multi-parameter) universality classes. In particular, if the only next-to-principal maximum is $t_i = -1$, we get

$$\mathcal{R}_{N,k} \;=\; \begin{cases} \lambda_{N,2}/\lambda_{N,k} & \text{for } k \text{ even} \\ \lambda_{N,2}/(\lambda_{N,k} + B) & \text{for } k \text{ odd} \end{cases} \qquad (4.50)$$

where $0 < B < \infty$ is a parameter that interpolates the limiting mass ratio between the $N$-vector and the $RP^{N-1}$ universality classes. [Explicitly: $B = 2A_i/(c_+A_+)$.]

For **theories of Type II** we set $t_+ \equiv t_1 = 1$ and $t_- \equiv t_2 = -1$; the expansion of $F_{N,k}$ is then given by

$$F_{N,k} \;\approx\; e^{J\tilde{h}_{max}} \left\{ A_+\, J^{-\alpha}\left(1 - \frac{c_+\,d_{k,+}}{J^{1/n}}\right) + (-1)^k\, A_-\, J^{-\alpha}\left(1 - \frac{c_-\,d_{k,-}}{J^{1/n}}\right) + \right.$$
$$\left. +\; o(J^{-\alpha-1/n}) \;+\; \sum_{i=3}^{M} A_i\, \frac{C_k^{N/2-1}(t_i)}{C_k^{N/2-1}(1)} \left[J^{-\alpha_i} + o(J^{-\alpha_i})\right] \right\} \qquad (4.51)$$

where $n \equiv n_+ = n_- = (N-1)/2\alpha$; here $d_{k,+}$ and $c_+$ are given by formulae (4.45) and (4.44), and $d_{k,-}$ and $c_-$ can be obtained from the same formulae by simply substituting $\tilde{h}^{(n)}(1)$ with $(-1)^n \tilde{h}^{(n)}(-1)$. Defining $\beta \equiv \min_{3 \leq i \leq M} \alpha_i$, we obtain that, for even values of $k$, the masses are given by equation (4.46) with $A_1$ replaced by $A_+ + A_-$ and

$$\Lambda(J) \;=\; \frac{A_+\,c_+ + A_-\,c_-}{A_+ + A_-}\, J^{-1/n} \qquad (4.52)$$

(as before, only the dominant term should be kept) while, for odd values of $k$, we have

$$-1 \;<\; \lim_{J\to+\infty} v_{N,k}(J) \;=\; \frac{A_+ - A_-}{A_+ + A_-} \;<\; 1 \,. \qquad (4.53)$$

Therefore the odd-spin sector of the theory is always non-critical, while the even-spin masses go to zero at the same rate for all even $k$. It follows that the mass ratios $\mathcal{R}_{N,k} \equiv m_{N,2}/m_{N,k}$ are zero for odd $k$ and nonzero for even $k$. If $\beta > \alpha + 1/n$, we reobtain the $RP^{N-1}$ universality class (4.27), while in the other cases an infinite number of new universality classes appear.

---

[37]See case (b) of the definition of the $RP^{N-1}$ universality class in Section 4.1.2.

### 4.1.4 An Example in More Detail

Finally, to examine more closely the possible universality classes, let us consider the special case in which there are only two possible maxima, namely those at $t = \pm 1$. This generalizes the two cases studied in Section 4.1.2: in that section we required $\tilde{h}'(\pm 1) \neq 0$ [i.e. $n_\pm = 1$], while now we lift this restriction. Let $n_+$ and $n_-$ be the orders of the first non-vanishing derivatives at $t = +1$ and $t = -1$ respectively; we will suppose $n_+ \geq n_-$ (since we are considering only theories of types I and II). Then, from the previous discussion we find four cases:

For Hamiltonians of Type I, namely $n_- < n_+$, formula (4.46) becomes

$$m_{N,k}(J) \approx \frac{c_+ \lambda_{N,k}}{J^{1/n_+}} + \frac{[1-(-1)^k]A_-}{J^{\beta-\alpha} A_+} \tag{4.54}$$

with $\alpha \equiv \alpha_+$ and $\beta \equiv \alpha_-$.

There are therefore three possibilities:

(a) If $n_- < [(N-1)/(N+1)] n_+$, we get

$$m_{N,k}(J) \approx \Lambda(J) \lambda_{N,k} \tag{4.55}$$

for all $k$, with $\Lambda(J)$ given by

$$\Lambda(J) = c_+ J^{-1/n_+} . \tag{4.56}$$

Therefore, the model belongs to the $N$-vector universality class.

(b) If $n_+ > n_- > [(N-1)/(N+1)] n_+$, we then have

$$m_{N,k}(J) \approx \begin{cases} \Lambda(J) \lambda_{N,k} & \text{for } k \text{ even} \\ 2 (A_-/A_+) J^{-(\beta-\alpha)} & \text{for } k \text{ odd} \end{cases} \tag{4.57}$$

with $\Lambda(J)$ given by (4.56).

Therefore all masses go to zero as $J \to +\infty$, but with different rates, so that in the limit the odd-spin masses are infinitely larger than the even-spin masses. The limiting mass ratios $\mathcal{R}_{N,k}$ are those of the $RP^{N-1}$ universality class.

(c) If $n_- = [(N-1)/(N+1)] n_+$, both terms in (4.54) contribute at the same order. We obtain

$$m_{N,k} \approx \begin{cases} \Lambda(J) \lambda_{N,k} & \text{for } k \text{ even} \\ \Lambda(J) (\lambda_{N,k} + B) & \text{for } k \text{ odd} \end{cases} \tag{4.58}$$

where $\Lambda(J)$ is given by (4.56) and

$$B = \frac{2 A_-}{c_+ A_+} \tag{4.59}$$

is a positive constant. So we get an infinite number of different continuum-limit theories, parametrized by $B$. Notice that $0 < B < \infty$; therefore, the $N$-vector and the $RP^{N-1}$ universality classes are not included as particular cases but only as the limiting cases for $B \to 0$ and $B \to +\infty$, respectively.

For Hamiltonians of Type II, namely $n_- = n_+$, we have:

(d) The masses are given by equation (4.53) for $k$ odd and by

$$m_{N,k} \approx \Lambda(J)\, \lambda_{N,k} \tag{4.60}$$

for $k$ even, where $\Lambda(J)$ is as in (4.52). This case clearly belongs to the $RP^{N-1}$ universality class.

### 4.1.5 Interpretation

We want now to interpret these results in another framework. In one dimension a continuum field theory is simply a continuous-time Markov process on the target manifold. Now, the generator of a continuous-time Markov process is the convex combination of a diffusion part (a second-order elliptic differential operator) and a jump part (a positive kernel).[38] Physically, this means that the "particle" diffuses for a while according to the specified differential operator, and then, at exponentially distributed random times, jumps according to the specified probability kernel. On the sphere $S^{N-1}$ for $N \geq 3$, the only $SO(N)$-invariant second-order elliptic operator is the Laplace-Beltrami operator (and multiples thereof); thus, the only $SO(N)$-invariant diffusion on $S^{N-1}$ is standard Brownian motion (with an arbitrary coefficient, corresponding to a rescaling of time).[39] On the other hand, there is an *infinite-dimensional* family of possible $SO(N)$-invariant jump kernels $K$: indeed, one can specify an arbitrary probability distribution of jump angles $\theta \in [0,\pi]$ (and $SO(N)$-invariance then determines $K$ uniquely, for $N \geq 3$). Each one of these quantum Hamiltonians $\widehat{H} = a\mathcal{L} + K$ ($a \geq 0$) defines a legitimate continuum $\sigma$-model.

Moreover, for each such quantum Hamiltonian $\widehat{H}$ and each $t > 0$, the integral kernel $e^{-t\widehat{H}}(\boldsymbol{\sigma}, \boldsymbol{\sigma}')$ is a smooth $O(N)$-invariant function of $\boldsymbol{\sigma}$ and $\boldsymbol{\sigma}'$ (and thus a function of $\boldsymbol{\sigma} \cdot \boldsymbol{\sigma}'$); it can therefore be realized as $e^{\mathcal{V}_t(\boldsymbol{\sigma} \cdot \boldsymbol{\sigma}')}$ for a suitable smooth potential $\mathcal{V}_t$. Thus, by taking some sequence $t \downarrow 0$, we see that each continuum $\sigma$-model can be realized as a continuum limit of lattice $\sigma$-models (i.e. discrete-time $O(N)$-invariant random walks on $S^{N-1}$) each of which has a smooth step distribution $e^{\mathcal{V}(\boldsymbol{\sigma} \cdot \boldsymbol{\sigma}')}$.

We can now interpret formula (4.46): the continuum limit of this theory is a Markov process on $S^{N-1}$ which contains a jump part with jump angles $\theta_i = \arccos t_i$. The coefficients $A_i$ are related to the probability distribution of the jump angles. The typical configuration here, for large $J$, consists of ordered domains where $\boldsymbol{\sigma}_x \cdot \boldsymbol{\sigma}_{x+1} \approx 1$ separated by links where a *jump* occurs, that is where $\boldsymbol{\sigma}_x \cdot \boldsymbol{\sigma}_{x+1} \approx t_i$. Notice that these jumps must be sufficiently rare, otherwise they destroy the order and thus no criticality appears (this occurs in case 4 of our classification, i.e. when $\alpha_i = \alpha$), but not too rare, otherwise they are unable to change the critical behavior of the system (this is the case when $\alpha_i > \beta$). Jumps of $\pi$ (which are simply spin flips) play a special role: the spin-$k$ correlations for $k$ *even* are insensitive to spin flips, and thus they remain critical irrespective of the frequency of such flips. In particular, for theories of type II these spin flips are *infinitely* rapid, and the continuum limit is best considered as a Markov process on $RP^{N-1} \equiv S^{N-1}/Z_2$.

---

[38] See [34], Example 1.2.1 (p. 6), Theorem 2.2.1 (p. 48) and Theorem 2.2.2 (p. 51).

[39] This is true also for $N = 2$ if one demands $O(N)$-invariance and not just $SO(N)$-invariance.

## 4.2 Finite-Size-Scaling Limit

### 4.2.1 Generalities on the Finite-Size-Scaling Limit

We want now to discuss the finite-size-scaling limit for theories of types I and II (see Section 4.1.3 for definitions). This limit is given by $L \to \infty$, $J \to +\infty$ [hence $\xi_{N,k}^{(\#)}(J;\infty) \to \infty$, where $\xi_{N,k}^{(\#)}$ denotes any one of the correlation lengths $\xi_{N,k}^{(2nd)}$ or $\xi_{N,k}^{(exp)}$ introduced earlier] in such a way that $\xi_{N,k}^{(\#)}(J;\infty)/L$ remains fixed.[40] We therefore define the scaling variables

$$z_k = z_k(J;L) \equiv \frac{\xi_{N,k}^{(exp)}(J;\infty)}{L} \,. \qquad (4.61)$$

When considering correlation functions in $x$-space, we also scale $x$, i.e. we will consider $x = \overline{x}L$ with $0 \leq \overline{x} \leq 1$ fixed. The corresponding correlations represent the correlations of a continuum theory in a periodic box of width 1.

Everywhere in this section, for theories of Type II or those theories of Type I belonging to the $RP^{N-1}$ universality class [see case I(b2) in Section 4.1.3, and case (b) in Section 4.1.4 [41]], $k$ must of course be even, and in all the formulae below only even values of $l$ and $m$ are to be included in the sums.

As we have seen in Section 4.1.1, the theory displays critical behavior only if the masses $m_{N,k}$ go to zero in the limit $J \to +\infty$. Therefore, in this section, we assume that

$$v_{N,k}(J) = 1 - \widetilde{a}_{N,k}\, \Lambda(J) + o(\Lambda(J)) \,, \qquad (4.62)$$

where $\Lambda(J)$ is a non-universal scale factor (assumed strictly positive) which goes to zero as $J \to +\infty$. The quantities $\widetilde{a}_{N,k}$ characterize the universality class of the theory and are completely defined modulo an overall constant which can be absorbed into $\Lambda(J)$. Since $v_{N,k}(J) < 1$, we have $\widetilde{a}_{N,k} > 0$. For the $N$-vector universality class, the coefficient $\widetilde{a}_{N,k}$ can be simply defined by

$$\widetilde{a}_{N,k} = \lambda_{N,k} \,, \qquad (4.63)$$

---

[40]Similarly to what we did in Section 4.1.1, we could consider a sequence $\langle \cdot \rangle^{(n)}$ of finite-volume lattice models with linear lattice sizes $L^{(n)} \to \infty$. A finite-size-scaling limit (= finite-volume continuum limit) yielding a continuum box of side $L^{(cont)}$ ($0 < L^{(cont)} < \infty$) is defined by rescaling lengths by factors $\Xi^{(n)} \equiv L^{(n)}/L^{(cont)}$ ($\to \infty$) and rescaling field strengths by factors $\zeta_{N,k}^{(n)}$ such that the spin-$k$ two-point functions have well-defined limits. Without loss of generality we can set $L^{(cont)} = 1$.

[41]In these two cases, the masses of the even and odd sectors go to zero with different rates. As explained in the remark at the end of Section 4.1.1, $\Lambda(J)$ is chosen to be of the order of the smallest mass of the theory (that of the even sector) and it can be seen that for $k$ odd $v_{N,k}(J)^L$ goes to zero exponentially in the finite-size-scaling limit. Therefore, the odd sector does not contribute to the finite-size-scaling functions, just as for case (a) of the $RP^{N-1}$ universality class.

where $\lambda_{N,k}$ are the eigenvalues of the Laplace-Beltrami operator on the sphere. For the $RP^{N-1}$ universality class the same holds for even $k$. For the other universality classes, which include jump processes, the coefficients $\widetilde{a}_{N,k}$ can be easily derived from (4.43) for theories of Type I and its analogue (4.51) for theories of Type II.

Now, from (4.61)–(4.62) and (3.24)/(3.26), it follows that, for large $J$, we have

$$z_k \;\approx\; 1/[\widetilde{a}_{N,k}\, L \Lambda(J)] \;. \tag{4.64}$$

Therefore, instead of considering the limit $L, J \to \infty$ at $z_k$ fixed, we will equivalently consider the more convenient limit at $L\,\Lambda(J) \equiv \gamma$ fixed, since the parameter $\gamma$ will appear naturally in our formulae. Let us then define the variables

$$\overline{z}_k \;=\; \overline{z}_k(J;L) \;\equiv\; \frac{1}{\widetilde{a}_{N,k}\, L\, \Lambda(J)} \;=\; \frac{1}{\widetilde{a}_{N,k}\, \gamma} \;. \tag{4.65}$$

(To leading order[42] we have $z_k \approx \overline{z}_k$. The distinction between $z_k$ and $\overline{z}_k$ will become relevant only when we consider *corrections* to finite-size scaling.) Our approach will be to compute various quantities as a function of $\gamma$, and then use (4.65) to re-express everything as a function of $\overline{z}_1$ or $\overline{z}_2$. The reason for this last step is that functions of $\gamma$ are universal only modulo a scale factor [corresponding to the arbitrariness of $\Lambda(J)$], while functions of physical continuum quantities (such as the $\overline{z}_k$) are universal *tout court*.

### 4.2.2 Computation of the Finite-Size-Scaling Functions

We want to compute the following finite-size-scaling functions:[43]

$$\widetilde{Z}_N^{(0)}(\gamma) \;\equiv\; \lim_{\substack{L, J \to \infty \\ \gamma \text{ fixed}}} Z_N(J;L)/F_{N,0}(J)^L \tag{4.66}$$

$$G_{N,k}^{(0)}(\overline{x}, \gamma) \;\equiv\; \lim_{\substack{L, J \to \infty \\ \gamma \text{ fixed}}} G_{N,k}(\overline{x}L, J; L) \tag{4.67}$$

$$\chi_{N,k}^{(0)}(\gamma) \;\equiv\; \lim_{\substack{L, J \to \infty \\ \gamma \text{ fixed}}} \frac{\chi_{N,k}(J;L)}{L} \tag{4.68}$$

$$\xi_{N,k}^{(2nd)(0)}(\gamma) \;\equiv\; \lim_{\substack{L, J \to \infty \\ \gamma \text{ fixed}}} \frac{\xi_{N,k}^{(2nd)}(J;L)}{L} \tag{4.69}$$

---

[42] More precisely we have

$$\lim_{J \to +\infty} \frac{z_k(J;L)}{\overline{z}_k(J;L)} \;=\; \lim_{J \to +\infty} \xi_{N,k}^{(exp)}\, \widetilde{a}_{N,k}\, \Lambda(J) \;=\; 1 \;.$$

[43] The superscript $^{(0)}$ indicates "leading order". The first corrections to these finite-size-scaling functions will be computed in Section 4.3.

Concerning $\chi_{N,k}$ and $\xi_{N,k}^{(2nd)}$, it is often convenient to look at the ratios

$$R_{\chi;N,k}(J;L) \equiv \frac{\chi_{N,k}(J;L)}{\chi_{N,k}(J;\infty)} \tag{4.70}$$

$$R_{\xi;N,k}(J;L) \equiv \frac{\xi_{N,k}^{(2nd)}(J;L)}{\xi_{N,k}^{(2nd)}(J;\infty)} \tag{4.71}$$

Note that by (3.22)/(3.23) we have

$$\chi_{N,k}(J;\infty) \approx \frac{2}{\tilde{a}_{N,k}\,\Lambda(J)} = 2L\,\overline{z}_k \tag{4.72}$$

$$\xi_{N,k}^{(2nd)}(J;\infty) \approx \xi_{N,k}^{(exp)}(J;\infty) \approx \frac{1}{\tilde{a}_{N,k}\,\Lambda(J)} = L\,\overline{z}_k \tag{4.73}$$

hence the ratios have well-behaved finite-size-scaling limits:

$$R_{\chi;N,k}^{(0)}(\gamma) \equiv \lim_{\substack{L,J \to \infty \\ \gamma \text{ fixed}}} R_{\chi;N,k}(J;L) = \frac{1}{2}\tilde{a}_{N,k}\,\gamma\,\chi_{N,k}^{(0)}(\gamma) \tag{4.74}$$

$$R_{\xi;N,k}^{(0)}(\gamma) \equiv \lim_{\substack{L,J \to \infty \\ \gamma \text{ fixed}}} R_{\xi;N,k}(J;L) = \tilde{a}_{N,k}\,\gamma\,\xi_{N,k}^{(2nd)(0)}(\gamma) \tag{4.75}$$

The computation of the finite-size-scaling functions (4.66)–(4.69) is straightforward. In the limit $L,J \to \infty$ with $\gamma$ fixed, we have

$$v_{N,k}(J)^L = v_{N,k}(J)^{\gamma/\Lambda(J)} \approx \exp\left\{\frac{\gamma}{\Lambda(J)}\log\left[1 - \tilde{a}_{N,k}\,\Lambda(J)\right]\right\} \approx \exp\left(-\gamma\,\tilde{a}_{N,k}\right). \tag{4.76}$$

Inserting this limit in the exact expressions (3.13)–(3.15) from Section 3.1, we obtain

$$\tilde{Z}_N^{(0)}(\gamma) = \sum_{l=0}^{\infty} \mathcal{N}_{N,l}\,e^{-\gamma\tilde{a}_{N,l}} \tag{4.77}$$

$$G_{N,k}^{(0)}(\overline{x},\gamma) = \frac{1}{\tilde{Z}_N^{(0)}(\gamma)}\sum_{l,m=0}^{\infty}\frac{\mathcal{C}_{N;k,l,m}^2}{\mathcal{N}_{N,k}}\,e^{-\gamma\tilde{a}_{N,l}}\,e^{-\gamma\overline{x}\Delta_{N;l,m}} \tag{4.78}$$

$$\chi_{N,k}^{(0)}(\gamma) = \frac{2}{\gamma\,\tilde{Z}_N^{(0)}(\gamma)}\sum_{l,m=0}^{\infty}\frac{\mathcal{C}_{N;k,l,m}^2}{\mathcal{N}_{N,k}}\,\frac{e^{-\gamma\tilde{a}_{N,l}}}{\Delta_{N;l,m}} \tag{4.79}$$

$$\xi_{N,k}^{(2nd)(0)}(\gamma) = \frac{1}{\gamma}\left\{\frac{\sum_{l,m=0}^{\infty}\left[B_{N;k,l,m}(\gamma)/\Delta_{N;l,m}^2\right]}{\sum_{l,m=0}^{\infty}B_{N;k,l,m}(\gamma)}\right\}^{1/2} \tag{4.80}$$

with[44]

$$\Delta_{N;l,m} \equiv \widetilde{a}_{N,m} - \widetilde{a}_{N,l} \qquad (4.81)$$

$$B_{N;k,l,m}(\gamma) \equiv \mathcal{C}^2_{N;k,l,m} \, e^{-\gamma \widetilde{a}_{N,l}} \, \frac{\gamma \, \Delta_{N;l,m}}{4\pi^2 + \gamma^2 \, \Delta^2_{N;l,m}} \qquad (4.82)$$

Let us notice that (4.77) can be rewritten as

$$\widetilde{Z}^{(0)}_N(\gamma) = \text{Tr} \, \exp(-\gamma \widehat{H}), \qquad (4.83)$$

where $\widehat{H}$ is the operator that generates the continuous-time Markov process corresponding to that universality class. For Type-I theories (except the case belonging to the $RP^{N-1}$ universality class), the trace is taken in the space $L^2(S^{N-1})$; while for Type-II theories (and for the case of Type I which falls in the $RP^{N-1}$ universality class), the trace is taken in the space $L^2(RP^{N-1})$, which is isomorphic to $L^2(S^{N-1})_{even} = \bigoplus^{\infty}_{k=0, \, k \, even} E_{N,k}$ and consists of the *even* functions on $S^{N-1}$. Physically, (4.83) expresses the fact that the finite-size-scaling limit corresponds to the continuum theory in a finite periodic box.

Notice that since the coefficients $\widetilde{a}_{N,k}$ are uniquely defined by the universality class of the theory, modulo a $k$-independent rescaling [which depends on the explicit definition of the scaling factor $\Lambda(J)$ but does not affect the products $\widetilde{a}_{N,k} \, \gamma$], these finite-size-scaling functions are universal modulo a rescaling of $\gamma$.

### 4.2.3 An Interesting Family of Universality Classes

Let us examine in more detail the finite-size-scaling curves for $R_{\chi;N,k}(J;L)$. In particular, we want to study their dependence on the different universality classes described in Section 4.1. As can be seen from the explicit expression of $R^{(0)}_{\chi;N,k}(\gamma)$, the finite-size-scaling curve is determined completely by $\{\widetilde{a}_{N,l}\}$. Therefore, we consider a family of universality classes parametrized by a continuous variable $B$, with $\widetilde{a}_{N,l}$ given by

$$\widetilde{a}_{N,l} = \begin{cases} \lambda_{N,l} & \text{for } l \text{ even} \\ \lambda_{N,l} + B & \text{for } l \text{ odd} \end{cases} \qquad (4.84)$$

(This family of universality classes was found in part (c) of the example in Section 4.1.4, and will be also found for the two-parameter Hamiltonians treated in Section 5.) We can get the $N$-vector universality class by choosing $B = 0$, and the $RP^{N-1}$ universality class by taking the limit $B \to \infty$.

---

[44] As mentioned in footnote 21 above, the expression (3.15) for $\chi_{N,k}$ requires some exegesis whenever $v_{N,l} = v_{N,m}$; and correspondingly (4.79)/(4.80) require exegesis whenever $\widetilde{a}_{N,l} = \widetilde{a}_{N,m}$. In such cases the combination $e^{-\gamma \widetilde{a}_{N,l}}/\Delta_{N;l,m}$, which occurs in (4.79) and in the numerator of (4.80), should be interpreted as $\gamma e^{-\gamma \widetilde{a}_{N,l}}/2$. This can be seen by going back to (3.14); it can also be obtained by the "quick-and-dirty" method of symmetrizing in $l$ and $m$, treating the $\widetilde{a}_{N,l}$ as if they were independent variables, and using l'Hôpital's rule. Note that in the $N$-vector and $RP^{N-1}$ universality classes this problem occurs only when $l = m$ (and hence $k$ is even). However, in the more general case (4.84), for certain values of $B$ one may have $\widetilde{a}_{N,l} = \widetilde{a}_{N,m}$ for $l \neq m$ (but only where $l - m$ and $k$ are odd).

Let us first look at the isovector sector ($k = 1$). In Figure 1 we plot $R^{(0)}_{\chi;N,1}(\gamma)$ for various values of the parameter $B$, using as an example $N = 4$. The graphs are drawn not as functions of $\gamma$, but as functions of the more "natural" variables $\bar{z}_k \equiv 1/(\tilde{a}_{N,k}\gamma)$ defined in equation (4.65). In Figure 1(a) we plot versus $\bar{z}_1$, while in Figure 1(b) we plot versus $\bar{z}_2$; different aspects of the behavior can be observed in these two plots.

A few interesting features that can be seen in the graphs for $N = 4$, and that can in fact be proven easily for arbitrary $N$, are:

(i) In the limit $\gamma \to 0$ (i.e. $\bar{z}_1, \bar{z}_2 \to \infty$), we have $\lim_{\gamma \to 0} R^{(0)}_{\chi;N,k}(\gamma) = 0$ (for finite $B$). More precisely, an expansion for small $\gamma$ of $R^{(0)}_{\chi;N,k}(\gamma; B)$ for arbitrary $k$ gives

$$\begin{aligned} R^{(0)}_{\chi;N,k}(\gamma; B) &= \tilde{a}_{N,k} \frac{\gamma}{2} [1 + O(\gamma)] \\ &= \frac{1}{2\bar{z}_k} + O\left(\frac{1}{\bar{z}_k^2}\right), \end{aligned} \quad (4.85)$$

independent of $B$. This behavior is observed in Figure 1(a), where the dashed curve represents (4.85).

(ii) For $0 \le B \le 2$ the curve is decreasing at small $\bar{z}_1$ (or $\bar{z}_2$), while for $B > 2$ it is increasing: this can be seen from a large-$\gamma$ expansion of $R^{(0)}_{\chi;N,1}(\gamma)$.

(iii) $\lim_{B \to \infty} R^{(0)}_{\chi;N,1}(\gamma; B) = 1$ for all fixed $\gamma > 0$ (i.e. all fixed $\bar{z}_2 < \infty$). This behavior is observed in Figure 1(b).

Let us next look at the isotensor sector ($k = 2$). In Figure 2 we plot the ratio $R^{(0)}_{\chi;N,2}(\gamma)$ as a function of $\bar{z}_2$ for three different values of the parameter $B$, for the case $N = 4$. Figures 2(a) and (b) show the same curves, but emphasizing different ranges of the variable $\bar{z}_2$. A few features that can be seen in the graphs for $N = 4$, and that can also be checked from the explicit formulae for general $N$, are the following:

(i) The curves are monotonically decreasing functions of the family parameter $B$ for each fixed value of the abscissa $\bar{z}_2$. [We can write

$$\begin{aligned} R^{(0)}_{\chi;N,k}(\gamma; B) &= R^{(0)}_{\chi;N,k}(\gamma; 0) - [R^{(0)}_{\chi;N,k}(\gamma; 0) - R^{(0)}_{\chi;N,k}(\gamma; \infty)] \\ &\quad \times \frac{\tilde{Z}^{(0)E}_N(\gamma)}{\tilde{Z}^{(0)}_N(\gamma)} \sum_{s=0}^{\infty} \left(1 - e^{-\gamma B}\right)^{s+1} \left(\frac{\tilde{Z}^{(0)O}_N(\gamma)}{\tilde{Z}^{(0)}_N(\gamma)}\right)^s \end{aligned} \quad (4.86)$$

where $\tilde{Z}^{(0)E}_N$ (or $\tilde{Z}^{(0)O}_N$) is $\tilde{Z}^{(0)}_N$ with the sum restricted to even (or odd) $l$, and all the $\tilde{Z}_N$'s are evaluated at $B = 0$. This proves the monotonicity in $B$ for fixed $\gamma$.]

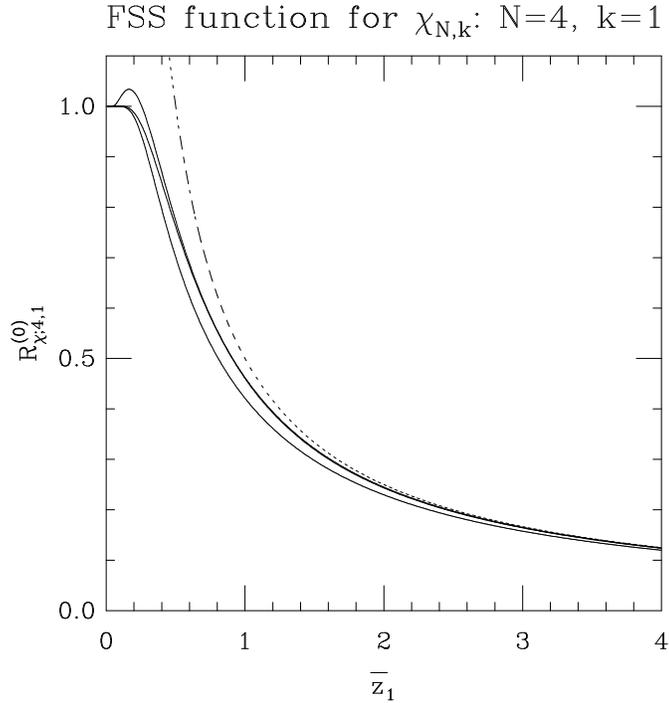

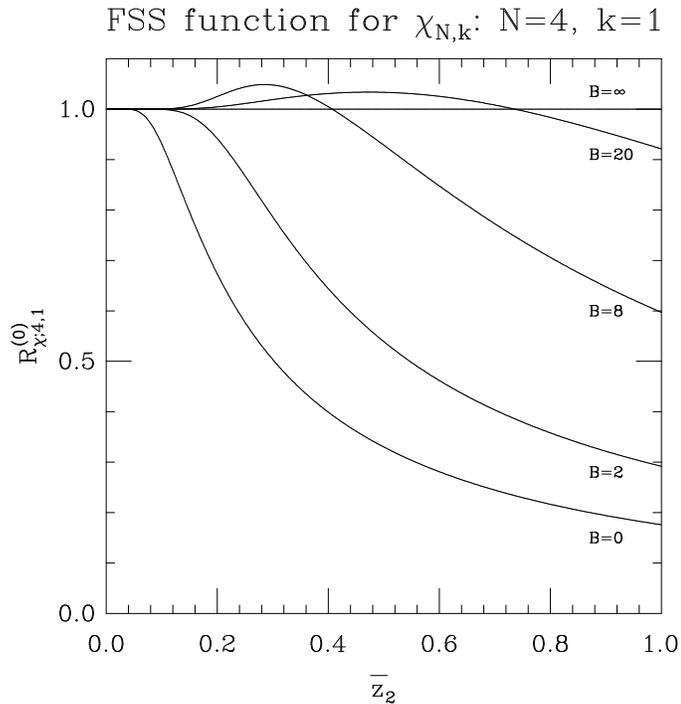

Figure 1: Graph of the ratio $R^{(0)}_{\chi;N,1}$ as a function of (a) $\bar{z}_1$ and (b) $\bar{z}_2$, for the case $N = 4$, for the family (4.84) of universality classes. In (a) the lowest curve corresponds to $B = 0$, which is the $N$-vector universality class; the highest curve is $B = 20$; the third solid curve is the limit $B \to +\infty$; and the dashed curve is the asymptotic behavior (4.85). In (b), the lowest curve is $B = 0$; the next three curves are $B = 2$, $B = 8$ and $B = 20$, respectively; and the straight line is the limit $B \to +\infty$.

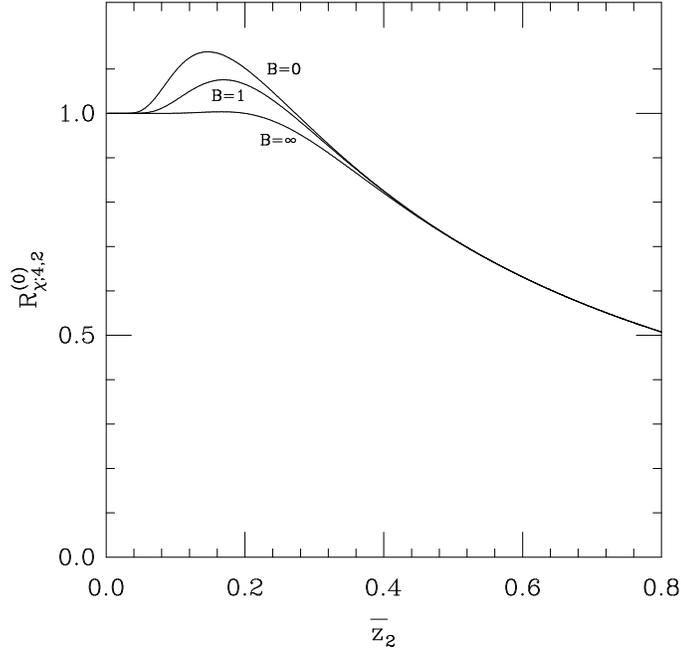

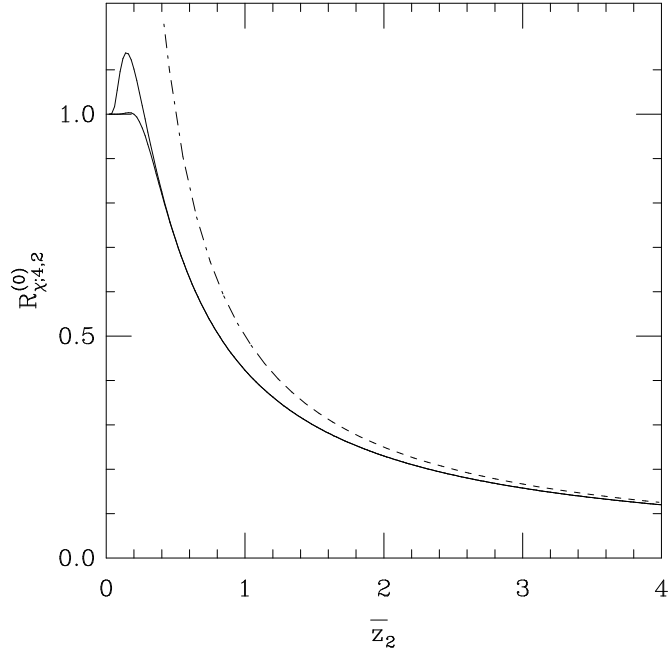

Figure 2: Graph of the ratio $R^{(0)}_{\chi;N,2}$ as a function of $\bar{z}_2$, for the case $N = 4$, for the family (4.84) of universality classes. In (a) the highest curve is $B = 0$, and corresponds to the $N$-vector universality class; the lowest curve is $B = \infty$, and corresponds to the the $RP^{N-1}$ universality class; the curve in-between is $B = 1$. In (b) the higher curve is obtained for $B = 0$ and the other for $B = +\infty$; the dashed curve is the asymptotic behavior (4.85).

(ii) The curves coincide *exponentially rapidly* for large $\overline{z}_2$ (i.e. small $\gamma$). Indeed, for $k$ even, the behavior for small $\gamma$ is (see Appendix B.6)

$$R^{(0)}_{\chi;N,k}(\gamma;B) \;=\; \frac{\lambda_{N,k}\,\gamma}{2}\, P_{N,k}(\gamma)\left[1 + O(e^{-\pi^2/4\gamma})\right]\,, \qquad (4.87)$$

where $P_{N,k}(\gamma)$ is a polynomial *independent of B*. (More precisely, in Appendix B we shall prove this behavior only for $k = 2$, but we *conjecture* that it holds for all even $k$.) This is why the dependence on $B$ disappears in Figure 2(b) long before the curves show the asymptotic behavior (4.85). Physically, the behavior (4.87) reflects the fact that the universality classes (4.84) are equivalent at all orders of perturbation theory; the $B$-dependence is a wholly *nonperturbative* effect. A similar situation occurs in the *two*-dimensional $\sigma$-models [5,6,7].

(iii) The curve for the $RP^{N-1}$ case is *not* monotonically decreasing as a function of $\overline{z}_2$, but is slightly increasing for small $\overline{z}_2$. (In fact, an expansion for large $\gamma$ shows that the function is increasing at small $\overline{z}_2$ for *all* values of $B$.)

Finally, we show the kind of finite-size-scaling plot that one usually considers in Monte Carlo simulations: here the infinite-volume correlation lengths are not known, so instead of $z_k$ or $\overline{z}_k$ we would use the variable

$$x_k \;=\; x_k(J;L) \;\equiv\; \frac{\xi^{(2nd)}_{N,k}(J;L)}{L}\,. \qquad (4.88)$$

Moreover, in this case we cannot compare lattice size $L$ with $\infty$ [as requested in (4.70)]; rather, we must compare $L$ with (for example) $2L$ [11,13]. In Figure 3(a,b) we show the analogues of Figure 1(a,b): that is, we plot the finite-size-scaling curves for the ratio $\chi_{N,1}(J;L)/\chi_{N,1}(J;2L)$ as a function of $x_1$ and $x_2$, respectively, for various values of $B$. The FSS curve for this ratio is given by $\chi^{(0)}_{N,1}(\gamma)/[2\chi^{(0)}_{N,1}(2\gamma)]$ plotted parametrically versus

$$\overline{x}_k \;\equiv\; \lim_{\substack{L,J \to \infty \\ \gamma \text{ fixed}}} x_k(J;L) \;=\; \xi^{(2nd)(0)}_{N,k}(\gamma)\,. \qquad (4.89)$$

In Figure 4 we show the analogous plot for the isotensor sector.

It is interesting to compare the curves in Figures 3(a) and 4 with those coming from a Monte Carlo study of a similar family of universality classes in *two* dimensions [5,6,7]. The curves are qualitatively very similar, although of course they are quantitatively different.

Now let us compare the finite-size-scaling curves to the explicit solution at finite $L$ and $J$. We show in Figure 5 the graph of the finite-size-scaling function of the spin-1 susceptibility for the $N = 4$ $N$-vector universality class [namely the lowest curve in Figure 1(a)] together with some points calculated from the exact expression of $R_{\chi;N,1}(J;L)$, for several values of $L$. More precisely, we plot:

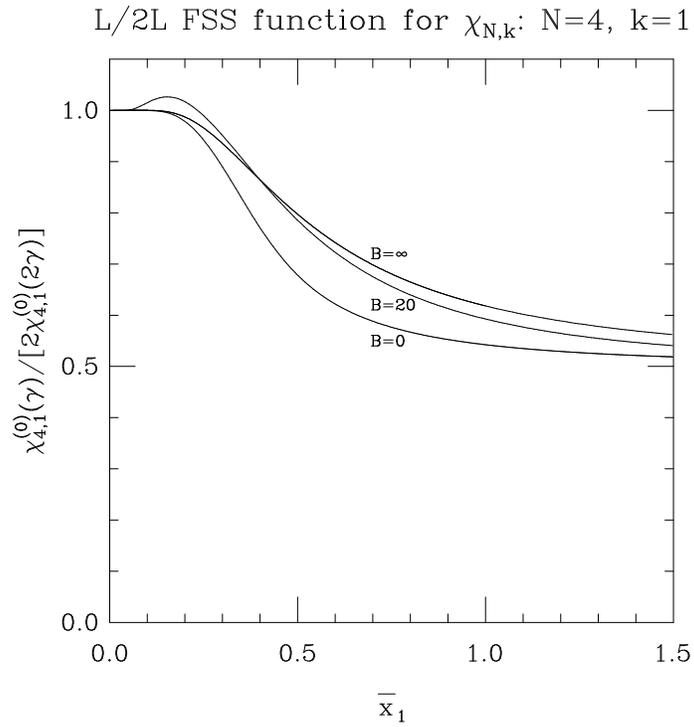

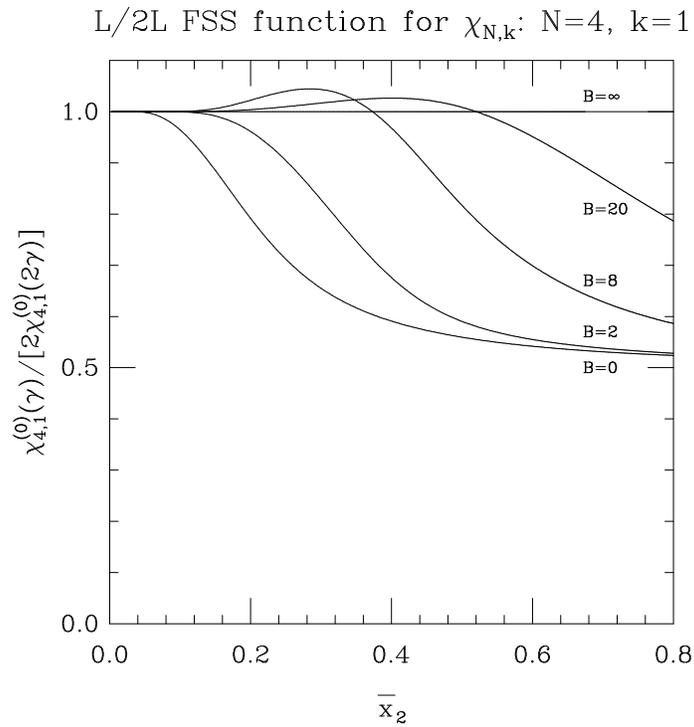

Figure 3: Graph of the ratio $\chi_{N,1}^{(0)}(\gamma)/[2\chi_{N,1}^{(0)}(2\gamma)]$ as a function of (a) $\overline{x}_1$ and (b) $\overline{x}_2$, for the case $N = 4$, for the family (4.84) of universality classes.

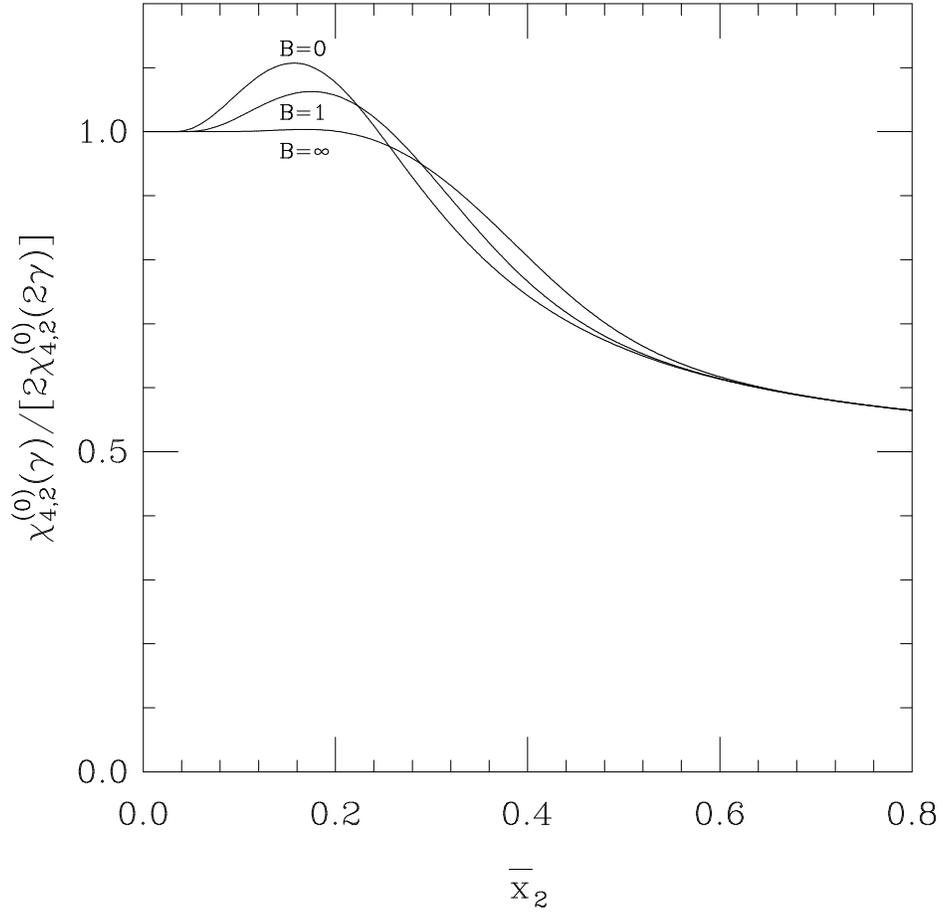

Figure 4: Graph of the ratio $\chi_{N,2}^{(0)}(\gamma)/[2\chi_{N,2}^{(0)}(2\gamma)]$ as a function of $\overline{x}_2$, for the case $N = 4$, for the family (4.84) of universality classes.

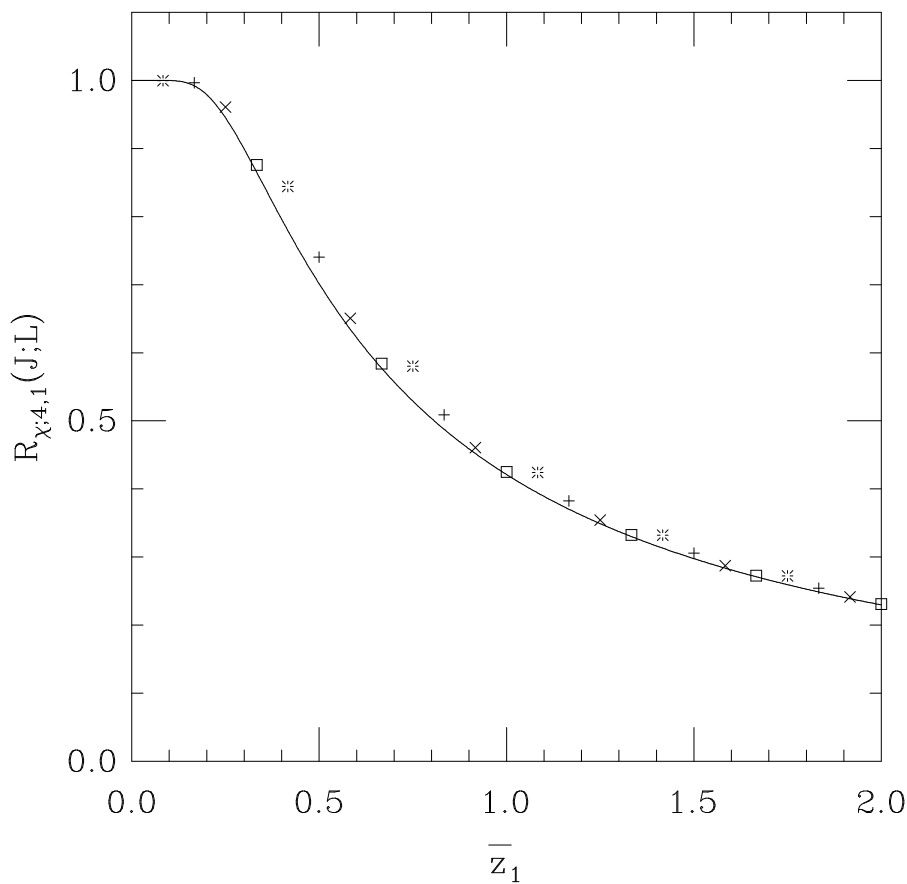

Figure 5: Graph of $R_{\chi;N,1}(J;L)$ as a function of $\overline{z}_1 = (\tilde{a}_{N,1}\gamma)^{-1}$ for the one-dimensional $N = 4$ $N$-vector model. Symbols indicate: $L = 4$ ($*$), 8 ($+$), 16 ($\times$), 32 ($\square$). The corresponding finite-size-scaling function is also plotted.

- **Points**: exact values of $R_{\chi;N,1}(J;L)$ plotted versus $\overline{z}_1 \equiv 1/(\tilde{a}_{N,1}L\Lambda(J))$ for the $N$-vector *model* (3.2), using the formulae from Section 3.1 and $v_{N,k}(J)$ given by the Bessel functions in (2.32). Here $\Lambda(J) = 1/(2J)$ and $\tilde{a}_{N,k} = \lambda_{N,k}$.

- **Curve**: the finite-size-scaling function $R^{(0)}_{\chi;N,1}(\gamma)$ for the $N$-vector *universality class*, as a function of $\overline{z}_1 = (\tilde{a}_{N,1}\gamma)^{-1}$.

Notice that for small values of $L$ there are significant corrections to finite-size scaling. These corrections will be discussed in the next section, and we will show that they are of order $1/L$, or equivalently, of order $\Lambda(J)$.

Let us now return to the original (and most natural) scaling variables $z_k \equiv \xi^{(exp)}_{N,k}(J;\infty)/L$. The finite-size-scaling curves are of course the same, since $z_k$ and $\overline{z}_k$ coincide at leading order. However, the meaning of the *points* in the plot is different, and the corrections to finite-size scaling may differ in the two variables. In Figure 6 we show the same data points as in Figure 5, but plotted versus $z_1$ instead of $\overline{z}_1$. In Figure 7 we make the analogous plot for the "$L/2L$" FSS function plotted versus $x_1$ (which is a close relative of $z_1$). Clearly, the corrections to finite-size scaling are considerably smaller if we use variables $z_k$ or $x_k$ rather than the variables $\overline{z}_k$. In Section 4.3 we will show that the $1/L$ corrections *vanish* in the variables $z_k$ or $x_k$; the leading correction appears to be of order $1/L^2$.

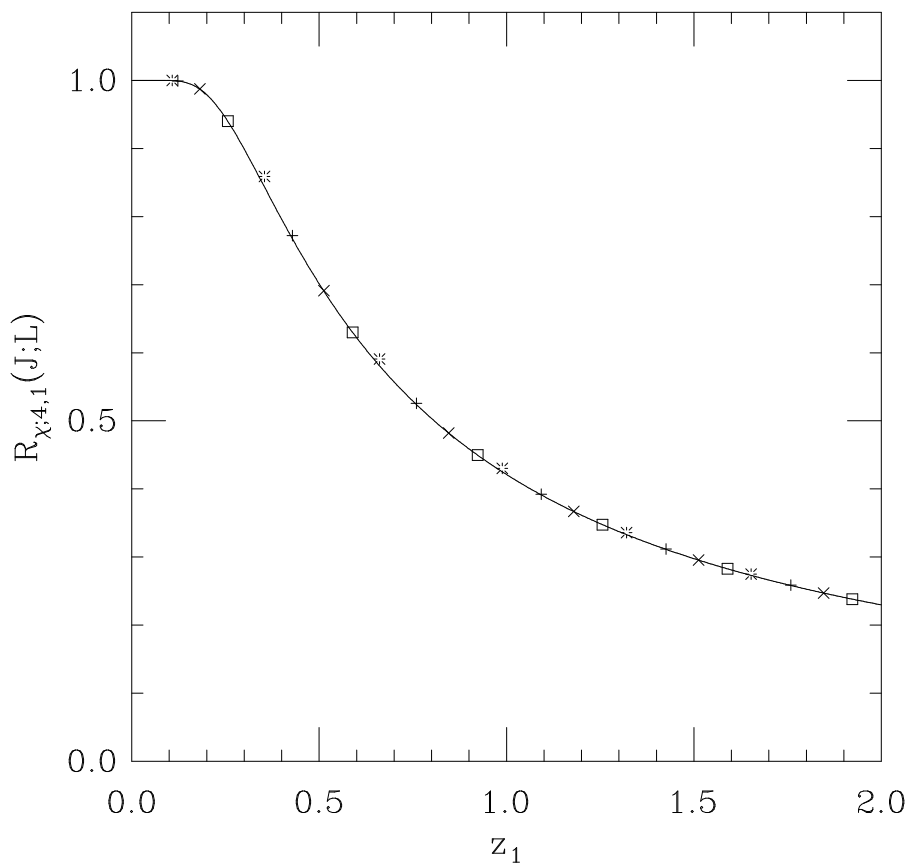

Figure 6: Graph of $R_{\chi;N,1}(J;L)$ as a function of $z_1 = \xi^{(exp)}_{N,1}(J;\infty)/L$ for the one-dimensional $N = 4$ $N$-vector model. Symbols indicate: $L = 4$ ($*$), 8 ($+$), 16 ($\times$), 32 ($\square$). The corresponding finite-size-scaling function is also plotted.

The fact that the plot in terms of $z_k$ or $x_k$ shows better agreement with the finite-size-scaling curve than the plot in terms of $\overline{z}_k$ can be interpreted as a manifestation of the difference between "scaling" and "asymptotic scaling". As used by lattice quantum field theorists, these terms mean the following (see e.g. [7]): "Scaling" denotes the convergence to the continuum limit for dimensionless ratios of long-distance observables and for the relations between such observables. "Asymptotic scaling", by contrast, denotes the convergence to the asymptotic predictions (e.g. as $J \to \infty$) for the relation between long-distance observables (such as $\chi$ or $\xi$ or combinations thereof) and the "bare" parameters in the Hamiltonian (i.e. $J$).[45] Clearly, asymptotic scaling (to a given degree of accuracy) implies scaling (to the same degree of accuracy), but not conversely; otherwise put, the corrections to asymptotic scaling may be much larger than the corrections to scaling. Now, $R_{\chi;N,k}(J;L) \equiv \chi_{N,k}(J;L)/\chi_{N,k}(J;\infty)$ and $z_k(J;L) \equiv \xi_{N,k}^{(exp)}(J;\infty)/L$ are examples of dimensionless ratios of long-distance observables, and the rapid convergence to the continuum limit in Figure 6 is an example of rapid "scaling". On the other hand, $\overline{z}_k(J;L) \equiv 1/[\widetilde{a}_{N,k} L \Lambda(J)]$ is an example of a bare parameter (by virtue of its explicit dependence on $J$), and the slower convergence to the continuum limit in Figure 5 is an example of the less-rapid onset of "asymptotic scaling". In this model the corrections to asymptotic scaling are of order $1/L$, while the corrections to scaling appear to be of order $1/L^2$.

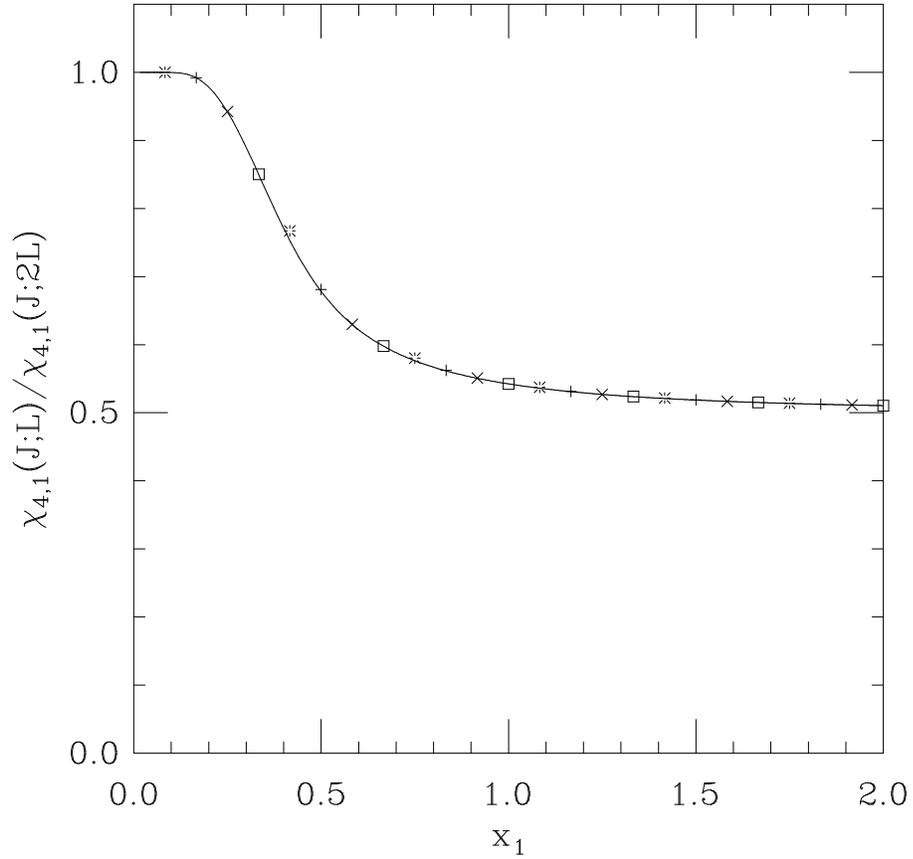

Figure 7: Graph of $\chi_{N,1}(J;L)/\chi_{N,1}(J;2L)$ as a function of $x_1 = \xi_{N,1}^{(2nd)}(J;L)/L$ for the one-dimensional $N$-vector model with $N = 4$. Symbols indicate: $L = 4$ ($*$), 8 ($+$), 16 ($\times$), 32 ($\square$). The corresponding finite-size-scaling function is also plotted.

## 4.3 Corrections to Finite-Size Scaling

In this section we want to compute the corrections to the finite-size-scaling functions. We assume a large-$J$ expansion of the form[46]

$$v_{N,k}(J) = 1 - \widetilde{a}_{N,k} \Lambda(J) + \widetilde{b}_{N,k} \Lambda_{corr}(J) + o(\Lambda_{corr}(J)) , \qquad (4.90)$$

where $\Lambda_{corr}(J)/\Lambda(J)$ goes to zero for $J \to +\infty$.

In the limit $J \to +\infty$, $L \to \infty$ at $L \Lambda(J) \equiv \gamma$ fixed we have

$$\begin{aligned} v_{N,k}(J)^L &= \exp(-\gamma \widetilde{a}_{N,k}) \left[ 1 + \gamma \widetilde{b}_{N,k} \frac{\Lambda_{corr}(J)}{\Lambda(J)} - \frac{\gamma}{2} \widetilde{a}_{N,k}^2 \Lambda(J) \right. \\ &\qquad\qquad\qquad\qquad \left. + o\left(\Lambda(J), \frac{\Lambda_{corr}(J)}{\Lambda(J)}\right) \right] \\ &= \exp(-\gamma \widetilde{a}_{N,k}) \left[ 1 - \gamma \overline{b}_{N,k} \overline{\Lambda}(J) + o\left(\overline{\Lambda}(J)\right) \right] , \qquad (4.91) \end{aligned}$$

where $\overline{\Lambda}(J)$ is the more slowly decreasing of $\Lambda_{corr}(J)/\Lambda(J)$ and $\Lambda(J)$, and $-\overline{b}_{N,k}$ is the corresponding coefficient. (Of course, if $\Lambda(J)$ and $\Lambda_{corr}(J)/\Lambda(J)$ are of the same order, then $-\overline{b}_{N,k}$ is given by the sum of the two coefficients.) Plugging this

---

[45] In place of the bare parameters, one may alternatively use *short-distance* quantities such as the energy $E$, inasmuch as they play a similar physical role.

[46] In Section 4.1.2 we computed $\widetilde{a}_{N,k}$ and $\widetilde{b}_{N,k}$ for two simple cases of Hamiltonians $\widetilde{h}(t)$:

(i) $t = 1$ is the only absolute maximum and $\widetilde{h}'(1) > 0$. (This generalizes the $N$-vector model.)

(ii) $t = \pm 1$ are the only absolute maxima, $\widetilde{h}(t)$ is an *even* function and $\widetilde{h}'(1) > 0$. (This is the *symmetric* subcase of what we called in Section 4.1.2 the "second simple case": see footnote 33 and the text following it. It generalizes the $RP^{N-1}$ model.) As discussed before, we look only at the even-$k$ sector in this case.

In both simple cases we obtain the same coefficients $\widetilde{a}_{N,k}$, $\widetilde{b}_{N,k}$ [cf. (4.19)–(4.21)]; and we have $\Lambda(J) \sim 1/J$, $\Lambda_{corr}(J) \sim 1/J^2$.

expression in (3.13)–(3.15) we immediately obtain the corrections to the finite-size-scaling functions. Notice that the procedure is straightforward, and by adding more terms in the expansion (4.90) we can compute the corrections to any arbitrary order. For example for the susceptibility we obtain

$$\chi_{N,k}(J;L) = L\left[\chi_{N,k}^{(0)}(\gamma) + \chi_{N,k}^{(1)}(\gamma)\,\overline{\Lambda}(J) + o\left(\overline{\Lambda}(J)\right)\right] \quad (4.92)$$

where

$$\chi_{N,k}^{(1)}(\gamma) = \frac{2}{\gamma\,\widetilde{Z}_N^{(0)}(\gamma)} \sum_{l,m=0}^{\infty} \frac{\mathcal{C}_{N;k,l,m}^2}{\mathcal{N}_{N,k}} \frac{e^{-\gamma\widetilde{a}_{N,l}}}{\Delta_{N;l,m}} \left[\frac{\overline{b}_{N,l} - \overline{b}_{N,m}}{\Delta_{N;l,m}} - \gamma\,\overline{b}_{N,l}\right.$$
$$\left. + \frac{\gamma}{\widetilde{Z}_N^{(0)}(\gamma)} \sum_{n=0}^{\infty} \mathcal{N}_{N,n} e^{-\gamma\widetilde{a}_{N,n}}\,\overline{b}_{N,n}\right]. \quad (4.93)$$

In the same way, the ratio $R_{\chi;N,k}(J;L)$ defined in (4.70) is given by

$$R_{\chi;N,k}(J;L) = R_{\chi;N,k}^{(0)}(\gamma) + \overline{\Lambda}(J)\,R_{\chi;N,k}^{(1)}(\gamma) + o\left(\overline{\Lambda}(J)\right) \quad (4.94)$$

with

$$R_{\chi;N,k}^{(1)}(\gamma) = \frac{\gamma}{2}\widetilde{a}_{N,k}\left[\chi_{N,k}^{(1)}(\gamma) + \frac{\overline{b}_{N,k}}{\widetilde{a}_{N,k}}\chi_{N,k}^{(0)}(\gamma)\right] \quad (4.95)$$

where we have used (4.74) and the expansion

$$\chi_{N,k}(J;\infty) = \frac{2}{\widetilde{a}_{N,k}\,\Lambda(J)}\left[1 - \frac{\overline{b}_{N,k}}{\widetilde{a}_{N,k}}\overline{\Lambda}(J) + o\left(\overline{\Lambda}(J)\right)\right]. \quad (4.96)$$

These formulae simplify considerably in case $\overline{b}_{N,k}$ has the simple structure

$$\overline{b}_{N,k} = \widetilde{a}_{N,k}\,c_N \quad (4.97)$$

for some coefficient $c_N$. In this case we obtain the simple formulae

$$\chi_{N,k}^{(1)}(\gamma) = c_N\,\gamma\,\frac{d}{d\gamma}\chi_{N,k}^{(0)}(\gamma) \quad (4.98)$$

$$R_{\chi;N,k}^{(1)}(\gamma) = \frac{\gamma}{2}\widetilde{a}_{N,k}\,c_N\,\frac{d}{d\gamma}\left[\gamma\,\chi_{N,k}^{(0)}(\gamma)\right] \quad (4.99)$$

In particular, the two simple cases mentioned above (see footnote 46) satisfy this requirement with

$$c_N = (N+1)\,r + 1\,. \quad (4.100)$$

Indeed, we have here $\Lambda_{corr}(J) = \Lambda(J)^2$, so that $\overline{\Lambda}(J) = \Lambda(J) \equiv 1/[2\widetilde{h}'(1)J]$ and $\overline{b}_{N,k} = \frac{1}{2}\widetilde{a}_{N,k}^2 - \widetilde{b}_{N,k}$; the claim then follows from (4.19)–(4.21). Note that $R_{\chi;N,1}^{(1)}(\gamma)$ here depends on $r$ only through the global factor $(N+1)r+1$.

We now study in more detail the classes satisfying (4.97).

In Figure 8 we show the correction to finite-size scaling for the spin-1 susceptibility for the $N = 4, 8$ $N$-vector universality class ($c_N$ given by (4.100) with $r = 0$). We plot as a function of $\overline{z}_1$ [defined in (4.65)]

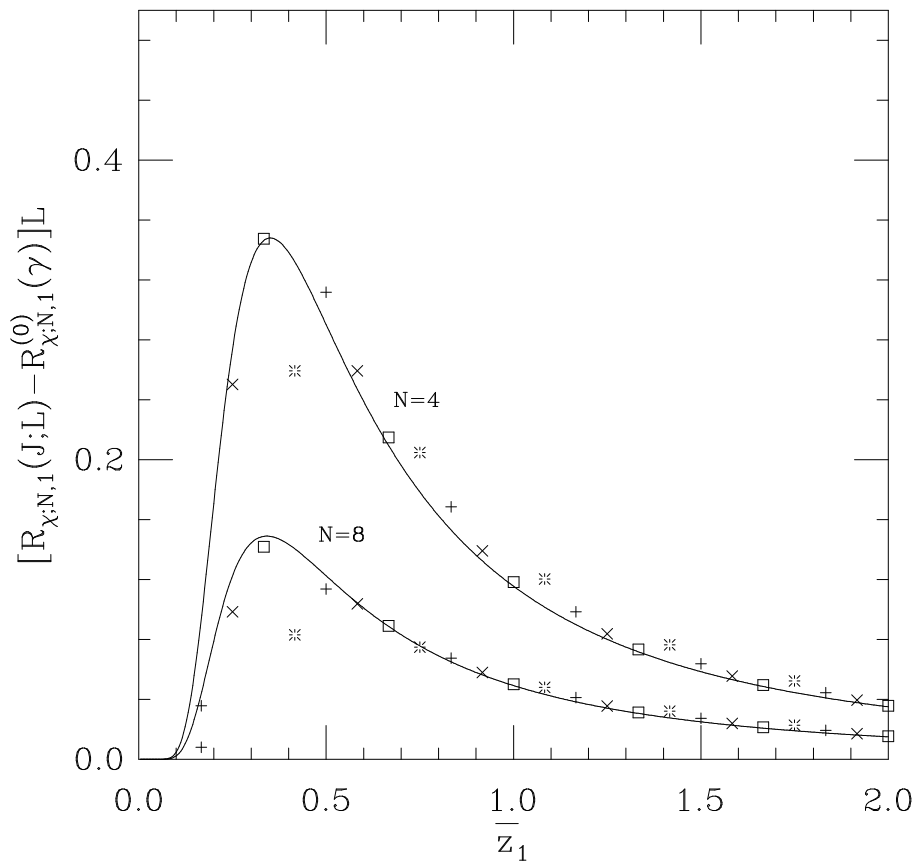

Figure 8: Corrections $[R_{\chi;N,1}(J;L) - R^{(0)}_{\chi;N,1}(\gamma)]L$ to the finite-size-scaling function of $R_{\chi;N,1}(J;L)$ for the one-dimensional $N$-vector model as a function of $\overline{z}_1$. The higher and lower curves correspond to $N = 4$ and $N = 8$, respectively. Symbols indicate $L = 4$ ($*$), 8 ($+$), 16 ($\times$), 32 ($\square$). The function (4.99) (for these two values of $N$) is also plotted.

- **Points:** the difference

$$[R_{\chi;N,1}(J;L) - R^{(0)}_{\chi;N,1}(\gamma)]L \qquad (4.101)$$

for the $N$-vector model (with $N = 4, 8$ and for different lattice sizes) and

- **Curves:** the corresponding limiting curves $R^{(1)}_{\chi;N,1}(\gamma)$ given by (4.99)/(4.100) with $r = 0$.

Now we can compare the corrections shown in Figure 8 to the corrections for the case in which we choose $z_1 = \xi^{(\#)}_{N,1}(J;\infty)/L$ as the variable in the abscissa. As was done in the previous section, we must replace $\gamma$ in formula (4.94) by its expansion in terms of $z_k$ to the desired order, given by

$$\gamma \;=\; \frac{1}{\widetilde{a}_{N,k}\,\overline{z}_k}\left[1 \,-\, \frac{\overline{b}_{N,k}}{\widetilde{a}_{N,k}}\,\overline{\Lambda}(J) \,+\, o\left(\overline{\Lambda}(J)\right)\right] \;. \tag{4.102}$$

We get

$$R_{\chi;N,1}(J;L) \;=\; R^{(0)}_{\chi;N,1}\left(\frac{1}{\widetilde{a}_{N,k}\,\overline{z}_k}\right) \,+\, o\left(\overline{\Lambda}(J)\right) \;. \tag{4.103}$$

That is, there are *no* corrections at order $\overline{\Lambda}(J)$ in this case; in other words, the corrections of order $1/L$ found for Figure 8 are not present here! Empirically it appears that the leading corrections are in fact of order $1/L^2$: see Figure 9. (The limiting curve shown was evaluated numerically by taking a very large value of $L$.) The same holds for the corrections to finite-size scaling written in terms of the variable $x_k$ defined in the previous section, as shown in Figure 10. As discussed in Section 4.2.3, this is a manifestation of the difference between "scaling" and "asymptotic scaling".

## 5 A Two-Parameter Family of Hamiltonians

In the previous section we have investigated the continuum limits arising from a one-parameter family of interactions. One might imagine that, by considering many-parameter families of Hamiltonians and taking appropriate trajectories in the multi-parameter space, one could find additional continuum limits. We have investigated this problem for a two-parameter family of interactions given by

$$h(\boldsymbol{\sigma}\cdot\boldsymbol{\tau}) \;=\; J_V\,\widetilde{h}_V(\boldsymbol{\sigma}\cdot\boldsymbol{\tau}) \,+\, J_T\,\widetilde{h}_T(\boldsymbol{\sigma}\cdot\boldsymbol{\tau}) \;. \tag{5.1}$$

We will not study the problem for generic $\widetilde{h}_V$ and $\widetilde{h}_T$, but will restrict our discussion to the case in which $\widetilde{h}_V$ is an odd function and has a unique maximum at 1 while $\widetilde{h}_T$ is an even function and has maxima at $\pm 1$. Moreover, we will assume $\widetilde{h}'_V(1) > 0$ and $\widetilde{h}'_T(1) > 0$ and we will consider only the case $J_V, J_T > 0$. This generalizes the mixed isovector/isotensor model

$$h(\boldsymbol{\sigma}\cdot\boldsymbol{\tau}) \;=\; J_V\,\boldsymbol{\sigma}\cdot\boldsymbol{\tau} \,+\, \frac{J_T}{2}\,(\boldsymbol{\sigma}\cdot\boldsymbol{\tau})^2 \tag{5.2}$$

studied in [5,6,7].

We want now to find the critical points of these theories. Since in dimension $d = 1$ no phase transition can occur for finite values of the couplings, we must investigate the limit in which at least one of the two couplings tends to infinity. It is trivial to see that in the limit $J_T \to +\infty$ with $J_V$ fixed and finite, one recovers the $RP^{N-1}$ universality class; while in the limit $J_V \to +\infty$ with $J_T$ fixed and finite, one reobtains the $N$-vector universality class. It therefore remains only to investigate the case in which both $J_V$ and $J_T$ go to infinity.

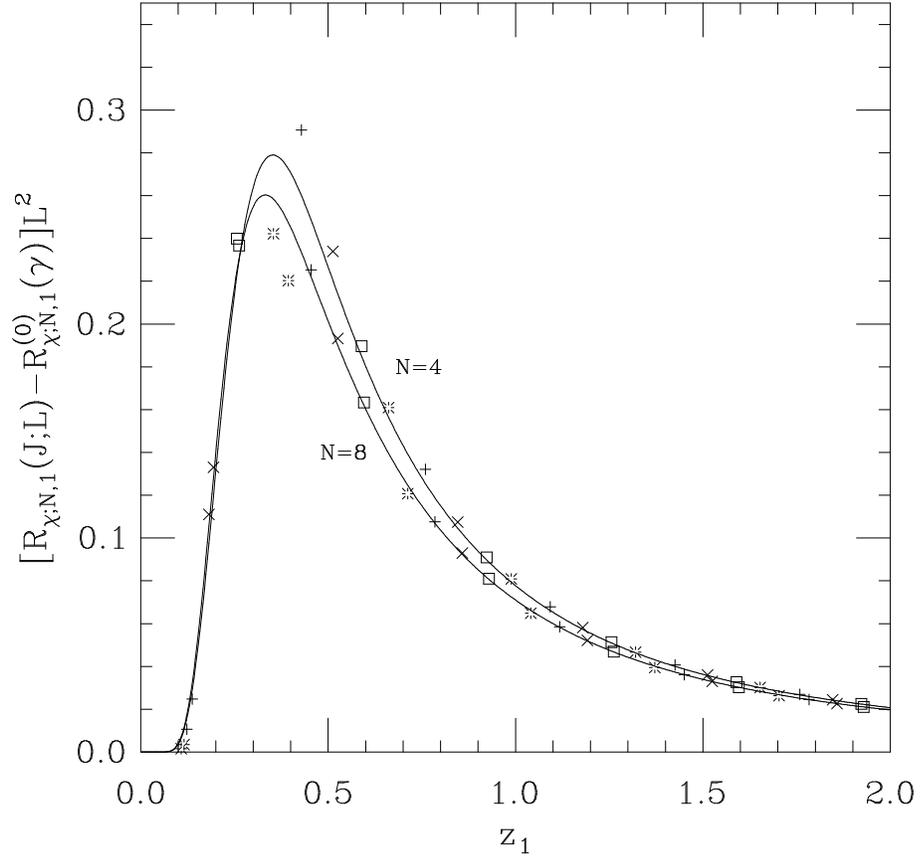

Figure 9: Corrections $[R_{\chi;N,1}(J;L) - R^{(0)}_{\chi;N,1}(\gamma)]L^2$ to the finite-size-scaling function of $R_{\chi;N,1}(J;L)$ for the one-dimensional $N$-vector model as a function of $z_1$. The higher and lower curves correspond to $N = 4$ and $N = 8$, respectively. Symbols indicate: $L = 4$ ($\ast$), 8 (+), 16 (×), 32 ($\square$). The corresponding limiting curve (numerically evaluated for a large value of $L$) is shown for both cases.

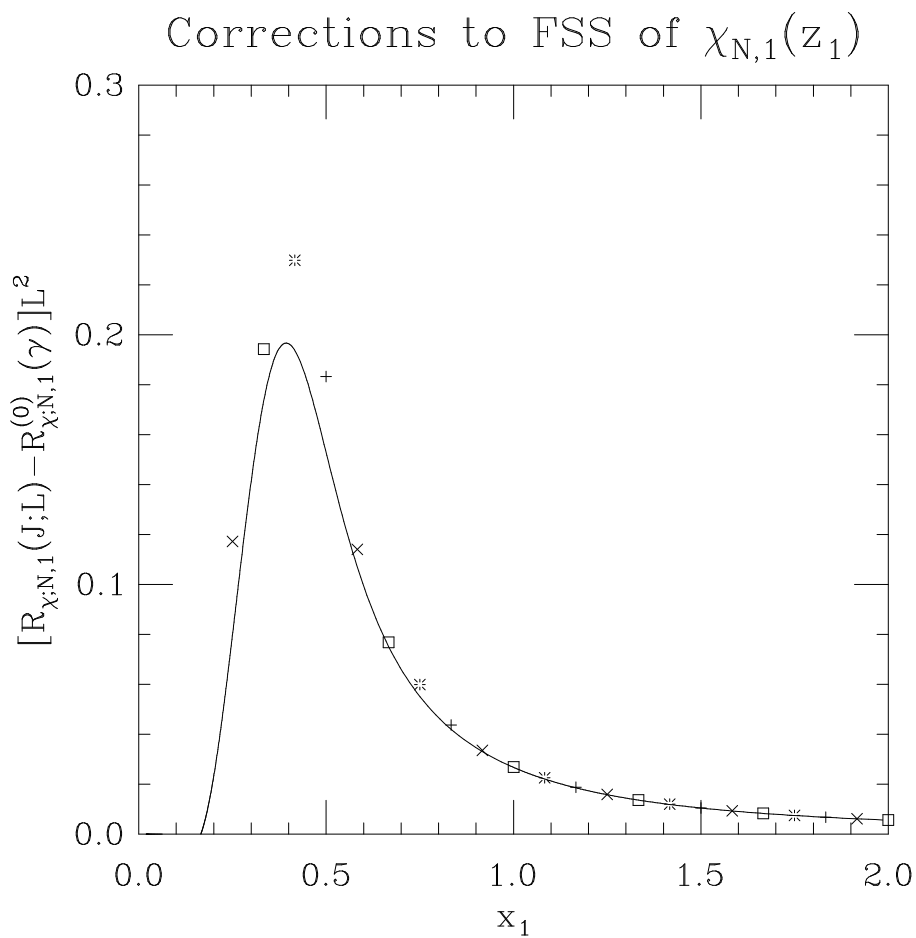

Figure 10: Graph of $\{[\chi_{N,1}(J;L)/\chi_{N,1}(J;2L)]-[\chi^{(0)}_{N,1}(\gamma)/\chi^{(0)}_{N,1}(2\gamma)]\}L^2$ as a function of $x_1$ for the one-dimensional $N=4$ $N$-vector model. Symbols indicate: $L=4$ ($*$), 8 ($+$), 16 ($\times$), 32 ($\square$). The corresponding limiting curve (numerically evaluated for a large value of $L$) is also shown.

(a) Let us first consider trajectories such that $J_T/J_V \to 0$ as $J_V, J_T \to +\infty$. In this case, from (2.27), we get

$$F_{N,k}(J_V, J_T) = f_N(J_V, J_T)\left[1 - \frac{a^V_{N,k}}{J_V} + O\left(\frac{1}{J_V^2}, \frac{J_T}{J_V^2}\right)\right] \tag{5.3}$$

where

$$f_N(J_V, J_T) = \frac{e^{h(1)}}{[2\pi h'(1)]^{1/2}} \Gamma\left(\frac{N}{2}\right) \left(\frac{h'(1)}{2}\right)^{1-N/2} \tag{5.4}$$

$$a^V_{N,k} = \frac{1}{2\tilde{h}'_V(1)}\left[\lambda_{N,k} + \frac{1}{4}(N-1)(N-3) - \frac{N^2-1}{4} r_V\right] \tag{5.5}$$

where $\lambda_{N,k}$ are the eigenvalues of the Laplace-Beltrami operator on the sphere and

$$r_V \equiv \tilde{h}''_V(1)/\tilde{h}'_V(1) \,. \tag{5.6}$$

Thus, in this limit $\tilde{h}_T$ is an irrelevant perturbation, and we get the $N$-vector universality class.

(b) Next let us consider trajectories such that $J_T/J_V = \alpha$ with $0 < \alpha < \infty$. In this case we can rewrite (5.1) as

$$\tilde{h}(\boldsymbol{\sigma} \cdot \boldsymbol{\tau}) = J_V\left[\tilde{h}_V(\boldsymbol{\sigma} \cdot \boldsymbol{\tau}) + \alpha \tilde{h}_T(\boldsymbol{\sigma} \cdot \boldsymbol{\tau})\right] \,. \tag{5.7}$$

This is a one-parameter family of interactions with Hamiltonian $\tilde{h}$ which has a unique maximum at $t = 1$. Thus also in this case we get the $N$-vector universality class.

(c) Finally, let us consider trajectories such that $J_T/J_V \to \infty$. We get from (2.27)

$$F_{N,k}(J_V, J_T) = f_N(J_V, J_T)\left\{1 - \frac{a^T_{N,k}}{J_T} + O\left(\frac{1}{J_T^2}, \frac{J_V}{J_T^2}\right) + \right.$$
$$\left. + (-1)^k \exp\left[-2 J_V \tilde{h}_V(1)\right]\left[1 + O\left(\frac{J_V}{J_T}\right)\right]\right\} \tag{5.8}$$

where $f_N(J_V, J_T)$ is defined in (5.4),

$$a^T_{N,k} = \frac{1}{2\tilde{h}'_T(1)}\left[\lambda_{N,k} + \frac{1}{4}(N-1)(N-3) - \frac{N^2-1}{4} r_T\right] \tag{5.9}$$

and

$$r_T \equiv \tilde{h}''_T(1)/\tilde{h}'_T(1) \,. \tag{5.10}$$

It follows that

$$v_{N,k}(J_V, J_T) \approx 1 - \frac{\tilde{a}_{N,k}}{J_T} - \left[1 - (-1)^k\right] \exp[-2 J_V \tilde{h}_V(1)] \,, \tag{5.11}$$

where
$$\tilde{a}_{N,k} = \frac{\lambda_{N,k}}{2\,\tilde{h}'_T(1)} \ . \tag{5.12}$$

To go further we must distinguish three different cases according to the relative size of the two correction terms in (5.11), i.e. according to the behavior of the product $J_T \exp[-2J_V \tilde{h}_V(1)]$.

(i) Let us first consider trajectories for which $J_T \exp[-2J_V \tilde{h}_V(1)]$ goes to zero. In this case the exponential term in (5.11) goes to zero faster than the $1/J_T$ term and can thus be dropped. We reobtain in this way the $N$-vector universality class.

(ii) In the opposite case, when $J_T \exp[-2J_V \tilde{h}_V(1)] \to +\infty$, the leading behavior is given by
$$v_{N,k}(J_V, J_T) \approx \begin{cases} 1 - \tilde{a}_{N,k}/J_T & \text{for } k \text{ even} \\ 1 - 2\,\exp[-2J_V \tilde{h}_V(1)] & \text{for } k \text{ odd} \end{cases} \tag{5.13}$$

so that
$$m_{N,k} \approx \begin{cases} \tilde{a}_{N,k}/J_T & \text{for } k \text{ even} \\ 2\,\exp[-2J_V \tilde{h}_V(1)] & \text{for } k \text{ odd} \end{cases} \tag{5.14}$$

Thus for all $k$ odd
$$\mathcal{R}_{N,k} \equiv \frac{m_{N,2}}{m_{N,k}} \approx \frac{\tilde{a}_{N,2}}{2J} \exp[-2J_V \tilde{h}_V(1)] \to 0 \ , \tag{5.15}$$

so that these limits belong to the $RP^{N-1}$ universality class.

(iii) Finally let us suppose
$$J_T \exp[-2J_V \tilde{h}_V(1)] \to \frac{B}{4\tilde{h}'_T(1)} \tag{5.16}$$

where $B$ is a constant. In this case we get
$$v_{N,k}(J_V, J_T) \approx \begin{cases} 1 - \tilde{a}_{N,k}/J_T & \text{for } k \text{ even} \\ 1 - \left[\tilde{a}_{N,k} + \frac{B}{4\tilde{h}'_T(1)}\right]/J_T & \text{for } k \text{ odd} \end{cases} \tag{5.17}$$

and
$$m_{N,k} \approx \frac{1}{2\tilde{h}'_T(1)J_T} \times \begin{cases} \lambda_{N,k} & \text{for } k \text{ even} \\ \lambda_{N,k} + B & \text{for } k \text{ odd} \end{cases} \tag{5.18}$$

Thus, for $0 < B < \infty$ we find again the intermediate universality class (4.58) — interpolating between the $N$-vector and the $RP^{N-1}$ universality classes — which appeared already for one-parameter Hamiltonians with maxima at $t = \pm 1$.

Let us notice that in this last case the even-spin correlation functions are equal to those of the $N$-vector model, while the odd-spin ones are a product of an Ising correlation and the corresponding $N$-vector correlation. This family of theories is parametrized by $B$ ($0 \leq B \leq \infty$) and all the limiting mass ratios $\mathcal{R}_{N,k}$ are determined in terms of $B$ as in equation (4.50). Equivalently, we can choose any one of these ratios (with $k$ odd) to characterize the universality class; for instance we can use the ratio

$$\mathcal{R}_{N,1} \equiv \frac{m_{N,2}}{m_{N,1}} = \frac{\xi_{N,1}^{(exp)}}{\xi_{N,2}^{(exp)}} \,. \tag{5.19}$$

In the continuum limit we have

$$\mathcal{R}_{N,1} = \frac{2N}{N-1+B} \,. \tag{5.20}$$

Thus each theory is labeled by the ratio $\mathcal{R}_{N,1}$, which can assume any value from 0 to $2N/(N-1)$. Notice one special feature of $d=1$: the maximum value of $m_{N,2}/m_{N,1}$ is not 2 but rather is larger. This is due to the fact that in (spacetime) dimension $d=1$ scattering states cannot exist, so the usual inequality $m_{N,2} < 2m_{N,1}$ does not apply.

# A  Properties of Hyperspherical Harmonics

## A.1  Calculation of $\mathcal{N}_{N,k} \equiv \dim E_{N,k}$

Let us begin by computing the dimension of the linear space $E_{N,k}$ consisting of the completely symmetric and traceless tensors of rank $k$ over $\mathbb{R}^N$. This can be done by computing the dimension of the space of *all* completely symmetric tensors of rank $k$, and then subtracting from it the number of independent trace conditions that have to be imposed to ensure the tracelessness of these tensors. The number of linearly independent symmetric tensors is given by $\binom{N+k-1}{k}$ (the number of ways of placing $k$ prisoners in $N$ cells), and the number of traces is given by $\binom{N+k-3}{k-2}$ (the same binomial as before but considering only $k-2$ indices; of course this simply vanishes if $k < 2$). Therefore we obtain

$$\mathcal{N}_{N,k} \equiv \dim E_{N,k} = \binom{N+k-1}{k} - \binom{N+k-3}{k-2} \tag{A.1a}$$

$$= \frac{\Gamma(N+k)}{k!\,\Gamma(N)} - \frac{\Gamma(N+k-2)}{(k-2)!\,\Gamma(N)} \tag{A.1b}$$

$$= \frac{N+2k-2}{k!}\,\frac{\Gamma(N+k-2)}{\Gamma(N-1)} \tag{A.1c}$$

[with the interpretation $(-2)! = (-1)! = \infty$ in (A.1b)]. This proves formula (2.2).

We shall take (A.1b)/(A.1c) as the *definition* of $\mathcal{N}_{N,k}$ for $N$ noninteger. (By contrast, we shall always consider $k$ to be an *integer* $\geq 0$.) Note that for each fixed integer $k \geq 0$, $\mathcal{N}_{N,k}$ is a polynomial of degree $k$ in $N$; in particular, it is well-defined and finite for all real $N$. Note also that for each fixed $N$ (not necessarily integer), we have $\mathcal{N}_{N,k} \approx 2k^{N-2}/\Gamma(N-1)$ as $k \to \infty$.

The simple identity

$$\mathcal{N}_{N+2,k-1} = \frac{k(N+k-2)}{N(N-1)}\mathcal{N}_{N,k} \tag{A.2}$$

will play an important role in Appendix B.

Finally, for *integer* $N \geq 3$ we have the following formula:

$$\mathcal{N}_{N,k} = \frac{2}{(N-2)!} \times \begin{cases} \prod_{r=0}^{(N-4)/2}\left[\left(k+\frac{N-2}{2}\right)^2 - r^2\right] & \text{for } N \text{ even} \geq 4 \\ \left(k+\frac{N-2}{2}\right)\prod_{r=0}^{(N-5)/2}\left[\left(k+\frac{N-2}{2}\right)^2 - \left(r+\frac{1}{2}\right)^2\right] & \text{for } N \text{ odd} \geq 3 \end{cases} \tag{A.3}$$

In particular, for even (resp. odd) $N \geq 3$, $\mathcal{N}_{N,k}$ is an even (resp. odd) polynomial in the "shifted index" $k+\frac{N-2}{2}$. Note, however, that for $N = 2$, $\mathcal{N}_{N,k}$ is *not* a polynomial in $k$: for $k \geq 1$ we have $\mathcal{N}_{N,k} = 2$, in agreement with (A.3), but $\mathcal{N}_{N,0} = 1 \neq 2$.

## A.2  Some Basic Formulae

Let us now compute the integral of a product of an even number of $\boldsymbol{\sigma}$'s (an odd number gives trivially zero). Let us introduce, for an arbitrary vector $A_\alpha$, the quantity

$$I_k(A) = \int d\Omega(\boldsymbol{\sigma}) \; (A \cdot \boldsymbol{\sigma})^{2k} \;. \tag{A.4}$$

As $d\Omega(\boldsymbol{\sigma})$ is rotationally invariant, we have $I_k(RA) = I_k(A)$ for every $R \in O(N)$, so $I_k(A)$ depends only on $|A|$. Moreover, $I_k$ is manifestly a homogeneous function of degree $2k$. Hence we must have $I_k(A) = J_k \left[A^2\right]^k$ for some constant $J_k$. Now, as $\boldsymbol{\sigma}^2 = 1$, we get from (A.4)

$$\frac{\partial}{\partial A_\alpha}\frac{\partial}{\partial A_\alpha}I_k(A) = 2k(2k-1) I_{k-1}(A) \tag{A.5}$$

A recursion relation for $J_k$ immediately follows:

$$J_k = \frac{2k-1}{N+2k-2} J_{k-1} \tag{A.6}$$

Using $J_0 = 1$ we obtain the general solution

$$J_k = \frac{\Gamma\left(k+\frac{1}{2}\right)\Gamma\left(\frac{N}{2}\right)}{\Gamma\left(\frac{1}{2}\right)\Gamma\left(\frac{N}{2}+k\right)} \tag{A.7}$$

Taking then $2k$ derivatives with respect to $A$ in (A.4) we obtain the well-known result

$$\int d\Omega(\boldsymbol{\sigma}) \ \sigma^{\alpha_1} \cdots \sigma^{\alpha_{2k}} \ = \ \frac{\Gamma\left(\frac{N}{2}\right)}{2^k \ \Gamma\left(\frac{N}{2}+k\right)} \ \left(\delta^{\alpha_1 \alpha_2} \cdots \delta^{\alpha_{2k-1} \alpha_{2k}} + \ \ldots \right) \tag{A.8}$$

where the terms in parentheses correspond to the $(2k-1)!!$ different pairings of the indices.

Let us now prove the orthogonality relation (2.10). This is completely equivalent to proving that for arbitrary completely symmetric and traceless tensors $T_{N,k}$ and $U_{N,l}$ we have

$$\left[\int d\Omega(\boldsymbol{\sigma}) \ Y_{N,k}^{\alpha_1 \ldots \alpha_k}(\boldsymbol{\sigma}) \ Y_{N,l}^{\beta_1 \ldots \beta_l}(\boldsymbol{\sigma})\right] T_{N,k}^{\alpha_1 \ldots \alpha_k} \ U_{N,l}^{\beta_1 \ldots \beta_l} \ = \ \delta_{kl} \ T_{N,k} \cdot U_{N,k} \ . \tag{A.9}$$

To see this, let us first use the definition (2.3) and let us notice that the "Traces" terms do not give any contribution due to the tracelessness of $T_{N,k}$ and $U_{N,l}$. Thus the l.h.s. in (A.9) becomes simply

$$\mu_{N,k} \ \mu_{N,l} \int d\Omega(\boldsymbol{\sigma}) \ \sigma^{\alpha_1} \ldots \sigma^{\alpha_k} \sigma^{\beta_1} \ldots \sigma^{\beta_l} \ T_{N,k}^{\alpha_1 \ldots \alpha_k} \ U_{N,l}^{\beta_1 \ldots \beta_l} \ . \tag{A.10}$$

Then let us use (A.8). The only non-vanishing contributions come from those terms which do not contain $\delta^{\alpha_i \alpha_j}$ or $\delta^{\beta_i \beta_j}$; such terms exist only if $l = k$. In this last case there are $k!$ equivalent contractions and we end up with

$$\delta_{kl} \ \mu_{N,k}^2 \ \left[\frac{\Gamma\left(\frac{N}{2}\right)}{2^k \ \Gamma\left(\frac{N}{2}+k\right)} \ k!\right] \ T_{N,k} \cdot U_{N,k} \ = \ \delta_{kl} \ T_{N,k} \cdot U_{N,k} \ . \tag{A.11}$$

We thus obtain the orthogonality relation (2.10) for the $Y$'s, provided that they are normalized as in (2.4).

The general formula for the hyperspherical harmonics[47] can be obtained by using the fact that they are completely symmetric and traceless. The complete symmetry, together with the needed transformation properties under $SO(N)$, implies an expansion of the form

$$Y_{N,k}^{\alpha_1 \ldots \alpha_k}(\boldsymbol{\sigma}) \ = \ \mu_{N,k} \sum_{s=0}^{\lfloor k/2 \rfloor} A_{N,k;s} \ P_{(k;s)}^{\alpha_1 \ldots \alpha_k}(\boldsymbol{\sigma}) \ , \tag{A.12}$$

where

$$P_{(k;s)}^{\alpha_1 \ldots \alpha_k}(\boldsymbol{\sigma}) \ \equiv \ \delta^{\alpha_1 \alpha_2} \ldots \delta^{\alpha_{2s-1} \alpha_{2s}} \ \sigma^{\alpha_{2s+1}} \ldots \sigma^{\alpha_k} + \text{permutations} \tag{A.13}$$

and the number of permutations necessary to make $P_{(k;s)}$ completely symmetric is $(2s-1)!! \binom{k}{2s}$. Now we impose the tracelessness. We first note that

$$\delta_{\alpha_1 \alpha_2} \ P_{(k;s)}^{\alpha_1 \ldots \alpha_k}(\boldsymbol{\sigma}) \ = \ P_{(k-2;s)}^{\alpha_3 \ldots \alpha_k}(\boldsymbol{\sigma}) + (N + 2k - 2s - 2) \ P_{(k-2;s-1)}^{\alpha_3 \ldots \alpha_k}(\boldsymbol{\sigma}) \ , \tag{A.14}$$

---

[47]This result is obtained in [27]. Note that in their notation $P$ includes all the $k!$ permutations, i.e. it is $[(k-2s)! \, s! \, 2^s]^{-1}$ times our $P$.

and therefore we get, from $\delta_{\alpha_1\alpha_2}Y^{\alpha_1...\alpha_k}_{N,k}(\boldsymbol{\sigma}) = 0$, the recursion relation

$$A_{N,k;s-1} + (N + 2k - 2s - 2) A_{N,k;s} = 0 . \tag{A.15}$$

Imposing the normalization $A_{N,k;0} = 1$, we find

$$A_{N,k;s} = \frac{(-1)^s}{2^s} \frac{\Gamma\left(\frac{N}{2} + k - s - 1\right)}{\Gamma\left(\frac{N}{2} + k - 1\right)} . \tag{A.16}$$

Thus we can write

$$Y^{\alpha_1...\alpha_k}_{N,k}(\boldsymbol{\sigma}) = \mu_{N,k} \sum_{s=0}^{\lfloor k/2 \rfloor} \frac{(-1)^s}{2^s} \frac{\Gamma\left(\frac{N}{2} + k - s - 1\right)}{\Gamma\left(\frac{N}{2} + k - 1\right)} P^{\alpha_1...\alpha_k}_{(k;s)}(\boldsymbol{\sigma}) . \tag{A.17}$$

Let us now discuss the relation between the hyperspherical harmonics and the Gegenbauer polynomials. From Section 2 we know that $Y^{1...1}_{N,k}(\boldsymbol{\sigma})$ is the restriction to the unit sphere of a degree-$k$ harmonic polynomial. Moreover it depends only on $\sigma^1$, so that the polynomial can be written as $r^k P_k(x_1/r)$ where $r = |\mathbf{x}|$. Requiring the polynomial to satisfy Laplace's equation we get for $P_k(x)$ the equation

$$(1 - x^2) \frac{d^2 P_k}{dx^2} - x (N - 1) \frac{dP_k}{dx} + k (N + k - 2) P_k = 0 \tag{A.18}$$

The regular solution of this equation[48] is the Gegenbauer polynomial $C^{N/2-1}_k(x)$. The normalization is fixed by the requirement that

$$Y^{1...1}_{N,k}(\boldsymbol{\sigma}) = \mu_{N,k}(\sigma^1)^k + \text{lower-order terms} \tag{A.19}$$

We thus get[49]

$$Y^{1...1}_{N,k}(\boldsymbol{\sigma}) = \frac{\frac{N}{2} + k - 1}{\Gamma\left(\frac{N}{2}\right)\mu_{N,k}} \left[\Gamma\left(\frac{N}{2} - 1\right) C^{\frac{N}{2}-1}_k(\sigma^1)\right] \tag{A.20}$$

which, using the fact that

$$C^{N/2-1}_k(1) = \binom{N + k - 3}{k} \tag{A.21}$$

gives (2.23).

Note that we could have derived (A.17) by using (2.23) and the expansion of the Gegenbauer polynomials [see [19], formula 14 (p. 294)].

---

[48]See [31], p. 1031.

[49]Note that this formula is well-defined in the limit $N \to 2$. See footnote 14.

## A.3 The Projector onto Symmetric Traceless Tensors

Using properties 1 to 4 of the projector $I_{N,k}^{\alpha_1...\alpha_k;\beta_1...\beta_k}$ we can derive its general expression. We start by noting that the most general form satisfying the symmetry properties 1 and 2 is

$$I_{N,k}^{\alpha_1...\alpha_k;\beta_1...\beta_k} = \sum_{s=0}^{\lfloor k/2 \rfloor} B_{N,k;s}\, Q_{(k;s)}^{\alpha_1...\alpha_k;\beta_1...\beta_k}, \qquad (A.22)$$

where

$$Q_{(k;s)}^{\alpha_1...\alpha_k;\beta_1...\beta_k} \equiv \delta^{\alpha_1\alpha_2}...\delta^{\alpha_{2s-1}\alpha_{2s}}\, \delta^{\beta_1\beta_2}...\delta^{\beta_{2s-1}\beta_{2s}}\, \delta^{\alpha_{2s+1}\beta_{2s+1}}...\delta^{\alpha_k\beta_k}$$
$$+ \text{permutations} \qquad (A.23)$$

[i.e. there are $s$ $\delta$'s among the $\alpha$'s, $s$ among the $\beta$'s, and $k-2s$ connecting the $\alpha$'s with the $\beta$'s] and the number of permutations necessary to make $Q_{(k;s)}$ completely symmetric is given by

$$\left(\frac{k!}{s!\,2^s}\right)^2 \frac{1}{(k-2s)!}. \qquad (A.24)$$

Now notice that a consequence of properties 3 and 4 is

$$I_{N,k}^{\alpha_1...\alpha_k;\beta_1...\beta_k}\, Y_{N,k}^{\beta_1...\beta_k} = \mu_{N,k}\, I_{N,k}^{\alpha_1...\alpha_k;\beta_1...\beta_k}\, \sigma^{\beta_1}...\sigma^{\beta_k} = Y_{N,k}^{\alpha_1...\alpha_k} \qquad (A.25)$$

If we substitute in this expression the general formula (A.17) for $Y_{N,k}^{\alpha_1...\alpha_k}$ and formula (A.22), we obtain

$$B_{N,k;s} = A_{N,k;s}\, \frac{2^s\, s!}{k!}. \qquad (A.26)$$

Therefore we get the general expression

$$I_{N,k}^{\alpha_1...\alpha_k;\beta_1...\beta_k} = \sum_{s=0}^{\lfloor k/2 \rfloor} (-1)^s \frac{s!}{k!} \frac{\Gamma\left(\frac{N}{2}+k-s-1\right)}{\Gamma\left(\frac{N}{2}+k-1\right)}\, Q_{(k;s)}^{\alpha_1...\alpha_k;\beta_1...\beta_k} \qquad (A.27)$$

We must now check the (A.27) satisfies properties 3 and 4. Property 4 follows immediately:

$$I_{N,k}^{\alpha_1...\alpha_k;\beta_1...\beta_k}\, T_{N,k}^{\beta_1...\beta_k} = \frac{1}{k!}\, Q_{(k;0)}^{\alpha_1...\alpha_k;\beta_1...\beta_k}\, T_{N,k}^{\beta_1...\beta_k} = T_{N,k}^{\alpha_1...\alpha_k}, \qquad (A.28)$$

where in the first step we used the tracelessness of $T_{N,k}$ and in the second its symmetry. In order to prove property 3 let us introduce

$$\widehat{P}_{(k;s)}^{\alpha_1...\alpha_k}(\mathbf{u}) \equiv (u^2)^s\, P_{(k;s)}^{\alpha_1...\alpha_k}(\mathbf{u}) \qquad (A.29)$$

$$\widehat{Y}_{N,k}^{\alpha_1...\alpha_k}(\mathbf{u}) \equiv \mu_{N,k} \sum_{s=0}^{\lfloor k/2 \rfloor} A_{N,k;s}\, \widehat{P}_{(k;s)}^{\alpha_1...\alpha_k}(\mathbf{u}) \qquad (A.30)$$

where $\mathbf{u}$ is an arbitrary vector. We note that

$$\frac{\partial}{\partial u_{\beta_1}}...\frac{\partial}{\partial u_{\beta_k}}\, \widehat{P}_{(k;s)}^{\alpha_1...\alpha_k}(\mathbf{u}) = 2^s\, s!\, Q_{(k;s)}^{\alpha_1...\alpha_k;\beta_1...\beta_k} \qquad (A.31)$$

and therefore we can write

$$I_{N,k}^{\alpha_1...\alpha_k;\beta_1...\beta_k} = \frac{1}{\mu_{N,k}\,k!}\,\frac{\partial}{\partial u_{\beta_1}}\,\cdots\,\frac{\partial}{\partial u_{\beta_k}}\,\widehat{Y}_{N,k}^{\alpha_1...\alpha_k}(\mathbf{u})\;. \qquad (A.32)$$

Also, from (A.29), we obtain

$$\delta_{\alpha_1\alpha_2}\,\widehat{P}_{(k;s)}^{\alpha_1...\alpha_k}(\mathbf{u}) = u^2\,\widehat{P}_{(k-2;s)}^{\alpha_3...\alpha_k}(\mathbf{u}) + (N+2k-2s-2)\,\widehat{P}_{(k-2;s-1)}^{\alpha_3...\alpha_k}(\mathbf{u})\;; \qquad (A.33)$$

from this and (A.15) it follows that

$$\delta_{\alpha_1\alpha_2}\,\widehat{Y}_{N,k}^{\alpha_1...\alpha_k}(\mathbf{u}) = 0 \qquad (A.34)$$

and therefore property 3 is satisfied.

Finally, using (2.25), we can prove the trace formula (2.16). Indeed from the orthogonality relations (2.10), summing over all indices we have

$$I_{N,k}^{\alpha_1...\alpha_k;\alpha_1...\alpha_k} = \int d\Omega(\boldsymbol{\sigma})\;Y_{N,k}(\boldsymbol{\sigma})\cdot Y_{N,k}(\boldsymbol{\sigma}) \qquad (A.35)$$

The scalar product in the r.h.s. is rotationally invariant and as such it does not depend on $\boldsymbol{\sigma}$. Choosing $\boldsymbol{\sigma} = \mathbf{w} \equiv (1, 0, \ldots, 0)$ and using (2.25) we get

$$I_{N,k}^{\alpha_1...\alpha_k;\alpha_1...\alpha_k} = Y_{N,k}(\mathbf{w})\cdot Y_{N,k}(\mathbf{w}) = Y_{N,k}^{1...1}(\mathbf{w})\,\mu_{N,k} = \mathcal{N}_{N,k} \qquad (A.36)$$

## A.4 Expansions in Terms of Hyperspherical Harmonics

We want now to discuss the convergence of the expansion (2.18). We will begin by showing the following result: given a generic (real) tensor $T_{N,k}^{\alpha_1...\alpha_k}$, the hyperspherical harmonics satisfy the inequality

$$[\,T_{N,k}\cdot Y_{N,k}(\boldsymbol{\sigma})\,]^2 \leq (T_{N,k}\cdot T_{N,k})\,\mathcal{N}_{N,k}\;. \qquad (A.37)$$

Indeed, using Schwarz's inequality and (2.25), we get

$$\begin{aligned}|\,T_{N,k}\cdot Y_{N,k}(\boldsymbol{\sigma})\,|^2 &\leq (T_{N,k}\cdot T_{N,k})\,[\,Y_{N,k}(\boldsymbol{\sigma})\cdot Y_{N,k}(\boldsymbol{\sigma})\,]\\ &= (T_{N,k}\cdot T_{N,k})\,[\,Y_{N,k}(\mathbf{w})\cdot Y_{N,k}(\mathbf{w})\,]\\ &= (T_{N,k}\cdot T_{N,k})\,\mathcal{N}_{N,k}\end{aligned} \qquad (A.38)$$

Moreover, equality in (A.37) is possible only for those $\boldsymbol{\sigma}$ for which

$$T_{N,k}^{\alpha_1...\alpha_k} = \gamma\,Y_{N,k}^{\alpha_1...\alpha_k}(\boldsymbol{\sigma}) \qquad (A.39)$$

for some constant $\gamma$. This requires first of all $T_{N,k}^{\alpha_1...\alpha_k}$ to be symmetric and traceless. The constant $\gamma$ is easily obtained squaring the previous relation:

$$\gamma^2 = \frac{(T_{N,k}\cdot T_{N,k})}{\mathcal{N}_{N,k}}\;. \qquad (A.40)$$

Now let us consider the special case $T_{N,k}^{\alpha_1...\alpha_k} = Y_{N,k}^{\alpha_1...\alpha_k}(\mathbf{w})$ with $\mathbf{w} \equiv (1, 0, \ldots, 0)$ and $k \geq 1$. Equality in (A.37) is possible only if

$$Y_{N,k}^{\alpha_1...\alpha_k}(\boldsymbol{\sigma}) = \pm Y_{N,k}^{\alpha_1...\alpha_k}(\mathbf{w}) . \tag{A.41}$$

We will now prove that if $N \geq 3$, this implies $\boldsymbol{\sigma} = \pm \mathbf{w}$. Let us firstly notice that if $\boldsymbol{\sigma}$ satisfies (A.41), then every $\boldsymbol{\sigma}' = R\boldsymbol{\sigma}$ with $R \in SO(N)$ such that $R\mathbf{w} = \mathbf{w}$ also satisfies (A.41). Now if $\boldsymbol{\sigma} \neq \mathbf{w}$ there exists an index $\alpha \neq 1$ such that $\sigma^\alpha \neq 0$. If $N \geq 3$, we can consider rotations $R$ in the $(\alpha,\beta)$-plane (with $\beta \neq 1$) and generate solutions $\boldsymbol{\sigma}'$ with $\sigma'^\alpha$ assuming any value between $-\sigma^\alpha$ and $\sigma^\alpha$. This means that (A.41) with $\alpha_1 = \ldots = \alpha_k = \alpha$ has an infinite number of solutions, which is impossible as this is a polynomial equation in $\sigma^\alpha$. Thus for $N \geq 3$ we must have $\boldsymbol{\sigma} = \pm \mathbf{w}$. This result can easily be rephrased in terms of Gegenbauer polynomials: since $C_k^{N/2-1}(1) > 0$, it implies that

$$|C_k^{N/2-1}(x)| < C_k^{N/2-1}(1) \tag{A.42}$$

for $-1 < x < 1$ and $k \geq 1$.

For $N = 2$ the previous result is not true. Indeed in this case every $\boldsymbol{\sigma} = (\cos \pi j/k, \sin \pi j/k)$ with $j = 1, 2, \ldots, k$ is a solution of (A.41). To show this let us notice that $\dim E_{2,k} = 2$ for all $k$ so that there are only two independent equations to satisfy. Using complex indices $\pm = 1 \pm i2$ and noticing that, since $\delta^{++} = \delta^{--} = 0$, we obtain

$$Y_{2,k}^{+\cdots+} = \mu_{2,k}\, \sigma^+ \ldots \sigma^+ = \mu_{2,k}\, e^{ik\theta} \tag{A.43a}$$

$$Y_{2,k}^{-\cdots-} = \mu_{2,k}\, \sigma^- \ldots \sigma^- = \mu_{2,k}\, e^{-ik\theta} \tag{A.43b}$$

where $\boldsymbol{\sigma} = (\cos\theta, \sin\theta)$. Therefore, equations (A.41) are equivalent to

$$e^{ik\theta} = \pm 1 \tag{A.44a}$$

$$e^{-ik\theta} = \pm 1 \tag{A.44b}$$

which proves the result.

To discuss the convergence of the series (2.18) let us firstly notice that

$$\sum_{k=0}^{\infty} \widetilde{f}_k^{\alpha_1...\alpha_k}\, \widetilde{f}_k^{\alpha_1...\alpha_k} = \sum_{k=0}^{\infty} \int d\Omega(\boldsymbol{\sigma})\, d\Omega(\boldsymbol{\tau})\, f(\boldsymbol{\sigma})\, f(\boldsymbol{\tau})\, Y_{N,k}^{\alpha_1...\alpha_k}(\boldsymbol{\sigma})\, Y_{N,k}^{\alpha_1...\alpha_k}(\boldsymbol{\tau}) . \tag{A.45}$$

Using the completeness relation (2.20) we get

$$\sum_{k=0}^{\infty} \widetilde{f}_k^{\alpha_1...\alpha_k}\, \widetilde{f}_k^{\alpha_1...\alpha_k} = \int d\Omega(\boldsymbol{\sigma})\, |f(\boldsymbol{\sigma})|^2 , \tag{A.46}$$

which is the Plancherel identity for harmonic analysis in $S^{N-1}$. Now let us consider, instead of $f$, the function $\mathcal{L}^n f$ where $\mathcal{L}$ is the Laplace-Beltrami operator. In this case $\widetilde{f}_k^{\alpha_1...\alpha_k}$ is replaced by $\lambda_{N,k}^n \widetilde{f}_k^{\alpha_1...\alpha_k}$, and thus we obtain

$$\sum_{k=0}^{\infty} \lambda_{N,k}^{2n}\, \widetilde{f}_k^{\alpha_1...\alpha_k}\, \widetilde{f}_k^{\alpha_1...\alpha_k} = \int d\Omega(\boldsymbol{\sigma})\, |\mathcal{L}^n f(\boldsymbol{\sigma})|^2 . \tag{A.47}$$

If now $f$ is $C^\infty$ function, the integral is finite for all $n$. Thus the sum on the l.h.s. is converging for all $n$. As $\lambda_{N,k} \sim k^2$ for $k \to \infty$ we get that, for every $n$, $k^{2n} \widetilde{f}_k^{\alpha_1...\alpha_k} \widetilde{f}_k^{\alpha_1...\alpha_k} \to 0$ for $k \to \infty$. This implies that all coefficients $\widetilde{f}_k^{\alpha_1...\alpha_k}$ decrease faster than any inverse power of $k$. To prove the convergence of the series (2.18) it is then enough to notice that $|Y_{N,k}^{\alpha_1...\alpha_k}| \leq (\mathcal{N}_{N,k})^{1/2}$ and that $\mathcal{N}_{N,k}$ behaves for large $k$ as $k^{N-2}$.

In general we can write

$$f(\boldsymbol{\sigma} \cdot \boldsymbol{\tau}) = \sum_{k,h=0}^{\infty} \widetilde{f}_{N;k,h}^{\alpha_1...\alpha_k;\beta_1...\beta_h} Y_{N,k}^{\alpha_1...\alpha_k}(\boldsymbol{\sigma}) Y_{N,h}^{\beta_1...\beta_h}(\boldsymbol{\tau}) . \qquad (A.48)$$

Invariance under rotations gives

$$T_{N,k}^{\alpha_1...\alpha_k;\gamma_1...\gamma_k}(R) T_{N,h}^{\beta_1...\beta_h;\delta_1...\delta_h}(R) \widetilde{f}_{N;k,h}^{\gamma_1...\gamma_k;\delta_1...\delta_h} = \widetilde{f}_{N;k,h}^{\alpha_1...\alpha_k;\beta_1...\beta_h} \qquad (A.49)$$

for every rotation $R \in SO(N)$. Then, by Schur's lemma,

$$\widetilde{f}_{N;k,h}^{\alpha_1...\alpha_k;\beta_1...\beta_h} = \delta_{kh} I_{N,k}^{\alpha_1...\alpha_k;\beta_1...\beta_h} F_{N,k} \qquad (A.50)$$

so that

$$f(\boldsymbol{\sigma} \cdot \boldsymbol{\tau}) = \sum_{k=0}^{\infty} F_{N,k} Y_{N,k}(\boldsymbol{\sigma}) \cdot Y_{N,k}(\boldsymbol{\tau}) . \qquad (A.51)$$

Let us now discuss the properties of the coefficients $F_{N,k}$ in (2.27). The second property follows immediately from the previous discussion. We want now to prove that, if $f(t)$ is positive for $t \in [-1,1]$, then $|F_{N,k}| < F_{N,0}$ for $k \geq 1$. Indeed from the definition and (A.42) we get

$$\begin{aligned} |F_{N,k}| &\leq \int_{-1}^{1} dt \ (1-t^2)^{(N-3)/2} f(t) \left| \frac{C_k^{N/2-1}(t)}{C_k^{N/2-1}(1)} \right| \\ &< \int_{-1}^{1} dt \ (1-t^2)^{(N-3)/2} f(t) \\ &= F_{N,0} . \end{aligned} \qquad (A.52)$$

## A.5 Clebsch-Gordan Coefficients

Let us now discuss the computation of the Clebsch-Gordan coefficients (2.36). For arbitrary completely symmetric and traceless tensors $T_{N,k}$, $U_{N,l}$ and $V_{N,m}$, we want to compute

$$\int d\Omega(\boldsymbol{\sigma}) \ [T_{N,k} \cdot Y_{N,k}(\boldsymbol{\sigma})] \ [U_{N,l} \cdot Y_{N,l}(\boldsymbol{\sigma})] \ [V_{N,m} \cdot Y_{N,m}(\boldsymbol{\sigma})] . \qquad (A.53)$$

Using the definition (2.3) of the hyperspherical harmonics this reduces to

$$\mu_{N,k} \ \mu_{N,l} \ \mu_{N,m} \int d\Omega(\boldsymbol{\sigma}) \ \big[ \sigma^{\alpha_1} \ldots \sigma^{\alpha_k} \sigma^{\beta_1} \ldots \sigma^{\beta_l} \sigma^{\gamma_1} \ldots \sigma^{\gamma_m} \\ \times T_{N,k}^{\alpha_1...\alpha_k} U_{N,l}^{\beta_1...\beta_l} V_{N,m}^{\gamma_1...\gamma_m} \big] \qquad (A.54)$$

From this we see that the integral vanishes if $k + l + m$ is odd. On the other hand if $k + l + m$ is even we can use (A.8). We must now compute how many scalars we can construct with the three tensors. It is easy to see that there is only one possible scalar, with the following structure: $i$ indices of $T_{N,k}$ are contracted with $i$ indices of $V_{N,m}$, $j$ indices of $V_{N,m}$ are contracted with $j$ indices of $U_{N,l}$ and $h$ indices of $U_{N,l}$ are contracted with $h$ indices of $T_{N,k}$. Here $i = (k + m - l)/2$, $j = (m - k + l)/2$ and $h = (l + k - m)/2$. Of course $i$, $j$ and $h$ must be positive and this is equivalent to $|l - k| \leq m \leq l + k$. We must then compute the combinatorial factor, i.e. in how many ways this scalar can be constructed. We find

$$\binom{k}{i} \binom{l}{h} \binom{m}{j} i! \, j! \, h! = \frac{k! \, l! \, m!}{i! \, j! \, h!} . \tag{A.55}$$

Thus the integral becomes

$$\frac{\mu_{N,k} \, \mu_{N,l} \, \mu_{N,m}}{\mu_{N,k+j}^2 \, (k+j)!} \frac{k! \, l! \, m!}{i! \, j! \, h!} \; T_{N,k}^{a_1 \ldots a_i b_1 \ldots b_h} \, U_{N,l}^{b_1 \ldots b_h c_1 \ldots c_j} \, V_{N,m}^{a_1 \ldots a_i c_1 \ldots c_j} . \tag{A.56}$$

Formula (2.36) immediately follows.

Finally we want to discuss the computation of $\mathcal{C}_{N;\,k,l,m}^2$. Using (2.36) and (2.34) we get

$$
\begin{aligned}
\mathcal{C}_{N;\,k,l,m}^2 &= \int d\Omega\,(\boldsymbol{\sigma}) \; \mathcal{C}_{N;\,k,l,m}^{\alpha_1 \ldots \alpha_k; \beta_1 \ldots \beta_l; \gamma_1 \ldots \gamma_m} \; Y_{N,k}^{\alpha_1 \ldots \alpha_k}(\boldsymbol{\sigma}) \; Y_{N,l}^{\beta_1 \ldots \beta_l}(\boldsymbol{\sigma}) \; Y_{N,m}^{\gamma_1 \ldots \gamma_m}(\boldsymbol{\sigma}) \\
&= \frac{\mu_{N,k} \, \mu_{N,l} \, \mu_{N,m}}{\mu_{N,k+j}^2 \, (k+j)!} \frac{k! \, l! \, m!}{i! \, j! \, h!} \\
&\quad \times \int d\Omega\,(\boldsymbol{\sigma}) \; Y_{N,k}^{\alpha_1 \ldots \alpha_i \beta_1 \ldots \beta_h}(\boldsymbol{\sigma}) \; Y_{N,l}^{\beta_1 \ldots \beta_h \gamma_1 \ldots \gamma_j}(\boldsymbol{\sigma}) \; Y_{N,m}^{\gamma_1 \ldots \gamma_j \alpha_1 \ldots \alpha_i}(\boldsymbol{\sigma}) \quad \text{(A.57)}
\end{aligned}
$$

where $i = (k + m - l)/2$, $j = (m - k + l)/2$ and $h = (l + k - m)/2$. The quantity which remains inside the integral is a scalar; as such it is $\boldsymbol{\sigma}$-independent and thus we can drop the integration.

To compute the remaining contraction let us use the general expression for the hyperspherical harmonics given in (A.17). Then a straightforward combinatorial exercise gives

$$P_{(k;s)}^{\alpha_1 \ldots \alpha_i \beta_1 \ldots \beta_h}(\boldsymbol{\sigma}) \; Y_{N,l}^{\beta_1 \ldots \beta_h \gamma_1 \ldots \gamma_j}(\boldsymbol{\sigma}) \; Y_{N,m}^{\gamma_1 \ldots \gamma_j \alpha_1 \ldots \alpha_i}(\boldsymbol{\sigma}) =$$

$$s! \binom{h}{s} \binom{i}{s} \sigma^{a_1} \ldots \sigma^{a_{h-s}} \sigma^{b_1} \ldots \sigma^{b_{i-s}} Y_{N,l}^{a_1 \ldots a_{h-s} c_1 \ldots c_{j+s}}(\boldsymbol{\sigma}) \; Y_{N,m}^{b_1 \ldots b_{i-s} c_1 \ldots c_{j+s}}(\boldsymbol{\sigma})$$

$$\tag{A.58}$$

Note that this gives zero if $s > h$ or $s > i$. Now let us define $x_k$ as

$$\sigma^{\alpha} \, Y_{N,k}^{\alpha \beta_1 \ldots \beta_{k-1}}(\boldsymbol{\sigma}) = x_k \, Y_{N,k-1}^{\beta_1 \ldots \beta_{k-1}}(\boldsymbol{\sigma}) \tag{A.59}$$

It follows that

$$\prod_{m=1}^{k} x_m = \sigma^{\alpha_1} \ldots \sigma^{\alpha_k} Y_{N,k}^{\alpha_1 \ldots \alpha_k}(\boldsymbol{\sigma}) = \frac{\mathcal{N}_{N,k}}{\mu_{N,k}} \tag{A.60}$$

and therefore we obtain

$$x_k = \frac{\mathcal{N}_{N,k}}{\mu_{N,k}} \frac{\mu_{N,k-1}}{\mathcal{N}_{N,k-1}} \tag{A.61}$$

Thus (A.58) becomes

$$s! \binom{h}{s}\binom{i}{s} \mathcal{N}_{N,j+s} \frac{\mathcal{N}_{N,l}}{\mu_{N,l}} \frac{\mathcal{N}_{N,m}}{\mu_{N,m}} \frac{\mu_{N,l-h+s}}{\mathcal{N}_{N,l-h+s}} \frac{\mu_{N,m-i+s}}{\mathcal{N}_{N,m-i+s}} \tag{A.62}$$

and we get the final result

$$\mathcal{C}^2_{N;\,k,l,m} = \frac{\mu^2_{N,k} \mathcal{N}_{N,l} \mathcal{N}_{N,m}}{\mu^2_{N,k+j} (k+j)!} \frac{k!\, l!\, m!}{i!\, j!\, h!} \times$$

$$\sum_{s=0}^{M} \frac{(-1)^s}{2^s} \frac{\Gamma\left(\frac{N}{2}+k-s-1\right)}{\Gamma\left(\frac{N}{2}+k-1\right)} s! \binom{h}{s}\binom{i}{s} \frac{\mu^2_{N,j+s}}{\mathcal{N}_{N,j+s}} \tag{A.63}$$

where $M = \min(\lfloor k/2 \rfloor, i, h)$ with $i = (k+m-l)/2$, $j = (m-k+l)/2$ and $h = (l+k-m)/2$. We remind the reader that (A.63) holds only when $k+l+m$ is even and $|l-k| \le m \le l+k$; in all other cases, $\mathcal{C}^2_{N;\,k,l,m} = 0$.

There is another way of computing $\mathcal{C}^2_{N;\,k,l,m}$. Using (2.35) we can write

$$\mathcal{C}^2_{N;\,k,l,m} = \int d\Omega(\boldsymbol{\sigma})d\Omega(\boldsymbol{\tau})\ [Y_{N,k}(\boldsymbol{\sigma})\cdot Y_{N,k}(\boldsymbol{\tau})]\,[Y_{N,l}(\boldsymbol{\sigma})\cdot Y_{N,l}(\boldsymbol{\tau})]\,[Y_{N,m}(\boldsymbol{\sigma})\cdot Y_{N,m}(\boldsymbol{\tau})] \tag{A.64}$$

The integrand is only a function of $\boldsymbol{\sigma}\cdot\boldsymbol{\tau}$. Thus, using the rotational invariance of the measure, we can fix one of the two spins to an arbitrary value. Let us set $\boldsymbol{\tau} = \mathbf{w} \equiv (1, 0, \ldots, 0)$. We obtain, after integrating in $d\Omega(\boldsymbol{\tau})$,

$$\mathcal{C}^2_{N;\,k,l,m} = \int d\Omega(\boldsymbol{\sigma})\ [Y_{N,k}(\boldsymbol{\sigma})\cdot Y_{N,k}(\mathbf{w})]\,[Y_{N,l}(\boldsymbol{\sigma})\cdot Y_{N,l}(\mathbf{w})]\,[Y_{N,m}(\boldsymbol{\sigma})\cdot Y_{N,m}(\mathbf{w})] \tag{A.65}$$

and by using (2.23) we end up with

$$\mathcal{C}^2_{N;\,k,l,m} = \frac{\mathcal{S}_{N-1}}{\mathcal{S}_N} \mathcal{N}_{N,k} \mathcal{N}_{N,l} \mathcal{N}_{N,m} \times$$

$$\times \int_{-1}^{1} dt\,(1-t^2)^{(N-3)/2} \frac{C_k^{N/2-1}(t)}{C_k^{N/2-1}(1)} \frac{C_l^{N/2-1}(t)}{C_l^{N/2-1}(1)} \frac{C_m^{N/2-1}(t)}{C_m^{N/2-1}(1)} \tag{A.66}$$

If one of the three indices $k, l, m$ is fixed to some specific value this integral is easily done using the recursion relations of the Gegenbauer polynomials and their orthogonality properties [31]. In this way we have checked the general formula (A.63) for $k = 1, 2$ and $l, m$ arbitrary.

## A.6  6−$j$ Symbols

In this section we will discuss briefly the 6−$j$ symbols. In dimension $d = 1$ they appear in the computation of the four-point function (not treated in this paper), and in higher dimensions they play a crucial role in deriving high-temperature expansions even for the two-point function.

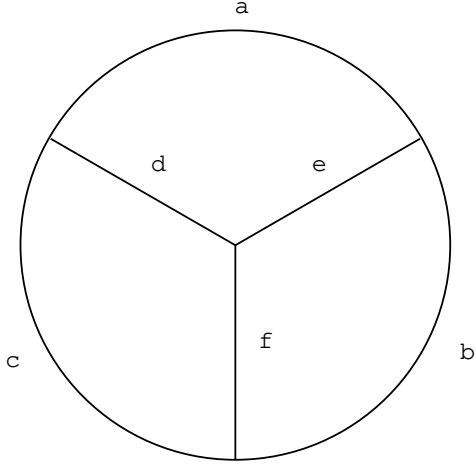

Figure 11: Graph showing the spin assignments in the 6–$j$ symbol $\mathcal{R}_N(a,b,c;d,e,f)$. Each vertex denotes a Clebsch-Gordan coefficient.

The 6–$j$ symbols (also called Racah symbols) are $O(N)$-scalars defined by

$$\mathcal{R}_N(a,b,c;d,e,f) \;=\; \mathcal{C}_{N;a,d,c}^{\alpha_1...\alpha_a;\beta_1...\beta_d;\gamma_1...\gamma_c}\, \mathcal{C}_{N;a,b,e}^{\alpha_1...\alpha_a;\delta_1...\delta_b;\epsilon_1...\epsilon_e}$$
$$\mathcal{C}_{N;d,e,f}^{\beta_1...\beta_d;\epsilon_1...\epsilon_e;\eta_1...\eta_f}\, \mathcal{C}_{N;b,c,f}^{\delta_1...\delta_b;\gamma_1...\gamma_c;\eta_1...\eta_f} \;. \quad (A.67)$$

See Figure 11 for a graphical representation. The tetrahedral symmetry which is enjoyed by the 6–$j$ symbols for $N=3$ [35] is trivially true also for generic $N$. A different conventional notation for $N=3$ is $\begin{Bmatrix} a & b & c \\ f & d & e \end{Bmatrix}$.

We have not yet been able to compute a general formula for the 6–$j$ symbols, but we have computed a very large class of special cases: among others, those in which one of the spins (say, $a$) takes the value 1 or 2, while the other five spins take arbitrary values. This class of special cases is sufficient for computing the high-temperature expansion of the $S^{N-1}$ $\sigma$-model in general dimension $d$ up to rather high order [28].

We begin by deriving a completeness relation for the 6–$j$ symbols. To do this, let us first prove two properties of the Clebsch-Gordan coefficients. Using their definition in terms of hyperspherical harmonics and the completeness relation (2.20) we can easily prove the crossing relation

$$\sum_p \mathcal{C}_{N;p,k,l}^{\alpha_1...\alpha_p;\beta_1...\beta_k;\gamma_1...\gamma_l}\, \mathcal{C}_{N;p,m,n}^{\alpha_1...\alpha_p;\delta_1...\delta_m;\epsilon_1...\epsilon_n} \;=\; \sum_p \mathcal{C}_{N;p,k,m}^{\alpha_1...\alpha_p;\beta_1...\beta_k;\delta_1...\delta_m}\, \mathcal{C}_{N;p,l,n}^{\alpha_1...\alpha_p;\gamma_1...\gamma_l;\epsilon_1...\epsilon_n} \;.$$
$$(A.68)$$

The second relation we need, which follows immediately from Schur's lemma, is

$$\mathcal{C}_{N;k,l,m}^{\alpha_1...\alpha_k;\beta_1...\beta_l;\gamma_1...\gamma_m} \, \mathcal{C}_{N;k,l,n}^{\alpha_1...\alpha_k;\beta_1...\beta_l;\delta_1...\delta_n} \;=\; \delta_{m,n} \frac{1}{\mathcal{N}_{N,m}} I_{N,m}^{\gamma_1...\gamma_m;\delta_1...\delta_m} \, \mathcal{C}_{N;k,l,m}^2 \, . \tag{A.69}$$

Inserting these two relations in (A.67) we get

$$\sum_a \mathcal{R}(a,b,c;d,e,f) \;=\; \frac{1}{\mathcal{N}_{N,f}} \mathcal{C}_{N;d,e,f}^2 \, \mathcal{C}_{N;b,c,f}^2 \, . \tag{A.70}$$

Next let us compute

$$\mathcal{K}_{N,p;k,l,m} \;\equiv\; \mathcal{C}_{N;k+p,l+p,m}^{\delta_1...\delta_p\alpha_1...\alpha_k;\delta_1...\delta_p\beta_1...\beta_l;\gamma_1...\gamma_m} \, \mathcal{C}_{N;k,l,m}^{\alpha_1...\alpha_k;\beta_1...\beta_l;\gamma_1...\gamma_m} \, . \tag{A.71}$$

Using (2.36) we get

$$\mathcal{K}_{N,p;k,l,m} \;=\; \frac{\mu_{N,k}\,\mu_{N,l}\,\mu_{N,m}}{\mu_{N,k+j}^2} \frac{k!\,l!\,m!}{i!\,j!\,h!} \times$$
$$\mathcal{C}_{N;k+p,l+p,m}^{\delta_1...\delta_p\alpha_1...\alpha_i\beta_1...\beta_h;\delta_1...\delta_p\beta_1...\beta_h\gamma_1...\gamma_j;\gamma_1...\gamma_j\alpha_1...\alpha_i} \, . \tag{A.72}$$

where $i = (m+k-l)/2$, $j = (m+l-k)/2$ and $h = (l+k-m)/2$. Now, using again (2.36) we also have

$$\mathcal{C}_{N;k+p,l+p,m}^2 \;=\; \frac{\mu_{N,k+p}\,\mu_{N,l+p}\,\mu_{N,m}}{\mu_{N,k+j+p}^2} \frac{(k+p)!\,(l+p)!\,m!}{i!\,j!\,(h+p)!} \times$$
$$\mathcal{C}_{N;k+p,l+p,m}^{\delta_1...\delta_p\alpha_1...\alpha_i\beta_1...\beta_h;\delta_1...\delta_p\beta_1...\beta_h\gamma_1...\gamma_j;\gamma_1...\gamma_j\alpha_1...\alpha_i} \, . \tag{A.73}$$

Comparing, we get

$$\mathcal{K}_{N,p;k,l,m} \;=\; \frac{\mu_{N,k}\,\mu_{N,l}}{\mu_{N,k+p}\,\mu_{N,l+p}} \frac{\mu_{N,k+j+p}^2\,(k+j+p)!}{\mu_{N,k+j}^2\,(k+j)!} \times$$
$$\frac{k!\,l!\,(h+p)!}{(k+p)!\,(l+p)!\,h!} \mathcal{C}_{N;k+p,l+p,m}^2 \, . \tag{A.74}$$

Using this result we can now compute $\mathcal{R}_N(p, k+p, l+p; l, k, m)$. Indeed, using (2.36) we get immediately

$$\mathcal{R}_N(p, k+p, l+p; l, k, m) \;=\; \frac{\mu_{N,p}^2\,\mu_{N,l}\,\mu_{N,k}}{\mu_{N,p+l}\,\mu_{N,p+k}} \mathcal{K}_{N,p;k,l,m} \, . \tag{A.75}$$

Let us now derive a few other particular cases which are relevant for high-temperature expansions.

Let us first consider the case in which one of the spins appearing in the 6–j symbols (say, $a$) is 1. In this case, using the tetrahedral symmetry, one can see that all non-vanishing symbols can be rewritten as $\mathcal{R}_N(1, k+1, l+1; l, k, m)$ or $\mathcal{R}_N(1, k+1, l-1; l, k, m)$, with $k,l$ arbitrary and $|k-l| \leq m \leq k+l$ in the first case, $\max(|k-l|, |k-l+2|) \leq m \leq k+l$ in the second one. The first quantity is a particular case of (A.75), while the second one can be computed using the completeness relation (A.70). Indeed we have

$$\mathcal{R}_N(1,k+1,l-1;l,k,m) = -\mathcal{R}_N(1,k+1,l+1;l,k,m) + \frac{1}{\mathcal{N}_{N,k}} \mathcal{C}^2_{N;1,k,k+1} \mathcal{C}^2_{N;k,l,m} \; . \tag{A.76}$$

Let us next consider the case in which one of the spins (say, $a$) is 2. Using the tetrahedral symmetry one can rewrite all the non-vanishing 6–$j$ symbols in one of the following forms:

$$\mathcal{A}^{(1)}_{N;k,l,m} = \mathcal{R}_N(2,k+2,l+2;l,k,m) \tag{A.77}$$

$$\mathcal{A}^{(2)}_{N;k,l,m} = \mathcal{R}_N(2,k,l+2;l,k,m) \tag{A.78}$$

$$\mathcal{A}^{(3)}_{N;k,l,m} = \mathcal{R}_N(2,k-2,l+2;l,k,m) \tag{A.79}$$

$$\mathcal{A}^{(4)}_{N;k,l,m} = \mathcal{R}_N(2,k,l;l,k,m) \tag{A.80}$$

with $k$ and $l$ arbitrary, $m < l+k$, and $m > |l-k|$ for $\mathcal{A}^{(1)}_{N;k,l,m}$ and $\mathcal{A}^{(4)}_{N;k,l,m}$, $m > \max(|l-k|,|l-k+2|)$ for $\mathcal{A}^{(2)}_{N;k,l,m}$, and $m > \max(|l-k|,|l-k+4|)$ for $\mathcal{A}^{(3)}_{N;k,l,m}$. Using the completeness relation (A.70) we can rewrite the last two quantities in terms of the others. Indeed

$$\mathcal{A}^{(3)}_{N;k,l,m} = -\mathcal{A}^{(1)}_{N;k,l,m} - \mathcal{A}^{(2)}_{N;k,l,m} + \frac{1}{\mathcal{N}_{N,l}} \mathcal{C}^2_{N;2,l,l+2} \mathcal{C}^2_{N;k,m,l} \tag{A.81}$$

$$\mathcal{A}^{(4)}_{N;k,l,m} = -\mathcal{A}^{(2)}_{N;l,k,m} - \mathcal{A}^{(2)}_{N;l,k-2,m} + \frac{1}{\mathcal{N}_{N,l}} \mathcal{C}^2_{N;2,l,l} \mathcal{C}^2_{N;k,m,l} \tag{A.82}$$

$\mathcal{A}^{(1)}_{N;k,l,m}$ is a particular case of (A.75). To compute $\mathcal{A}^{(2)}_{N;k,l,m}$, we first use (2.36) to get

$$\mathcal{A}^{(2)}_{N;k,l,m} = \frac{2k}{k+1} \frac{\mu^2_{N,2}\, \mu^2_{N,k}\, \mu_{N,l}}{\mu^2_{N,k+1}\, \mu_{N,l+2}} \mathcal{C}^{\alpha\beta\gamma_1\ldots\gamma_l;\alpha\delta_1\ldots\delta_{k-1};\eta_1\ldots\eta_m}_{N;l+2,k,m} \mathcal{C}^{\gamma_1\ldots\gamma_l;\beta\delta_1\ldots\delta_{k-1};\eta_1\ldots\eta_m}_{N;l,k,m} \; . \tag{A.83}$$

Then, using again (2.36) and (A.22)/(A.23)/(A.26) we get

$$\begin{aligned}\mathcal{A}^{(2)}_{N;k,l,m} = & \; \frac{2k}{k+1} \frac{\mu^2_{N,2}\, \mu^2_{N,k}\, \mu_{N,l}}{\mu^2_{N,k+1}\, \mu_{N,l+2}} \mathcal{C}^{\alpha\beta\gamma_1\ldots\gamma_l;\alpha\delta_1\ldots\delta_{k-1};\eta_1\ldots\eta_m}_{N;l+2,k,m} \times \\ & \left[ \frac{\mu_{N,k}\, \mu_{N,l}}{\mu_{N,k-1}\, \mu_{N,l+1}} \mathcal{C}^{\beta\gamma_1\ldots\gamma_l;\delta_1\ldots\delta_{k-1};\eta_1\ldots\eta_m}_{N;l+1,k-1,m} + \right. \\ & \left. (k-1) A_{N,k;1} \frac{\mu_{N,k}}{\mu_{N,k-2}} \delta^{\beta\delta_1} \mathcal{C}^{\gamma_1\ldots\gamma_l;\delta_2\ldots\delta_{k-1};\eta_1\ldots\eta_m}_{N;l,k-2,m} \right] \; . \end{aligned} \tag{A.84}$$

[where $A_{N,k;1}$ is defined in (A.16)], and thus

$$\begin{aligned}\mathcal{A}^{(2)}_{N;k,l,m} = & \; N(N+2) \frac{k}{k+1} \frac{\mu^3_{N,k}\, \mu_{N,l}}{\mu^2_{N,k+1}\, \mu_{N,k-1}\, \mu_{N,l+2}} \times \\ & \left[ \frac{\mu_{N,l}}{\mu_{N,l+1}} \mathcal{K}_{N,1;l+1,k-1,m} - \frac{\mu_{N,k-2}}{\mu_{N,k-1}} \mathcal{K}_{N,2;l,k-2,m} \right] \; . \end{aligned} \tag{A.85}$$

# B  Finite-Size-Scaling Functions for the Universality Classes (4.84)

In this appendix we want to study the finite-size-scaling functions for the one-parameter family of universality classes (4.84): this family is parametrized by a real number $B \in [0, +\infty]$, and interpolates between the $N$-vector universality class ($B = 0$) and the $RP^{N-1}$ universality class ($B = \infty$). In particular, we want to study the asymptotic behavior in the perturbative regime ($\gamma \to 0$), and show that in the *even-spin* sectors ($k = 2, 4, \ldots$) the finite-size-scaling functions are *independent of* $B$ modulo nonperturbative corrections of order roughly $e^{-\pi^2/4\gamma}$. (We will succeed here in doing this only for $k = 2$, but we *conjecture* that it is true for all even $k$.)

The basic idea can be seen in the simple case of the partition-function scaling function $\widetilde{Z}_N^{(0)}(\gamma)$, defined in (4.77). We have

$$\widetilde{Z}_N^{(0)}(\gamma; B) \;\equiv\; \sum_{l=0}^{\infty} \mathcal{N}_{N,l}\, e^{-\gamma \widetilde{a}_{N,l}} \tag{B.1a}$$

$$= \sum_{\substack{l=0 \\ l \text{ even}}}^{\infty} \mathcal{N}_{N,l}\, e^{-\gamma \lambda_{N,l}} \;+\; e^{-\gamma B} \sum_{\substack{l=0 \\ l \text{ odd}}}^{\infty} \mathcal{N}_{N,l}\, e^{-\gamma \lambda_{N,l}} \tag{B.1b}$$

$$= \frac{1 + e^{-\gamma B}}{2}\, \widetilde{Z}_N^+(\gamma) \;+\; \frac{1 - e^{-\gamma B}}{2}\, \widetilde{Z}_N^-(\gamma)\,, \tag{B.1c}$$

where we have defined

$$\widetilde{Z}_N^{\pm}(\gamma) \;\equiv\; \sum_{l=0}^{\infty} (\pm 1)^l\, \mathcal{N}_{N,l}\, e^{-\gamma \lambda_{N,l}}\,. \tag{B.2}$$

It is easy to see that $\widetilde{Z}_N^+(\gamma)$ is of order $\gamma^{-(N-1)/2}$ as $\gamma \to 0$: roughly speaking, for small $\gamma$ the sum over $l$ can be replaced by an integral. On the other hand, we shall show that $\widetilde{Z}_N^-(\gamma)$ is *exponentially small* as $\gamma \to 0$: more precisely, it is of order $e^{-\pi^2/4\gamma} \gamma^{-(N-\frac{3}{2})}$. Thus, the $B$-dependence of $\widetilde{Z}_N^{(0)}(\gamma; B)$ is given by the trivial prefactor $(1 + e^{-\gamma B})/2$, up to nonperturbative corrections of order roughly $e^{-\pi^2/4\gamma}$.

A similar result will be shown for the numerator of the susceptibility scaling function $\chi_{N,k}^{(0)}(\gamma; B)$ [see (4.79)] for $k = 2$, from which it will follow that $\chi_{N,2}^{(0)}(\gamma; B)$ is independent of $B$ modulo nonperturbative corrections of order roughly $e^{-\pi^2/4\gamma}$.

The crux of the matter will thus be to control the behavior of $\widetilde{Z}_N^{\pm}(\gamma)$ [and the analogous numerator functions] as $\gamma \to 0$. For $N = 2$ this is a simple consequence of the Poisson summation formula [(B.6) below]:

$$\widetilde{Z}_{N=2}^+(\gamma) \;\equiv\; \sum_{l=-\infty}^{\infty} e^{-\gamma l^2} \;=\; \left(\frac{\pi}{\gamma}\right)^{1/2} \sum_{m=-\infty}^{\infty} e^{-(\pi^2/\gamma) m^2} \tag{B.3}$$

$$\widetilde{Z}_{N=2}^-(\gamma) \;\equiv\; \sum_{l=-\infty}^{\infty} (-1)^l\, e^{-\gamma l^2} \;=\; \left(\frac{\pi}{\gamma}\right)^{1/2} \sum_{m=-\infty}^{\infty} e^{-(\pi^2/\gamma)(m+\frac{1}{2})^2} \tag{B.4}$$

Likewise, for $N = 3$ the Poisson summation formula controls $\widetilde{Z}^-$ (though not $\widetilde{Z}^+$):

$$\widetilde{Z}^-_{N=3}(\gamma) \equiv e^{\gamma/4} \sum_{l=-\infty}^{\infty} (-1)^l \left(l+\tfrac{1}{2}\right) e^{-\gamma(l+\frac{1}{2})^2} = e^{\gamma/4} \left(\frac{\pi}{\gamma}\right)^{3/2} \sum_{m=-\infty}^{\infty} (-1)^m \left(m+\tfrac{1}{2}\right) e^{-(\pi^2/\gamma)(m+\frac{1}{2})^2} \; . \tag{B.5}$$

We shall derive the analogous identities for general $N$, by two alternative methods:

(a) by developing generalizations of the Poisson summation formula (Sections B.1 and B.4); and

(b) by differentiating the identities for $N = 2$ or $N = 3$ with respect to $\gamma$ [e.g. (B.42)].

## B.1 Generalized Poisson Summation Formulae

Let $f \in \mathcal{S}(\mathbb{R})$ [i.e. $f$ is a function of one real variable that is infinitely differentiable and that together with all its derivatives vanishes at infinity faster than any inverse power], and define $\widehat{f}(t) = \int_{-\infty}^{\infty} e^{-itx} f(x) \, dx$. Then we have the well-known *Poisson summation formula*

$$\sum_{k=-\infty}^{\infty} f(k) = \sum_{l=-\infty}^{\infty} \widehat{f}(2\pi l) \tag{B.6}$$

and its (less well-known) one-sided generalization

$$f(0) + 2\sum_{k=1}^{\infty} f(k) = \sum_{l=-\infty}^{\infty} \widehat{f}(2\pi l) + \frac{i}{2\pi} \mathrm{P}\!\int_{-\infty}^{\infty} \cot \frac{t}{2} \, \widehat{f}(t) \, dt \; , \tag{B.7}$$

where P denotes Cauchy principal value at each of the singularities of the integrand. For a proof, see [36, pp. 31–32 and 64–65]. Here we will prove the following generalization of the one-sided Poisson formula: for any real $N$,

$$\sum_{k=0}^{\infty} \mathcal{N}_{N,k} \, f(k) = \frac{1}{2\pi} \int_{-\infty}^{\infty} K_N(t) \, \widehat{f}(t) \, dt \; , \tag{B.8}$$

where

$$K_N(t) \equiv \lim_{\epsilon \downarrow 0} \frac{1 + e^{i(t+i\epsilon)}}{(1 - e^{i(t+i\epsilon)})^{N-1}} \tag{B.9a}$$

$$= e^{-it(N-2)/2} \lim_{\epsilon \downarrow 0} \frac{2\cos\frac{t+i\epsilon}{2}}{(-2i\sin\frac{t+i\epsilon}{2})^{N-1}} \tag{B.9b}$$

$$\equiv e^{-it(N-2)/2} L_N(t) \tag{B.9c}$$

is a well-defined distribution in $\mathcal{S}'(\mathbb{R})$.[50] Note also the recursion formula

$$L_{N+2}(t) = \frac{1}{N(N-1)} \left[ -\frac{d^2}{dt^2} - \left(\frac{N-2}{2}\right)^2 \right] L_N(t) . \tag{B.10}$$

For $N$ integer $\geq 2$, we can make (B.8)/(B.9) more explicit:

$$\sum_{k=0}^{\infty} \mathcal{N}_{N,k} f(k) = \sum_{n=0}^{N-2} c_{N,n} \sum_{l=-\infty}^{\infty} \widehat{f}^{(n)}(2\pi l) + \frac{1}{2\pi} \mathrm{P}\!\int_{-\infty}^{\infty} \frac{1+e^{it}}{(1-e^{it})^{N-1}} \, \widehat{f}(t) \, dt \tag{B.11a}$$

$$= \sum_{\substack{n=0 \\ n+N \text{ even}}}^{N-2} \widetilde{c}_{N,n} \sum_{l=-\infty}^{\infty} (-1)^{Nl} \frac{d^n}{dt^n} \left[ e^{-it(N-2)/2} \widehat{f}(t) \right]\Big|_{t=2\pi l} +$$

$$\frac{1}{2\pi} \mathrm{P}\!\int_{-\infty}^{\infty} \frac{2\cos\frac{t}{2}}{(-2i\sin\frac{t}{2})^{N-1}} \, e^{-it(N-2)/2} \widehat{f}(t) \, dt \quad , \tag{B.11b}$$

where the $c_{N,n}$ and $\widetilde{c}_{N,n}$ are Laurent coefficients (see below), and of course $\widehat{f}^{(n)}$ denotes the $n^{th}$ derivative of $\widehat{f}$. For $N = 2$ we will have $c_{2,0} = \widetilde{c}_{2,0} = 1$, so that in this case (B.11) reduces to (B.7).

We start from the well-known identity

$$\sum_{k=0}^{\infty} \frac{\Gamma(N+k)}{\Gamma(N)} \frac{z^k}{k!} = (1-z)^{-N} , \tag{B.12}$$

valid for complex $z$ in the disc $|z| < 1$. (If $N$ is noninteger, we of course take the branch that equals 1 at $z = 0$.) Using the expression (A.1b) for $\mathcal{N}_{N,k}$, it follows immediately that

$$F_N(z) \equiv \sum_{k=0}^{\infty} \mathcal{N}_{N,k} z^k = (1-z)^{-N}(1-z^2)$$

$$= (1-z)^{-(N-1)}(1+z) . \tag{B.13}$$

We will use this identity to construct a "complex-variables" proof of (B.8)/(B.9) and (B.11). [It would be interesting to know whether there is a simple "real-variables" proof, as indeed there is for (B.6) and (B.7): see e.g. Ref. [36], pp. 31–32 and 64–65.]

---

[50]The existence of (B.9) in the sense of tempered distributions is a consequence of the following theorem [37, Theorem 2-10, pp. 62–63]: If $f$ is analytic in the strip $0 < \mathrm{Im}\, z < R$ and satisfies there the bound $|f(x+iy)| \leq C(1+|x|^p)y^{-r}$ for some $C, p, r < \infty$, then $\lim_{y \downarrow 0} f(x+iy)$ exists in $\mathcal{S}'(\mathbb{R})$. SKETCH OF PROOF: For a test function $g \in \mathcal{S}(\mathbb{R})$, define $h(y) = \int f(x+iy) \, g(x) \, dx$. We can compute the derivatives of $h$ using the analyticity of $f$ and integration by parts: $h^{(n)}(y) = (-i)^n \int f(x+iy) \, g^{(n)}(x) \, dx$. It follows that $|h^{(n)}(y)| \leq C_n \|g\|_{p,n} \, y^{-r}$ for a Schwartz norm $\|\cdot\|_{p,n}$. Starting from $n > r$ and using the fundamental theorem of calculus, it is easy to show that $\lim_{y \downarrow 0} h(y)$ exists, with uniform bounds in terms of a Schwartz norm of $g$. Q.E.D. See also [38, Section 12.2, Corollary 4, p. 192] for a similar result.

Let us begin by assuming that the function $f$, in addition to lying in $\mathcal{S}(\mathbb{R})$, satisfies the bounds $|f^{(n)}(x)| \leq C_n e^{-\delta|x|}$ for some constants $C_n < \infty$ and $\delta > 0$; later we will relax this assumption. It follows from this that $\widehat{f}(t)$ is analytic in the strip $|\operatorname{Im} t| < \delta$; moreover, in this strip $\widehat{f}$ vanishes faster than any inverse power of $|\operatorname{Re} t|$ when $|\operatorname{Re} t| \to \infty$. So we can use the representation

$$f(k) = \frac{1}{2\pi} \int_{-\infty}^{\infty} e^{itk} \widehat{f}(t)\, dt \,, \tag{B.14}$$

where the contour of integration runs slightly above the real axis (say, at $\operatorname{Im} t = \epsilon$ with $0 < \epsilon < \delta$). It follows that

$$\sum_{k=0}^{\infty} \mathcal{N}_{N,k} f(k) = \frac{1}{2\pi} \sum_{k=0}^{\infty} \mathcal{N}_{N,k} \int_{-\infty}^{\infty} (e^{it})^k \widehat{f}(t)\, dt \,. \tag{B.15}$$

This joint sum/integral is absolutely convergent (since $|e^{it}| = e^{-\epsilon} < 1$ and $\widehat{f}(t)$ decays rapidly at infinity), so we can interchange the summation and integration. Using (B.13), we obtain

$$\sum_{k=0}^{\infty} \mathcal{N}_{N,k} f(k) = \frac{1}{2\pi} \int_{-\infty}^{\infty} \frac{1 + e^{it}}{(1 - e^{it})^{N-1}} \widehat{f}(t)\, dt \,, \tag{B.16}$$

where the integration still runs at $\operatorname{Im} t = \epsilon$. Since the value of the integral is independent of $\epsilon$ (for $0 < \epsilon < \delta$), we can trivially take $\epsilon \downarrow 0$, thus proving (B.8)/(B.9) for functions $f$ satisfying the above restrictions.

It is easy to remove the assumption that $f$ and its derivatives decay exponentially. Just apply the foregoing result to $f_\alpha(x) \equiv f(x) e^{-\alpha x^2}$ and let $\alpha \downarrow 0$. Then $\widehat{f_\alpha}$ equals $\widehat{f}$ convoluted with a Gaussian $(4\pi\alpha)^{-1/2} e^{-t^2/4\alpha}$, and this Gaussian tends (in the sense of distributions) to a delta function as $\alpha \downarrow 0$; therefore, $\widehat{f_\alpha} \to \widehat{f}$ in $\mathcal{S}(\mathbb{R})$ as $\alpha \downarrow 0$. In particular, the right-hand side of (B.8), taken on $\widehat{f_\alpha}$, tends as $\alpha \downarrow 0$ to its value taken on $\widehat{f}$: this is an immediate consequence of the fact that $K_N(t)$ defines a distribution in $\mathcal{S}'(\mathbb{R})$. On the other hand, the left-hand side of (B.8) converges to its $\alpha = 0$ value by virtue of the dominated convergence theorem.

Let us now assume that $N$ is an integer $\geq 2$, and let us again temporarily assume that $f$ and its derivatives decay exponentially. Then the integral (B.16) at $\operatorname{Im} t = \epsilon$ can be written as the half-sum of the integrals taken over $\operatorname{Im} t = \pm\epsilon$ plus the half-difference. Now the half-sum is, by definition, precisely the principal-value integral in (B.11a)[51]; on the other hand, the half-difference is $-\pi i$ times the sum of the residues at the poles $t = 2\pi l$ ($l$ integer). Using the Laurent expansion

$$\frac{1 + e^{it}}{(1 - e^{it})^{N-1}} = \sum_{m=-(N-1)}^{\infty} a_{N,m} t^m \tag{B.17}$$

---

[51]Our "principal-value integral" is the same as the "canonical regularization" of Gel'fand and Shilov [36, Sections I.3 and I.4]. Note, in particular, equations (6), (7), (12) and (13) in Section I.4.4 (pp. 94–95).

around $t = 0$ (and of course an identical expansion at each pole $t = 2\pi l$, $l$ integer), we obtain (B.11a) with

$$c_{N,n} = -\frac{i}{2} \frac{a_{N,-n-1}}{n!} \,. \tag{B.18}$$

We can see that the leading term in the Laurent expansion is $a_{N,-(N-1)} = 2i^{N-1}$, and hence the highest-derivative coefficient in (B.11a) is $c_{N,N-2} = -i^N/(N-2)!$. In particular, for $N = 2$ we have $c_{2,0} = 1$, so that (B.11a) reproduces (B.7).

Equivalently, we can use the Laurent expansion

$$\frac{2\cos\frac{t}{2}}{(-2i\sin\frac{t}{2})^{N-1}} = \sum_{m=-(N-1)}^{\infty} \widetilde{a}_{N,m}\, t^m \tag{B.19}$$

around $t = 0$ (and of course an identical expansion multiplied by $(-1)^{Nl}$ at each pole $t = 2\pi l$, $l$ integer). Note that $\widetilde{a}_{N,m} \neq 0$ only when $m + N$ is odd. We therefore obtain (B.11b) with

$$\widetilde{c}_{N,n} = -\frac{i}{2} \frac{\widetilde{a}_{N,-n-1}}{n!} \,. \tag{B.20}$$

Note that $\widetilde{c}_{N,n} \neq 0$ only when $n+N$ is even. The leading terms are $\widetilde{a}_{N,-(N-1)} = 2i^{N-1}$ and hence $\widetilde{c}_{N,N-2} = -i^N/(N-2)!$. From (B.10) we can derive the recursion relation

$$\widetilde{a}_{N+2,m} = -\frac{1}{N(N-1)}\left[(m+2)(m+1)\widetilde{a}_{N,m+2} + \left(\frac{N-2}{2}\right)^2 \widetilde{a}_{N,m}\right]\,, \tag{B.21}$$

which together with the initial conditions $\widetilde{a}_{2,-1} = 2i$ and $\widetilde{a}_{3,-2} = -2$ yields all the coefficients. Unfortunately, we have been unable to find a closed-form solution for this recursion relation.

The assumption that $f$ and its derivatives decay exponentially can be removed as before, using the fact that both terms on the right-hand side of (B.11a)/(B.11b) define distributions in $\mathcal{S}'(\mathbb{R})$.

If $N$ is a real number $< 2$ (not necessarily integer), we can rewrite the kernel $K_N(t)$ in a somewhat more explicit form. Note that $\mathrm{Re}(1 - e^{i(t+i\epsilon)}) > 0$ for all $t$, hence $|\arg(1 - e^{i(t+i\epsilon)})| < \pi/2$. It follows that

$$K_N(t) = 2^{2-N}\, e^{i(N-1)\pi/2}\, \frac{|\cos\frac{t}{2}|}{|\sin\frac{t}{2}|^{N-1}}\, e^{i\varphi(t)} \,, \tag{B.22}$$

where

$$\varphi(t) = -\frac{N-2}{2} \times (t \bmod 2\pi) \tag{B.23}$$

and $t \bmod 2\pi$ is taken to lie in the interval $[0, 2\pi)$. Since $N < 2$, (B.22) defines a locally absolutely integrable function, hence is unambiguous as a distribution. Equivalently, we can write

$$L_N(t) \equiv e^{i\frac{N-2}{2}t}K_N(t) = 2^{2-N}\, e^{i(N-1)\pi/2}\, \frac{|\cos\frac{t}{2}|}{|\sin\frac{t}{2}|^{N-1}}\, e^{i\psi(t)} \,, \tag{B.24}$$

where

$$\psi(t) \equiv \varphi(t) + \frac{N-2}{2}t = \pi(N-2)\lfloor t/2\pi \rfloor \qquad (B.25)$$

and $\lfloor x \rfloor$ denotes the largest integer $\leq x$.

For $N \geq 2$ these formulae are ill-defined because $K_N(t)$ has *nonintegrable* singularities at $t = 2\pi l$ ($l$ integer). For *noninteger* $N > 2$, explicit formulae can be obtained by using the recursion formula (B.10) [starting from (B.24) at some $N < 2$] together with integration by parts. (More precisely, integration by parts is how one *defines* the derivative of a distribution!)

Finally, let us go back to (B.9b,c) and note an interesting property of the kernel $L_N$ (valid for all $N$): we claim that if we decompose $L_N(t)$ into its symmetric and antisymmetric parts around $t = \pi$,

$$L_N^{\pm}(t) \equiv \tfrac{1}{2}[L_N(t) \pm L_N(2\pi - t)] \;, \qquad (B.26)$$

then the symmetric part $L_N^+(t)$ *vanishes* on the interval $0 < t < 2\pi$ (i.e. it is supported outside this interval). PROOF: The numerator $\cos\frac{t+i\epsilon}{2}$ is obviously antisymmetric around $t = \pi$ in the limit $\epsilon \downarrow 0$. As for the denominator, the function values $\sin\frac{t+i\epsilon}{2}$ and $\sin\frac{(2\pi-t)+i\epsilon}{2}$ belong to the same Riemann sheet of the function $z^{N-1}$ *provided that* $0 < t < 2\pi$ (and *not* otherwise), so that in this case they tend as $\epsilon \downarrow 0$ to the same point on the Riemann surface. Therefore, the denominator is symmetric around $t = \pi$ in the limit $\epsilon \downarrow 0$, for $0 < t < 2\pi$ (and *only* there). Q.E.D.

Of course, the same argument can be made around any point $t = (2l+1)\pi$, $l$ integer: the symmetric part vanishes on the interval $2\pi l < t < 2\pi(l+1)$.

This symmetry/support property is of particular relevance in case the function $e^{-i\frac{N-2}{2}t}\widehat{f}(t)$ is symmetric around $t = \pi$ (as will be the case in our application below).

## B.2   Some Generalized Theta Functions

Now we want to apply the generalized Poisson summation formulae to analyze the asymptotic behavior as $\gamma \to 0$ of some generalized theta functions. Let us define

$$Z_{N,\theta,\alpha}(\gamma) \equiv \sum_{k=0}^{\infty} \mathcal{N}_{N,k}\, e^{ik\theta}\, e^{-\gamma(k+\alpha)^2} \;, \qquad (B.27)$$

which of course is periodic in $\theta$ with period $2\pi$. Applying (B.16) with $f(x) = e^{i\theta x}e^{-\gamma(x+\alpha)^2}$, we obtain

$$Z_{N,\theta,\alpha}(\gamma) = \frac{1}{2\pi}\left(\frac{\pi}{\gamma}\right)^{1/2} \int_{-\infty}^{\infty} \frac{1+e^{it}}{(1-e^{it})^{N-1}}\, e^{i\alpha(t-\theta)}\, e^{-(t-\theta)^2/4\gamma}\, dt \qquad (B.28a)$$

$$= \frac{1}{2\pi}\left(\frac{\pi}{\gamma}\right)^{1/2} e^{-i\alpha\theta} \int_{-\infty}^{\infty} \frac{2\cos\frac{t}{2}}{(-2i\sin\frac{t}{2})^{N-1}}\, e^{i(\alpha-\frac{N-2}{2})t}\, e^{-(t-\theta)^2/4\gamma}\, dt \qquad (B.28b)$$

$$= \frac{1}{2\pi}\left(\frac{\pi}{\gamma}\right)^{1/2} e^{-i\alpha\theta} \frac{2i}{N-2} \int_{-\infty}^{\infty} \frac{1}{(-2i\sin\frac{t}{2})^{N-2}}\, \frac{d}{dt}\left[e^{i(\alpha-\frac{N-2}{2})t}\, e^{-(t-\theta)^2/4\gamma}\right] dt \;, \qquad (B.28c)$$

where the contour of integration runs at $\text{Im}\, t = \epsilon > 0$; here (B.28c) is obtained from (B.28b) by integration by parts. Note that the formula becomes slightly simpler in the case $\alpha = (N-2)/2$.

Let us consider first the case of $N$ integer $\geq 2$. As before, the integral (B.28b) can be written as a principal-value integral plus $-\pi i$ times a sum of residues. To compute the residue contribution, we use the Laurent expansion (B.19), yielding

$$\text{residue contribution to } Z_{N,\theta,\alpha}(\gamma)$$
$$= \left(\frac{\pi}{\gamma}\right)^{1/2} e^{-i\alpha\theta} \sum_{n=0}^{N-2} \widetilde{c}_{N,n} \sum_{l=-\infty}^{\infty} (-1)^{Nl} \frac{d^n}{dt^n}\left[e^{i(\alpha - \frac{N-2}{2})t} e^{-(t-\theta)^2/4\gamma}\right]\bigg|_{t=2\pi l} . \tag{B.29}$$

The sum over $l$ is absolutely convergent, uniformly on compact subsets of the half-plane $\text{Re}\,\gamma > 0$, thanks to the rapid decay of $e^{-(t-\theta)^2/4\gamma}$ as $t \to \pm\infty$. Moreover, as $\gamma \to 0$ this sum is dominated by its leading term(s), namely the one(s) for which $|2\pi l - \theta|$ is smallest.

Concerning the principal-value integral, we first note that in certain cases it vanishes by symmetry: If $N$ is an integer, then $\cos\frac{t}{2}/(\sin\frac{t}{2})^{N-1}$ has parity $(-1)^{N-1}$; it follows that the combination

$$e^{i\alpha\theta} Z_{N,\theta,\alpha}(\gamma) + (-1)^N e^{-i(N-2-\alpha)\theta} Z_{N,-\theta,N-2-\alpha}(\gamma) \tag{B.30}$$

is given exactly by the sum of residues.[52] In particular, in two cases $Z_{N,\theta,\alpha}(\gamma)$ itself is given by the sum of residues (B.29):

(a) $\alpha = (N-2)/2$, $\theta = 0$, $N$ even:

$$Z_{N,0,\frac{N-2}{2}} = \left(\frac{\pi}{\gamma}\right)^{1/2} \sum_{n=0}^{N-2} \widetilde{c}_{N,n} \sum_{l=-\infty}^{\infty} \frac{d^n}{dt^n}\left[e^{-t^2/4\gamma}\right]\bigg|_{t=2\pi l} . \tag{B.31}$$

As $\gamma \to 0$, this sum equals its $l = 0$ term (which is of order $\gamma^{-(N-1)/2}$) up to nonperturbative corrections of order $e^{-\pi^2/\gamma}\gamma^{-(N-\frac{3}{2})}$. For $N = 2$ this reduces to (B.3).

(b) $\alpha = (N-2)/2$, $\theta = \pi$, $N$ integer (even or odd):

$$Z_{N,\pi,\frac{N-2}{2}} = \left(\frac{\pi}{\gamma}\right)^{1/2} (-i)^{N-2} \sum_{n=0}^{N-2} \widetilde{c}_{N,n} \sum_{l=-\infty}^{\infty} (-1)^{Nl} \frac{d^n}{dt^n}\left[e^{-(t-\pi)^2/4\gamma}\right]\bigg|_{t=2\pi l} . \tag{B.32}$$

As $\gamma \to 0$, this sum is exponentially small, of order $e^{-\pi^2/4\gamma}\gamma^{-(N-\frac{3}{2})}$. For $N = 2$ (resp. $N = 3$) this reduces to (B.4) [resp. (B.5)].

---

[52] If one makes the change of variables $t = s + \pi$ and then uses the oddness of the function $\sin\frac{s}{2}/(\cos\frac{s}{2})^{N-1}$ (for *all* integers $N$), one obtains the *same* combination (B.30).

Next let us consider the general case of $N$ real (not necessarily integer). Symmetrizing (B.28b) around $t = \pi$, we find

$$e^{i\alpha\theta} Z_{N,\theta,\alpha}(\gamma) \; + \; e^{i\pi N} e^{-i(N-2-\alpha)\theta} Z_{N,-\theta,N-2-\alpha}(\gamma) \; =$$
$$\frac{1}{2\pi} \left(\frac{\pi}{\gamma}\right)^{1/2} \int_{-\infty}^{\infty} L_N^+(t) \left[ e^{i(\alpha - \frac{N-2}{2})t} e^{-(t-\theta)^2/4\gamma} \; + \; e^{i(\alpha - \frac{N-2}{2})(2\pi - t)} e^{-(2\pi - t - \theta)^2/4\gamma} \right] dt \; ,$$
(B.33)

where $L_N^+(t)$ is the symmetric part of $L_N(t)$ around $t = \pi$ [cf. (B.26)]. As discussed in the preceding subsection, $L_N^+(t)$ is supported away from the interval $0 < t < 2\pi$, and is the $p^{th}$ derivative of a polynomially bounded function, where $p = \max(\lfloor N \rfloor, 0)$. The following lemma then implies that (B.33) and all its derivatives with respect to $\gamma$ are exponentially small whenever $\theta \neq 0 \bmod 2\pi$: more precisely, the $n^{th}$ derivative of (B.33) is bounded by const $\times \gamma^{-(p+\frac{1}{2}+2n)} e^{-\Theta^2/4\gamma}$, where

$$\Theta \; = \; \min_{k \in \mathbb{Z}} |\theta - 2\pi k| \; .$$
(B.34)

In particular, if $\alpha = (N-2)/2$ and $0 < \theta < 2\pi$, then $\text{Re}[e^{i(\frac{N-2}{2}\theta - \frac{\pi}{2}N)} Z_{N,\theta,\frac{N-2}{2}}(\gamma)]$ is $O(\gamma^{-p} e^{-\Theta^2/4\gamma})$ as $\gamma \downarrow 0$ ($\gamma$ real). For $\theta = \pi$, this says that $Z_{N,\pi,\frac{N-2}{2}}(\gamma)$ is $O(\gamma^{-p} e^{-\pi^2/4\gamma})$.

**Lemma.** Let $L(t)$ be a tempered distribution on $\mathbb{R}$, supported on $|t| \geq \Theta$, and define

$$F(\gamma) \; = \; \gamma^{-1/2} \int_{-\infty}^{\infty} L(t) \, e^{-t^2/4\gamma} \, dt$$
(B.35)

for $\text{Re}\,\gamma > 0$. Then there exist constants $p$ and $C_n$ such that

$$|F^{(n)}(\gamma)| \; \leq \; C_n \, |\gamma|^{-(p+\frac{1}{2}+2n)} \, e^{-\Theta^2 \, \text{Re}(1/4\gamma)}$$
(B.36)

for $n \geq 0$ and (say) $|\gamma| < 1$.

PROOF. For some $p \geq 0$, $L$ is the $p^{th}$ derivative of a polynomially bounded function $f$, i.e. $|f(t)| \leq C(1 + |t|^m)$. Thus, for each $n \geq 0$,

$$F^{(n)}(\gamma) \; = \; (-1)^p \int_{-\infty}^{\infty} f(t) \frac{\partial^p}{\partial t^p} \frac{\partial^n}{\partial \gamma^n} \left( \gamma^{-1/2} \, e^{-t^2/4\gamma} \right) dt \; .$$
(B.37)

Clearly

$$\left| \frac{\partial^p}{\partial t^p} \frac{\partial^n}{\partial \gamma^n} \left( \gamma^{-1/2} \, e^{-t^2/4\gamma} \right) \right| \; \leq \; \text{const}(n) \times (1 + |t|^{p+2n}) \, |\gamma|^{-(p+\frac{1}{2}+2n)} \, e^{-t^2 \, \text{Re}(1/4\gamma)} \; .$$
(B.38)

Integrating over $t$ then proves the lemma. ∎

Finally, let us consider the cases in which the principal-value integral (for $N$ integer) or the integral over $0 < t < 2\pi$ (for $N$ generic) does *not* vanish. In these cases we can obtain an asymptotic expansion for $Z_{N,\theta,\alpha}(\gamma)$ in powers of $\gamma$, by the usual method of expanding the integrand of (B.28b) around its peak at $t = \theta$. For $\theta \neq 0 \pmod{2\pi}$ the result is as follows:

$$Z_{N,\theta,\alpha}(\gamma) = e^{-i(N-2)\theta/2} e^{i\pi(N-1)/2} 2^{2-N} \frac{\cos\frac{\theta}{2}}{(\sin\frac{\theta}{2})^{N-1}} \times$$
$$\left[1 + \gamma\left\{\frac{3N-4}{4} - \left(\alpha - \frac{N-2}{2}\right)^2 - i\left(\alpha - \frac{N-2}{2}\right)\left[\tan\frac{\theta}{2} + (N-1)\cot\frac{\theta}{2}\right]\right.\right.$$
$$\left.\left. + \frac{N(N-1)}{4}\cot^2\frac{\theta}{2}\right\} + O(\gamma^2)\right] \quad (B.39)$$

This expansion can be proven rigorously by cutting $L_N$ (using a smooth partition of unity) into a part supported on the interval $\epsilon < t < 2\pi - \epsilon$ (here we suppose $0 < t < 2\pi$) and a part supported on the union of intervals $t < 2\epsilon$ and $t > 2\pi - 2\epsilon$. The integral over the first interval is then an ordinary integral of smooth functions, and the asymptotic expansion can be controlled by standard techniques; while the integral over the second region is exponentially small by virtue of the Lemma above.

An alternative way of deriving these formulae is to use a recursion formula yielding $Z_{N+2}$ in terms of $Z_N$; in this way, all integer values of $N$ ($\geq 2$) can be handled by differentiating the cases $N = 2$ and $N = 3$ with respect to $\gamma$, while all noninteger values of $N$ ($> 0$) can be handled by differentiating one of the cases in the interval $0 < N < 2$. The basis of this approach is the simple identity

$$\mathcal{N}_{N+2,k-1} = \frac{k(N+k-2)}{N(N-1)}\mathcal{N}_{N,k} \quad (B.40)$$

[see (A.2)]. It follows from this that

$$Z_{N+2,\theta,\alpha+1}(\gamma) = \frac{e^{-i\theta}}{N(N-1)}\left[-\frac{d}{d\gamma} + 2i\left(\alpha - \frac{N-2}{2}\right)\frac{d}{d\theta} - \alpha^2\right]Z_{N,\theta,\alpha}(\gamma). \quad (B.41)$$

In particular, for $\alpha = (N-2)/2$ we get

$$Z_{N+2,\theta,\frac{N}{2}}(\gamma) = \frac{e^{-i\theta}}{N(N-1)}\left[-\frac{d}{d\gamma} - \left(\frac{N-2}{2}\right)^2\right]Z_{N,\theta,\frac{N-2}{2}}(\gamma). \quad (B.42)$$

Clearly, if $Z_{N,\theta,\frac{N-2}{2}}(\gamma)$ is exponentially small together with all its derivatives, then the same holds for $Z_{N+2r,\theta,\frac{N+2r-2}{2}}(\gamma)$ for every positive integer $r$.

**Remark.** The equivalence of (B.10) and (B.42) comes from the fact that $\gamma^{-1/2}e^{-(t-\theta)^2/4\gamma}$ is a solution of the heat equation, hence $d/d\gamma$ and $d^2/dt^2$ act identically on it; and since these two operators commute, the same holds true for multiple applications

of these operators:

$$\left(\frac{d}{d\gamma}\right)^m \left(\frac{d^2}{dt^2}\right)^n \left(\gamma^{-1/2} e^{-(t-\theta)^2/4\gamma}\right) = \left(\frac{d}{d\gamma}\right)^{m+n} \left(\gamma^{-1/2} e^{-(t-\theta)^2/4\gamma}\right)$$

$$= \left(\frac{d^2}{dt^2}\right)^{m+n} \left(\gamma^{-1/2} e^{-(t-\theta)^2/4\gamma}\right) . \quad \text{(B.43)}$$

## B.3  The Partition-Function Scaling Function $\widetilde{Z}_N^{(0)}(\gamma; B)$

Recall from (B.1)/(B.2) that

$$\widetilde{Z}_N^{(0)}(\gamma; B) = \frac{1 + e^{-\gamma B}}{2} \widetilde{Z}_N^+(\gamma) + \frac{1 - e^{-\gamma B}}{2} \widetilde{Z}_N^-(\gamma) , \quad \text{(B.44)}$$

where

$$\widetilde{Z}_N^\pm(\gamma) \equiv \sum_{l=0}^\infty (\pm 1)^l \, \mathcal{N}_{N,l} \, e^{-\gamma \lambda_{N,l}} . \quad \text{(B.45)}$$

Now

$$\lambda_{N,l} \equiv l(N + l - 2) = \left(l + \frac{N-2}{2}\right)^2 - \left(\frac{N-2}{2}\right)^2 . \quad \text{(B.46)}$$

So the functions $\widetilde{Z}_N^\pm$ are precisely generalized theta functions of the type considered in the preceding subsection; indeed, they belong to the "simple" case $\alpha = (N-2)/2$:

$$\widetilde{Z}_N^+(\gamma) = e^{\gamma(N-2)^2/4} Z_{N,0,\frac{N-2}{2}}(\gamma) \quad \text{(B.47a)}$$

$$\widetilde{Z}_N^-(\gamma) = e^{\gamma(N-2)^2/4} Z_{N,\pi,\frac{N-2}{2}}(\gamma) \quad \text{(B.47b)}$$

It follows immediately from the results of the preceding subsection that

$$\widetilde{Z}_N^{(0)}(\gamma; B) = \frac{1 + e^{-\gamma B}}{2} \widetilde{Z}_N^+(\gamma) + O(e^{-\pi^2/4\gamma} \gamma^{-(N-\frac{3}{2})}) , \quad \text{(B.48)}$$

as claimed.

**Remark.** The duality formula (B.3)/(B.4) for ordinary theta functions is a special case of a modular transformation, and is connected with the theory of elliptic functions [39] [40, Chapter 13] [41]; it also has applications in string theory [42]. We wonder whether the corresponding formulae for integer $N \geq 3$ are telling us something deep about the Riemannian geometry of the sphere $S^{N-1}$. We are intrigued by the fact that the generalized theta functions arising from $\widetilde{Z}_N^\pm$ fall precisely into the "simple" case $\alpha = (N-2)/2$ — it can't be a mere coincidence! And we wonder why there is a convergent duality formula for $\widetilde{Z}_N^-$ for all integer $N$ [cf. (B.32)], but for $\widetilde{Z}_N^+$ only for *even* $N$ [cf. (B.31)]. Is this perhaps related to the fact that $-I \in SO(N)$ for $N$ even but not for $N$ odd? Or to the fact that the groups $SO(N)$ fall into different families of the Lie classification for $N$ even and $N$ odd? And can our results be generalized to symmetric spaces other than $S^{N-1}$?

## B.4 Some More Generalized Poisson Summation Formulae

To handle the numerator of the susceptibility scaling function, we will need to study sums of the form $\sum_k \mathcal{N}_{N,k} R(k) f(k)$, where $R$ is a rational function and $f \in \mathcal{S}(\mathbb{R})$. Unfortunately, such sums are *not* covered by the generalized Poisson formulae of Section B.1: the trouble is that $R$ is typically not a smooth function on all of $\mathbb{R}$, so it cannot be absorbed into $f$. Instead, we shall derive some further generalizations of the Poisson summation formula, in which $R$ is absorbed into the kernel $K_N$.

Let, therefore, $R$ be a rational function of the form

$$R(x) = \frac{P(x)}{Q(x)} = \frac{P(x)}{(x+\beta_1)\cdots(x+\beta_q)}, \tag{B.49}$$

where $P$ is a polynomial. Let $k_0$ be an nonnegative integer, chosen large enough so that none of the $\beta_i$ are equal to an integer $\leq -k_0$. (That is, $R$ does not have any poles at integers $\geq k_0$. When $k_0 = 0$ we shall omit it from the notation.) We shall then prove a formula of the form

$$\sum_{k=k_0}^{\infty} \mathcal{N}_{N,k} R(k) f(k) = \frac{1}{2\pi} \int_{-\infty}^{\infty} K_{N;R;k_0}(t)\,\widehat{f}(t)\,dt . \tag{B.50}$$

In fact, the derivation is virtually identical to that in Section B.1: We introduce the function

$$F_{N;R;k_0}(z) \equiv \sum_{k=k_0}^{\infty} \mathcal{N}_{N,k} R(k) z^k . \tag{B.51}$$

This series converges in the disc $|z| < 1$, but $F_{N;R;k_0}$ then has an analytic continuation to the whole $z$-plane cut along the ray $[+1, +\infty)$.[53] In particular, the only

---

[53]This follows from [40, Theorem 11.1.3, pp. 41–43], which states the following: Let $\mu$ be a finite complex measure on $[0,1]$, with $|\mu(\{1\})| < \rho$, and let $\mu_k \equiv \int x^k\,d\mu(x)$ be its moments. Let $k_1$ be any integer such that $\int x^{k_1} |d\mu(x)| < \rho$ (such an integer always exists). Let $G$ be an analytic function in the disc of radius $\rho$ centered at the origin. Let $f(z) = \sum_{k=k_1}^{\infty} c_k z^k$ be a function having an analytic continuation to a domain $\mathbf{A}$ which is starlike with respect to the origin. Then the function defined by $\sum_{k=k_1}^{\infty} G(\mu_k) c_k z^k$ likewise has an analytic continuation to $\mathbf{A}$.

We shall apply this theorem as follows: Let $\mathbf{A}$ be the cut plane $\mathbb{C} \setminus [+1, +\infty)$. Let

$$\begin{aligned} f(z) &= \sum_{k=k_1}^{\infty} \mathcal{N}_{N,k}\,z^k \\ &= (1-z)^{-(N-1)}(1+z) - \sum_{k=0}^{k_1-1} \mathcal{N}_{N,k}\,z^k , \end{aligned}$$

where we will choose $k_1$ later. Let $d\mu(x) = dx$, so that $\mu_k = 1/(k+1)$. Let

$$G(x) = \left(\frac{x}{1+(\beta-1)x}\right)^n ,$$

so that $G(\mu_k) = 1/(k+\beta)^n$. To apply the theorem, it suffices to take $k_1 > |\beta - 1| - 1$. But then we can add in "by hand" the terms $k_0 \leq k \leq k_1 - 1$, provided that none of these values of $k$ equals $-\beta$.

singularity of $F_{N;R;k_0}$ on the unit circle is at $z = 1$, and the growth of $F_{N;R;k_0}$ as this singularity is approached is bounded by a polynomial in $|1 - z|^{-1}$. We can therefore introduce the distribution $K_{N;R;k_0}$ by

$$K_{N;R;k_0}(t) \equiv \lim_{\epsilon \downarrow 0} F_{N;R;k_0}(e^{i(t+i\epsilon)}) , \qquad (B.52)$$

and its only singularities are at $t = 2\pi l$ ($l$ integer). The proof of (B.50) then follows exactly as in Section B.1.

If $k_0 = 0$, then $F_{N;R}$ can be written explicitly in terms of the generalized hypergeometric function $_{q+1}F_q$ (defined e.g. in [31, formula 9.14.1, p. 1045]):

$$\begin{aligned} F_{N;R}(z) & = & P\left(z \frac{d}{dz}\right) \left[\left(\prod_{i=1}^{q} \beta_i^{-1}\right) {}_{q+1}F_q(N, \beta_1, \ldots, \beta_q; \beta_1 + 1, \ldots, \beta_q + 1; z) \right. \\ & & \left. - z^2 \left(\prod_{i=1}^{q}(\beta_i + 2)^{-1}\right) {}_{q+1}F_q(N, \beta_1 + 2, \ldots, \beta_q + 2; \beta_1 + 3, \ldots, \beta_q + 3; z)\right] . \end{aligned}$$
(B.53)

For $q = 1$, this special case of the $_2F_1$ corresponds to incomplete beta functions.

We want now to derive some general properties of these functions. Let us thus introduce

$$\begin{aligned} f_q(\beta; z) & \equiv & \sum_{k=0}^{\infty} \frac{\Gamma(N+k)}{\Gamma(N)} \frac{z^k}{k!(k+\beta)^q} \\ & = & \beta^{-q} {}_{q+1}F_q(N, \beta, \ldots, \beta; \beta + 1, \ldots, \beta + 1; z) . \end{aligned} \qquad (B.54)$$

(Here $N$ is fixed, so we suppress it from the notation.) By making a shift $k \to k+1$ in the sums, it is easy to derive the recursion relation

$$\begin{aligned} f_q(\beta; z) & = & \beta^{-q}(1-z)^{1-N} + \left(1 - \frac{N-1}{\beta}\right) z f_q(\beta + 1; z) \\ & & - (N-1)z \sum_{n=1}^{q-1} \beta^{n-q-1} f_n(\beta + 1; z) . \end{aligned} \qquad (B.55)$$

Using this formula it is easy to get $F_{N;R}$ for the simplest nontrivial case $R(x) = 1/(x + \beta)$:

$$\begin{aligned} \sum_{k=0}^{\infty} \mathcal{N}_{N,k} \frac{z^k}{k + \beta} & = & \frac{1}{\beta}\left[1 + \frac{z(\beta + 1 - N)}{\beta + 1}\right](1 - z)^{1-N} \\ & & + z^2 \frac{(N-1)(N-2-2\beta)}{\beta(\beta+1)} f_1(\beta + 2; z) . \end{aligned} \qquad (B.56)$$

---

This proves the claim for $R(x) = 1/(x+\beta)^n$. A general denominator $Q(x)$ then follows by expansion in partial fractions, and a general rational function $R(x) = P(x)/Q(x)$ follows by application of the differential operator $P(z\,\partial/\partial z)$. Q.E.D.

A simplification occurs for $\beta = (N-2)/2$: the last term vanishes, and we have the explicit formula

$$\sum_{k=0}^{\infty} \mathcal{N}_{N,k} \frac{z^k}{k + \frac{N-2}{2}} = \frac{2}{N-2}(1-z)^{2-N} \ . \tag{B.57}$$

In the following we will be especially interested in the value of $f_q(\beta; z)$ at $z = -1$. For $q = 1$ a general formula can be obtained for $\beta = (N/2) + \text{integer}$. The starting point is Kummer's formula[54]

$$_2F_1(a, b; 1 + a - b; -1) = 2^{-a} \sqrt{\pi} \frac{\Gamma(1 + a - b)}{\Gamma\left(1 - b + \frac{a}{2}\right) \Gamma\left(\frac{1}{2} + \frac{a}{2}\right)} \ . \tag{B.58}$$

Setting $a = N$ and $b = N/2$, we get

$$_2F_1(N, \tfrac{N}{2}; \tfrac{N}{2} + 1; -1) = \frac{\sqrt{\pi}}{2^N} \frac{\Gamma\left(1 + \frac{N}{2}\right)}{\Gamma\left(\frac{1+N}{2}\right)} \ , \tag{B.59}$$

so that

$$f_1(\tfrac{N}{2}; -1) = \frac{\sqrt{\pi}}{2^N} \frac{\Gamma\left(\frac{N}{2}\right)}{\Gamma\left(\frac{1+N}{2}\right)} \ . \tag{B.60}$$

Using the recursion relations (B.55), it is then possible to compute $f_1(\beta; -1)$ for all $\beta = (N/2) + \text{integer}$.

In the following we will use two specific functions:

$$U_N(z) \equiv \sum_{k=0}^{\infty} \mathcal{N}_{N,k} \frac{z^k}{(k + \frac{N-2}{2})^2 - 1} \tag{B.61}$$

$$V_N(z) \equiv \sum_{k=0}^{\infty} \mathcal{N}_{N,k} \frac{z^k}{[(k + \frac{N-2}{2})^2 - 1]^2} \tag{B.62}$$

These series are well-defined provided that $N \neq 4, 2, 0, -2, \ldots$ . Simple algebraic manipulations yield

$$U_N(z) = \frac{1}{2} \left[ f_1(\tfrac{N}{2} - 2; z) - (z^2 + 1) f_1(\tfrac{N}{2}; z) + z^2 f_1(\tfrac{N}{2} + 2; z) \right] \ . \tag{B.63}$$

Using now the recursion relation (B.55) forward and backward to express everything in terms of $f_1(\frac{N}{2}; z)$, we obtain (specializing for simplicity to $z = -1$)

$$U_N(-1) = \frac{8(N-1)}{(N-2)(N-4)} f_1(\tfrac{N}{2}; -1)$$

$$= 2^{2-N} \sqrt{\pi} \frac{\Gamma\left(\frac{N-4}{2}\right)}{\Gamma\left(\frac{N-1}{2}\right)} \ . \tag{B.64}$$

---

[54]See [43, p. 107, equation (47)]. See also [44, p. 50, Theorem 8.6c].

In complete analogy we can compute $V_N(-1)$. Using the recursion relations (B.55) we get

$$\begin{aligned} V_N(-1) &= \frac{16(N-1)(N-3)}{(N-2)^2(N-4)^2} f_1(\tfrac{N}{2}+2;-1) \\ &= \frac{2^{4-N}\sqrt{\pi}}{(N-2)(N-4)} \frac{\Gamma\left(\frac{N-4}{2}\right)}{\Gamma\left(\frac{N-3}{2}\right)} \,. \end{aligned} \quad (B.65)$$

Notice that in principle one would expect here also a term proportional to $f_2(\tfrac{N}{2};-1)$; but for the specific combination which appears in $V_N(-1)$, the coefficient of this term vanishes. Notice, finally, that

$$(N-2)(N-4)V_N(-1) = 2(N-3)U_N(-1) \,. \quad (B.66)$$

A key cancellation in Section B.6 will rely on this identity but *not* on the specific values of $U_N(-1)$ and $V_N(-1)$.

## B.5 Some More Generalized Theta Functions

Let us now introduce some more generalized theta functions, which will play an important role in our treatment of the numerator of the susceptibility scaling function $\chi_{N,2}^{(0)}(\gamma;B)$. We define, for $\operatorname{Re}\gamma > 0$,

$$U_{N,\theta;k_0}(\gamma) \equiv \sum_{k=k_0}^{\infty} \mathcal{N}_{N,k}\, e^{ik\theta}\, \frac{e^{-\gamma(k+\frac{N-2}{2})^2}}{(k+\frac{N-2}{2})^2 - 1} \quad (B.67)$$

$$V_{N,\theta;k_0}(\gamma) \equiv \sum_{k=k_0}^{\infty} \mathcal{N}_{N,k}\, e^{ik\theta}\, \frac{e^{-\gamma(k+\frac{N-2}{2})^2}}{[(k+\frac{N-2}{2})^2 - 1]^2} \quad (B.68)$$

Here $k_0$ is an integer; if $\alpha \equiv (N-2)/2$ is an integer, then we require that $k_0 > 1 - \alpha$ in order to avoid zeros of the denominators. We will thus take $k_0 = 0$ except when $N$ is an even integer $\leq 4$. When $k_0 = 0$ we omit it from the notation.

We remark that the functions $U_{N,\theta}$ and $V_{N,\theta}$ satisfy recursion relations identical to (B.42).

Applying (B.50) with $f(x) = e^{i\theta x}e^{-\gamma(x+\alpha)^2}$, we obtain

$$U_{N,\theta}(\gamma) = \frac{1}{2\pi}\left(\frac{\pi}{\gamma}\right)^{1/2} \int_{-\infty}^{\infty} K_N^{(U)}(t)\, e^{i\alpha(t-\theta)}\, e^{-(t-\theta)^2/4\gamma}\, dt \quad (B.69)$$

$$V_{N,\theta}(\gamma) = \frac{1}{2\pi}\left(\frac{\pi}{\gamma}\right)^{1/2} \int_{-\infty}^{\infty} K_N^{(V)}(t)\, e^{i\alpha(t-\theta)}\, e^{-(t-\theta)^2/4\gamma}\, dt \quad (B.70)$$

with the obvious kernels $K_N^{(U)}$ and $K_N^{(V)}$. For $\theta \neq 0 \pmod{2\pi}$ we can then obtain an asymptotic expansion of $U_{N,\theta}(\gamma)$ and $V_{N,\theta}(\gamma)$ in powers of $\gamma$, using the Lemma of

Section B.2 to control the contribution of the singularity. At zeroth order we have

$$\lim_{\gamma \to 0} U_{N,\theta}(\gamma) = U_N(e^{i\theta}) \qquad (B.71)$$

$$\lim_{\gamma \to 0} V_{N,\theta}(\gamma) = V_N(e^{i\theta}) \qquad (B.72)$$

and in particular these limits are finite. For $\theta = \pi$ we have calculated these limits in (B.64)/(B.65).

When $\theta = \pi$ — which we will assume henceforth — much more can be said. The simplest approach is to use the differential equations

$$\left(\frac{d}{d\gamma} + 1\right) U_{N,\theta;k_0}(\gamma) = -Z_{N,\theta,\frac{N-2}{2};k_0}(\gamma)$$
$$\equiv -Z_{N,\theta,\frac{N-2}{2}}(\gamma) + \sum_{k=0}^{k_0-1} \mathcal{N}_{N,k}\, e^{ik\theta}\, e^{-\gamma(k+\frac{N-2}{2})^2} \qquad (B.73)$$

$$\left(\frac{d}{d\gamma} + 1\right) V_{N,\theta;k_0}(\gamma) = -U_{N,\theta;k_0}(\gamma) \qquad (B.74)$$

to reduce the problem to known results for $Z_{N,\pi,\frac{N-2}{2}}$. One can write immediately the solution of (B.73)/(B.74):

$$U_{N,\theta;k_0}(\gamma) = e^{-\gamma} U_{N,\theta;k_0}(0) - \int_0^{\gamma} e^{-(\gamma-\gamma')}\, Z_{N,\theta,\frac{N-2}{2};k_0}(\gamma')\, d\gamma' \qquad (B.75)$$

$$V_{N,\theta;k_0}(\gamma) = e^{-\gamma} V_{N,\theta;k_0}(0) - \int_0^{\gamma} e^{-(\gamma-\gamma')}\, U_{N,\theta;k_0}(\gamma')\, d\gamma' \qquad (B.76)$$

For $N \neq$ an even integer $\leq 4$, we can take $k_0 = 0$ and use the fact that $Z_{N,\pi,\frac{N-2}{2}}(\gamma)$ is exponentially small as $\gamma \downarrow 0$, to get

$$U_{N,\pi}(\gamma) = U_N(-1)\, e^{-\gamma} + \text{exponentially small terms} \qquad (B.77)$$
$$V_{N,\pi}(\gamma) = [V_N(-1) - U_N(-1)\, \gamma]\, e^{-\gamma} + \text{exponentially small terms} \qquad (B.78)$$

where $U_N(-1)$ and $V_N(-1)$ have been calculated in (B.64)/(B.65).

Next let us treat the case $N = 4$, taking $k_0 = 1$. We have the initial conditions $U_{4,\pi;1}(0) = -\frac{3}{4}$ and $V_{4,\pi;1}(0) = \frac{1}{16} - \frac{\pi^2}{24}$. Using the fact that $Z_{4,\pi,1}(\gamma)$ is exponentially small, we get

$$U_{4,\pi;1}(\gamma) = \left(-\frac{3}{4} + \gamma\right) e^{-\gamma} + \text{exponentially small terms} \qquad (B.79)$$

$$V_{4,\pi;1}(\gamma) = \left(\frac{1}{16} - \frac{\pi^2}{24} + \frac{3}{4}\gamma - \frac{1}{2}\gamma^2\right) e^{-\gamma} + \text{exponentially small terms}$$
$$(B.80)$$

Finally let us treat the case $N = 2$, taking $k_0 = 2$. We have the initial conditions $U_{2,\pi;2}(0) = \frac{1}{2}$ and $V_{2,\pi;2}(0) = \frac{\pi^2}{12} - \frac{5}{8}$. By the same logic we get

$$U_{2,\pi;2}(\gamma) = 1 - \left(\frac{1}{2} + 2\gamma\right) e^{-\gamma} + \text{exponentially small terms} \quad (B.81)$$

$$V_{2,\pi;2}(\gamma) = -1 + \left(\frac{\pi^2}{12} + \frac{3}{8} + \frac{1}{2}\gamma + \gamma^2\right) e^{-\gamma} +$$
$$\text{exponentially small terms} \quad (B.82)$$

Formulae (B.79)–(B.82) can alternatively be derived from (B.77)/(B.78) and (B.64)/(B.65). by taking the limits $N \to 2, 4$ starting from noninteger $N$.

## B.6  The Susceptibility Scaling Function $\chi^{(0)}_{N,2}(\gamma; B)$

Now we want to prove a formula for the numerator of the susceptibility scaling function $\chi^{(0)}_{N,k}(\gamma; B)$ — that is, for the sum appearing in (4.79) — analogous to that proven in Section B.3 for the partition-function scaling function $\widetilde{Z}^{(0)}_N(\gamma; B)$. We *conjecture* that such a formula is true for all even $k$, but we shall prove it here only for $k = 2$. We define

$$X^{\pm}_{N,k}(\gamma) \equiv \sum_{l,m=0}^{\infty} (\pm 1)^l \, \mathcal{C}^2_{N;\,k,l,m} \, \frac{e^{-\gamma \lambda_{N,l}}}{\Delta_{N;\,l,m}} \,. \quad (B.83)$$

(Note that the properties of the Clebsch-Gordan coefficients guarantee that, for $k$ even, $l$ and $m$ in this sum have the same parity. In particular, $\Delta_{N;\,l,m} = \lambda_{N,m} - \lambda_{N,l}$, independent of $B$.) We shall prove that $X^-_{N,2}(\gamma)$ is exponentially small as $\gamma \downarrow 0$, so that the numerator in $\chi^{(0)}_{N,2}(\gamma; B)$ becomes simply

$$\frac{1 + e^{-\gamma B}}{2 \mathcal{N}_{N,2}} X^+_{N,2}(\gamma) + \text{exponentially small terms} \,. \quad (B.84)$$

From this and (B.48) it follows immediately that

$$\chi^{(0)}_{N,2}(\gamma; B) = \frac{2}{\mathcal{N}_{N,2}} \frac{X^+_{N,2}(\gamma)}{\gamma \widetilde{Z}^+_N(\gamma)} + \text{exponentially small terms} \,. \quad (B.85)$$

In particular, $\chi^{(0)}_{N,2}(\gamma; B)$ is independent of $B$ modulo exponentially small terms.

We start by rewriting $X^-_{N,2}(\gamma)$ as[55]

$$X^-_{N,2}(\gamma) = \sum_{l=0}^{\infty}(-1)^l \left[\frac{\mathcal{C}^2_{N;\,2,l,l+2}}{\Delta_{N;\,l,l+2}} - \frac{\mathcal{C}^2_{N;\,2,l,l-2}}{\Delta_{N;\,l-2,l}}\right] e^{-\gamma \lambda_{N,l}} + \frac{\gamma}{2} \sum_{l=1}^{\infty}(-1)^l \mathcal{C}^2_{N;\,2,l,l} \, e^{-\gamma \lambda_{N,l}} \,,$$
(B.86)

---

[55] In the final term we have $l = m$, and so we must use the comment in footnote 44 to resolve the ambiguity in (4.79).

where we set $\mathcal{C}^2_{N;\,2,l,l-2} = 0$ if $l = 0, 1$. We shall deal with these two sums separately.

Let us suppose first that $N \neq 2, 4$. For the first sum in (B.86), using the formula (2.41) we get

$$\frac{N(N+2)}{16} e^{\gamma(N-2)^2/4} \sum_{l=0}^{\infty} (-1)^l \mathcal{N}_{N,l} \frac{(N-3)(l+\frac{N-2}{2})^2 - \frac{N^2}{2} + 2N - 1}{[(l+\frac{N-2}{2})^2 - 1]^2} e^{-\gamma(l+\frac{N-2}{2})^2}$$

$$= \frac{N(N+2)}{16} e^{\gamma(N-2)^2/4} \left(-(N-3)\frac{d}{d\gamma} - \frac{N^2}{2} + 2N - 1\right) V_{N;\pi}(\gamma) \,. \quad \text{(B.87)}$$

We can then use (B.77)/(B.78) to get

$$-\frac{N(N+2)}{32} e^{\gamma(N-2)^2/4} \Big[(N-2)(N-4)V_N(-1) - 2(N-3)U_N(-1)$$
$$- (N-2)(N-4)\gamma U_N(-1)\Big] + \text{exponentially small terms} \,. \quad \text{(B.88)}$$

Using the identity (B.66), we finally get

$$-\frac{N(N-4)(N^2-4)}{32} e^{\gamma(N-2)^2/4} \gamma U_N(-1) + \text{exponentially small terms} \,. \quad \text{(B.89)}$$

For the second sum appearing in (B.86), using (2.42) we get

$$\frac{N^2-4}{8} \gamma\, e^{\gamma(N-2)^2/4} \sum_{l=0}^{\infty} (-1)^l \mathcal{N}_{N,l} \frac{(l+\frac{N-2}{2})^2 - (\frac{N-2}{2})^2}{(l+\frac{N-2}{2})^2 - 1} e^{-\gamma(l+\frac{N-2}{2})^2}$$

$$= \frac{N^2-4}{8} \gamma\, e^{\gamma(N-2)^2/4} \left[-\frac{d}{d\gamma} - \left(\frac{N-2}{2}\right)^2\right] U_{N;\pi}(\gamma) \,. \quad \text{(B.90)}$$

Using (B.77) we get

$$\frac{N(N-4)(N^2-4)}{32} e^{\gamma(N-2)^2/4} \gamma U_N(-1) + \text{exponentially small terms} \,. \quad \text{(B.91)}$$

Collecting together (B.89) and (B.91), we conclude that $X^-_{N,2}(\gamma)$ is exponentially small as $\gamma \downarrow 0$.

Next let us discuss the case $N = 2$. Here $\mathcal{C}^2_{2;\,2,l,l+2} = 2$, $\mathcal{C}^2_{2;\,2,l,l-2} = 2$ for $l \geq 2$, $\mathcal{C}_{2;\,2,l,l} = 0$ for $l \neq 1$, and $\mathcal{C}_{2;\,2,1,1} = 2$. Thus

$$X^-_{2,2}(\gamma) = \frac{1}{2} - \frac{1}{4}e^{-\gamma} - \gamma e^{-\gamma} - \frac{1}{2}U_{2,\pi;2}(\gamma) \,, \quad \text{(B.92)}$$

so that, using (B.81), we get that $X^-_{2,2}(\gamma)$ is exponentially small.

Finally, for $N = 4$, using (2.41), we can write

$$\sum_{l=0}^{\infty} (-1)^l \left[\frac{\mathcal{C}^2_{4;\,2,l,l+2}}{\Delta_{4;\,l,l+2}} - \frac{\mathcal{C}^2_{4;\,2,l,l-2}}{\Delta_{4;\,l-2,l}}\right] e^{-\gamma\lambda_{4,l}} = \frac{9}{8}e^{-\gamma} + \frac{3}{2}U_{4,\pi;1}(\gamma) \,, \quad \text{(B.93)}$$

while, from (2.42), we get

$$\sum_{l=1}^{\infty} (-1)^l \mathcal{C}^2_{4;\,2,l,l}\, e^{-\gamma\lambda_{4,l}} = -3 + 3e^{-\gamma}\tilde{Z}^-_4(\gamma) \,. \quad \text{(B.94)}$$

As $\tilde{Z}^-_4(\gamma)$ is exponentially small, it follows from (B.79) that $X^-_{4,2}(\gamma)$ is exponentially small.

# C  Large-$N$ Limit

In this Appendix we want to discuss the $N \to \infty$ limit of the finite-size-scaling functions for the one-dimensional $N$-vector universality class. We will first discuss the derivation using the standard large-$N$ formalism; then we will show, in two particular cases (the spin-1 and spin-2 susceptibilities), how to recover these results through a direct evaluation of the $N \to \infty$ limit of the expressions given in Section 4.2.2 .

## C.1  Review of Results from Standard Large-$N$ Formalism

Let us thus start with the standard large-$N$ formalism [45]. Consider, on a one-dimensional lattice of length $L$ with periodic boundary conditions, the standard $N$-vector Hamiltonian
$$\mathcal{H}(\{\boldsymbol{\sigma}\}) = -J \sum_x \boldsymbol{\sigma}_x \cdot \boldsymbol{\sigma}_{x+1} \tag{C.1}$$
and the partition function
$$Z = \int \mathcal{D}\boldsymbol{\sigma}\, e^{-\mathcal{H}(\{\boldsymbol{\sigma}\})} . \tag{C.2}$$

As is well known, the $N \to \infty$ limit must be taken with $J/N$ fixed. We will therefore introduce a rescaled coupling $\widetilde{J} \equiv J/N$. It then turns out [45] that in the $N \to \infty$ limit all correlation functions can be expressed in terms of a mass parameter $m_L$ which is related to the coupling $\widetilde{J}$ by the gap equation
$$\widetilde{J} = \frac{1}{L} \sum_p \frac{1}{\hat{p}^2 + m_L^2} , \tag{C.3}$$
where $p = 2\pi n/L$, the sum runs over $n = 0, \ldots, L-1$, and $\hat{p} = 2\sin(p/2)$. The summation in (C.3) can be performed exactly, and thus one gets
$$\widetilde{J} = \frac{1}{m_L\sqrt{4 + m_L^2}} \coth\left(L \operatorname{arcsinh}\frac{m_L}{2}\right) . \tag{C.4}$$

We can now take the limit $N \to \infty$ at fixed $\widetilde{J} \equiv J/N$ and fixed $L$. All the two-point correlation functions (and indeed *all* the correlation functions) can be easily computed [45]: the result is
$$G_{\infty,k}(x, \widetilde{J}; L) = \left(\frac{1}{\widetilde{J}L} \sum_p \frac{e^{ipx}}{\hat{p}^2 + m_L^2}\right)^k \tag{C.5a}$$
$$= \frac{\cosh^k\left[(L-2x) \operatorname{arcsinh} m_L/2\right]}{\cosh^k\left[L \operatorname{arcsinh} m_L/2\right]} \qquad \text{for } 0 \le x < L \tag{C.5b}$$

From this expression one immediately gets for the susceptibilities:

$$\chi_{\infty,k}(\widetilde{J};L) = \frac{1}{(\widetilde{J}L)^k} \sum_{p_1,\ldots,p_k} L\,\delta(p_1 + \ldots + p_k) \prod_{i=1}^{k} \frac{1}{\hat{p}_i^2 + m_L^2} \quad \text{(C.6a)}$$

$$= \sum_{x=0}^{L-1} \frac{\cosh^k\left[(L-2x)\operatorname{arcsinh} m_L/2\right]}{\cosh^k\left[L\operatorname{arcsinh} m_L/2\right]} \quad \text{(C.6b)}$$

For $k = 1, 2$ one gets simpler expressions:

$$\chi_{\infty,1}(\widetilde{J};L) = \frac{1}{\widetilde{J} m_L^2} \quad \text{(C.7)}$$

$$\chi_{\infty,2}(\widetilde{J};L) = -\frac{1}{\widetilde{J}^2} \frac{\partial}{\partial m_L^2} \left[ \frac{1}{m_L \sqrt{4 + m_L^2}} \coth\left(L \operatorname{arcsinh} \frac{m_L}{2}\right) \right] \quad \text{(C.8)}$$

Analogously one can compute the correlation lengths. For example, in the spin-1 channel we get

$$\xi_{\infty,1}^{(2nd)}(\widetilde{J};L) = \frac{1}{m_L} \,. \quad \text{(C.9)}$$

Having taken the limit $N \to \infty$, $J \to \infty$ at fixed $\widetilde{J} \equiv J/N$ and fixed $L$, we can now take either one of two further limits:

(a) *The standard infinite-volume limit $L \to \infty$ at fixed $\widetilde{J}$.* This limit is trivial and corresponds simply to the substitution of all sums by the corresponding integrals and the parameter $m_L$ by $m_\infty$. In particular the gap equation becomes

$$\widetilde{J} = \int_{-\pi}^{\pi} \frac{dp}{2\pi} \frac{1}{\hat{p}^2 + m_\infty^2} = \frac{1}{m_\infty \sqrt{4 + m_\infty^2}} \,. \quad \text{(C.10)}$$

(b) *The finite-size-scaling limit $L \to \infty$, $\widetilde{J} \to \infty$ [hence $\xi \to \infty$] at fixed $\xi/L$.* From (C.9) we immediately see that this corresponds to considering the limit $L \to \infty$, $m_L \to 0$ with $m_L L \equiv \rho$ fixed. The variable $\rho$ is the natural one in the finite-size-scaling limit of the $N = \infty$ model, and all the finite-size-scaling functions will be expressed in terms of it.

Let us first derive the FSS function for the correlation length $\xi_{\infty,1}^{(2nd)}$. Equating the right-hand sides of (C.4) and (C.10) and taking the limit $m_\infty \to 0$, $m_L \to 0$, $L \to \infty$ with $\rho$ fixed, we get

$$\frac{\xi_{\infty,1}^{(2nd)}(L)}{\xi_{\infty,1}^{(2nd)}(\infty)} = \frac{m_\infty}{m_L} = \tanh \frac{\rho}{2} \,. \quad \text{(C.11)}$$

All the other FSS functions can be computed analogously. For the spin-1 and spin-2 susceptibilities we get

$$\frac{\chi_{\infty,1}(L)}{\chi_{\infty,1}(\infty)} = \tanh^2 \frac{\rho}{2} \quad \text{(C.12)}$$

$$\frac{\chi_{\infty,2}(L)}{\chi_{\infty,2}(\infty)} = \tanh^2 \frac{\rho}{2} + \frac{\rho}{2} \frac{\sinh \rho/2}{\cosh^3 \rho/2} \quad \text{(C.13)}$$

## C.2 Alternate Derivation from Hyperspherical-Harmonics Formalism

Let us now compare our results with those of Section 4.2.2. In that section we took first the finite-size-scaling limit $L \to \infty$, $J \to \infty$ at fixed $\gamma \equiv L\Lambda(J) \approx L/(2J)$ and fixed $N$; now we want to take the further limit $N \to \infty$, $\gamma \to 0$ at fixed $\widetilde{\gamma} \equiv N\gamma$.[56] We want to show that we recover the same results as in the preceding subsection; in other words, we want to show that the two limits *commute*.

We need first to find the relation between $\widetilde{\gamma}$ and $\rho$. This is easily obtained if one considers in (C.4) the limit $\widetilde{J} \to \infty$, $L \to \infty$, $m_L \to 0$ with $\widetilde{\gamma} \equiv L/(2\widetilde{J})$ and $\rho$ fixed. We get

$$\frac{1}{\widetilde{\gamma}} = \frac{1}{\rho} \coth\frac{\rho}{2} . \tag{C.14}$$

Let us begin by computing the limit $N \to \infty$, $\gamma \to 0$ at fixed $\widetilde{\gamma} \equiv N\gamma$ of the partition-function scaling function

$$\widetilde{Z}_N^{(0)}(\gamma) = \sum_{l=0}^{\infty} \mathcal{N}_{N,l}\, e^{-\gamma\lambda_{N,l}} \tag{C.15}$$

[cf. (4.77)]. Since $\gamma$ is tending to zero, it is natural to apply generalized Poisson summation formulae of Appendix B. Using (B.28c) we get

$$\widetilde{Z}_N^{(0)}(\gamma) = -\frac{i}{2\pi\widetilde{\gamma}}\left(\frac{\pi N}{\widetilde{\gamma}}\right)^{1/2} \frac{N}{N-2} \exp\left[\widetilde{\gamma}(N-2)^2/(4N)\right] \times$$
$$\int_{-\infty+i\epsilon}^{+\infty+i\epsilon} dt\, t\, e^{-Nt^2/(4\widetilde{\gamma})} \left(-2i\sin\frac{t}{2}\right)^{2-N} \tag{C.16}$$

where $\epsilon > 0$ is arbitrary (the integral is independent of $\epsilon$). The large-$N$ asymptotic expansion of this integral can be obtained by the standard saddle-point technique. We rewrite the integral as

$$\int_{-\infty+i\epsilon}^{+\infty+i\epsilon} dt\, t\, \left(-2i\sin\frac{t}{2}\right)^2 e^{Nf(t)} \tag{C.17}$$

with

$$f(t) = -\frac{t^2}{4\widetilde{\gamma}} - \log\left(-2i\sin\frac{t}{2}\right) . \tag{C.18}$$

We must now find a saddle point, i.e. a solution of $f'(t) = 0$ with $\text{Im}\, t > 0$. We find $t = i\rho$ where $\rho$ is the unique positive solution of (C.14). Expanding around the saddle point we get

$$\int_{-\infty+i\epsilon}^{+\infty+i\epsilon} dt\, t\, \left(-2i\sin\frac{t}{2}\right)^2 e^{Nf(t)} =$$
$$i\rho\left(2\sinh\frac{\rho}{2}\right)^2 e^{Nf(i\rho)} \left(-\frac{2\pi}{N}\frac{1}{f''(i\rho)}\right)^{1/2} [1 + O(1/N)] . \tag{C.19}$$

---

[56]This is clearly the correct scaling, since $\gamma \approx L/(2J) = (L/2N\widetilde{J})$.

Collecting everything together we get

$$\widetilde{Z}_N^{(0)}(\gamma) = \frac{\rho}{2} \widetilde{\gamma}^{-3/2} \left[ \frac{1}{4\widetilde{\gamma}} + \frac{1}{8 \sinh^2 \rho/2} \right]^{-1/2} \times$$
$$\exp\left( \frac{N\rho^2}{4\widetilde{\gamma}} + \frac{N\widetilde{\gamma}}{4} \right) \left( 2 \sinh \frac{\rho}{2} \right)^{2-N} [1 + O(1/N)] \qquad (C.20)$$

In order to compute the large-$N$ behavior of the finite-size-scaling functions for the susceptibilities, we must also evaluate the large-$N$ behavior of more general sums of the form

$$\sum_{k=0}^{\infty} \mathcal{N}_{N,k} \, R(k) \, e^{-\gamma[k+(N-2)/2]^2} \qquad (C.21)$$

where $R$ is a rational function of the form

$$R(x) = \frac{P(x)}{(x+\beta_1)\ldots(x+\beta_q)} \qquad (C.22)$$

and $P(x)$ is a polynomial. The coefficients of $P(x)$ and coefficients $\beta_i$ are in general $N$-dependent. These series can be handled using the generalized Poisson summation formula (B.50). As an example let us determine the large-$N$ behavior of the sum

$$\sum_{k=0}^{\infty} \mathcal{N}_{N,k} \frac{1}{(k+N/2+\alpha)^q} e^{-\widetilde{\gamma}[k+(N-2)/2]^2/N} \,. \qquad (C.23)$$

The $N$-dependence of the denominator is the one which appears in the finite-size-scaling functions of the susceptibilities. We want to compute its large-$N$ behavior for $\alpha$ and $\widetilde{\gamma}$ fixed. Using (B.50) with $f(x) = \exp[-(\widetilde{\gamma}/N)(x-(N-2)/2)^2]$, we can rewrite the sum as

$$\frac{1}{2\pi} \left( \frac{\pi N}{\widetilde{\gamma}} \right)^{1/2} \int_{-\infty+i\epsilon}^{+\infty+i\epsilon} dt \, e^{i(N-2)t/2} e^{-Nt^2/(4\widetilde{\gamma})} \times$$
$$\left[ f_q\left(\alpha + \frac{N}{2}; e^{it}\right) - e^{2it} f_q\left(\alpha + 2 + \frac{N}{2}; e^{it}\right) \right] \qquad (C.24)$$

where $f_q(\beta; z)$ is defined in (B.54). In order to compute the limit $N \to \infty$ of the integral we must discuss the large-$N$ expansion of $f_q(\alpha + N/2; z)$. The leading order is easily obtained if one notices that it is independent of $\alpha$. In this case one can use the recursion relation (B.55) to compute it. For $q = 1$ we get for $N \to \infty$

$$f_1\left(\alpha + \frac{N}{2}; z\right) = \frac{2}{N}(1-z)^{1-N} - z f_1\left(1 + \alpha + \frac{N}{2}; z\right) + O(1/N^2) \,. \qquad (C.25)$$

from which

$$f_1\left(\alpha + \frac{N}{2}; z\right) = \frac{2}{N} \frac{(1-z)^{1-N}}{1+z} [1 + O(1/N)] \,. \qquad (C.26)$$

A similar formula, which can be proved by induction, is valid for generic $q$:

$$f_q\left(\alpha + \frac{N}{2}; z\right) = \left(\frac{2}{N}\right)^q \frac{(1-z)^{q-N}}{(1+z)^q} [1 + O(1/N)] \,. \qquad (C.27)$$

Using this expansion one can rewrite (C.24) in the limit $N \to \infty$ as

$$-\frac{i}{\pi}\left(\frac{\pi N}{\widetilde{\gamma}}\right)^{1/2}\left(\frac{2}{N}\right)^q \int_{-\infty+i\epsilon}^{+\infty+i\epsilon} dt \, \sin t \left(\frac{1-e^{it}}{1+e^{it}}\right)^q e^{Nf(t)} \qquad (C.28)$$

where $f(t)$ is given in (C.18). The large-$N$ expansion of the remaining integral is then obtained using the same method used for the partition function. We get finally

$$\sum_{k=0}^{\infty} \mathcal{N}_{N,k} \frac{1}{(k+N/2+\alpha)^q} e^{-\widetilde{\gamma}[k+(N-2)/2]^2/N} = \qquad (C.29)$$

$$\widetilde{\gamma}^{-1/2}\left(\frac{2}{N}\right)^q \exp\left(\frac{N\rho^2}{4\widetilde{\gamma}}\right) \left[\frac{1}{4\widetilde{\gamma}} + \frac{1}{8\sinh^2 \rho/2}\right]^{-1/2} \times$$

$$\sinh\rho \left(\tanh\frac{\rho}{2}\right)^q \left(2\sinh\frac{\rho}{2}\right)^{-N} [1+O(1/N)] \,. \qquad (C.30)$$

Generic sums of the type (C.21) can be handled exactly in the same way using the generalized Poisson formula (B.50). In this case what one needs is the large-$N$ behavior of the kernel $F_{N;R}(z)$. To get explicit formulae we must specify the $N$-dependence of the coefficients $\beta_i$. We will assume $\beta_i = N/2 + \alpha_i$, which is the dependence of the sums appearing in the finite-size-scaling functions of the susceptibilities. Using the fact that the generic kernel is obtained by summing and taking derivatives with respect of $z$ of $f_q$, using (C.27) we see that generically the large-$N$ behavior of $F_{N;R}(z)$ is given by

$$F_{N;R}(z) = N^p (1-z)^{-N} \widetilde{F}_R(z) [1+O(1/N)] \,, \qquad (C.31)$$

where $\widetilde{F}_R(z)$ is a rational function of $z$ independent of $N$, and $p$ is an integer. Then we obtain the general formula

$$\sum_{k=0}^{\infty} \mathcal{N}_{N,k} R(k) e^{-\widetilde{\gamma}[k+(N-2)/2]^2/N} =$$

$$\widetilde{\gamma}^{-1/2} N^p \exp\left(\frac{N\rho^2}{4\widetilde{\gamma}}\right) \left[\frac{1}{4\widetilde{\gamma}} + \frac{1}{8\sinh^2 \rho/2}\right]^{-1/2} \times$$

$$\sinh\rho \, \widetilde{F}_R(e^{-\rho}) \left(2\sinh\frac{\rho}{2}\right)^{-N} [1+O(1/N)] \,. \qquad (C.32)$$

Thus the whole computation boils down to the computation of $\widetilde{F}_R(z)$.

In some cases it is possible to simplify the computation by using a relation between the large-$N$ behavior of different series. Indeed let us differentiate (C.32) with respect to $\widetilde{\gamma}$. Keeping only the leading-$N$ contributions we get

$$-\frac{1}{N}\sum_{k=0}^{\infty} \mathcal{N}_{N,k} R(k) \left(k+\frac{N-2}{2}\right)^2 e^{-\widetilde{\gamma}[k+(N-2)/2]^2/N} =$$

$$-\frac{N\rho^2}{4\widetilde{\gamma}^2}\left(\sum_{k=0}^{\infty} \mathcal{N}_{N,k} R(k) e^{-\widetilde{\gamma}[k+(N-2)/2]^2/N}\right) [1+O(1/N)] \,, \qquad (C.33)$$

where we have used the relation (C.14) to eliminate the terms proportional to $d\rho/d\tilde{\gamma}$. In the large-$N$ limit we can of course substitute $(k+\frac{N-2}{2})^2$ by $(k+\frac{N}{2}+\alpha)(k+\frac{N}{2}+\beta)$, where $\alpha$ and $\beta$ are arbitrary. Thus we get the relation

$$\sum_{k=0}^{\infty} \mathcal{N}_{N,k} R(k) \left(k+\frac{N}{2}+\alpha\right)\left(k+\frac{N}{2}+\beta\right) e^{-\tilde{\gamma}[k+(N-2)/2]^2/N} = $$
$$\left(\frac{N\rho}{2\tilde{\gamma}}\right)^2 \left(\sum_{k=0}^{\infty} \mathcal{N}_{N,k} R(k) e^{-\tilde{\gamma}[k+(N-2)/2]^2/N}\right) [1+O(1/N)] . \qquad \text{(C.34)}$$

This formula will allow us to compute the large-$N$ behavior of the sums appearing in the numerators of the spin-1 and spin-2 susceptibilities.

Let us start with the spin-1 case. Using (2.39) we get

$$\sum_{lm} \mathcal{C}^2_{N;1,l,m} \frac{e^{-\gamma \lambda_{N,l}}}{\Delta_{N;l,m}} = \frac{N(N-3)}{4} \sum_{l} \mathcal{N}_{N,l} \frac{e^{-\gamma \lambda_{N,l}}}{[l+N/2-1/2][l+N/2-3/2]} \qquad \text{(C.35)}$$

Using (C.34) with $\alpha = -1/2$, $\beta = -3/2$ we get

$$\sum_{lm} \mathcal{C}^2_{N;1,l,m} \frac{e^{-\gamma \lambda_{N,l}}}{\Delta_{N;l,m}} = \left(\frac{\tilde{\gamma}}{\rho}\right)^2 \tilde{Z}^{(0)}_N(\gamma) [1+O(1/N)] \qquad \text{(C.36)}$$

and thus

$$R^{(0)}_{\chi;\infty,1}(\gamma) = \left(\frac{\tilde{\gamma}}{\rho}\right)^2 = \tanh^2 \frac{\rho}{2} , \qquad \text{(C.37)}$$

which coincides with (C.12).

In order to evaluate the large-$N$ limit of the finite-size-scaling function for the spin-2 susceptibility we need to evaluate the series (B.67) and (B.68). Using again (C.34) we get

$$\sum_{k=0}^{\infty} \mathcal{N}_{N,k} \frac{1}{\left((k+(N-2)/2)^2-1\right)^q} e^{-\tilde{\gamma}[k+(N-2)/2]^2/N} = $$
$$\left(\frac{2\tilde{\gamma}}{N\rho}\right)^{2q} \left(\sum_{k=0}^{\infty} \mathcal{N}_{N,k} e^{-\tilde{\gamma}[k+(N-2)/2]^2/N}\right) [1+O(1/N)] . \qquad \text{(C.38)}$$

Then using (B.86), (B.87) and (B.90) we get

$$\sum_{l,m=0}^{\infty} \mathcal{C}^2_{N;2,l,m} \frac{e^{-\gamma \lambda_{N,l}}}{\Delta_{N;l,m}} = \frac{N}{4} \left[\tanh^2 \frac{\rho}{2} + \frac{\tilde{\gamma}}{2} \frac{1}{\cosh^2 \rho/2}\right] \tilde{Z}^{(0)}_N(\gamma) [1+O(1/N)] \qquad \text{(C.39)}$$

and thus

$$R^{(0)}_{\chi;\infty,2}(\gamma) = \tanh^2 \frac{\rho}{2} + \frac{\tilde{\gamma}}{2} \frac{1}{\cosh^2 \rho/2} . \qquad \text{(C.40)}$$

Using (C.14), one immediately sees that (C.40) agrees with (C.13).

# Acknowledgments

We would like to thank Erhard Seiler for suggesting that we study the new universality classes of [5,6,7] in dimension $d = 1$. We also thank Bob Mawhinney and Paolo Rossi for pointing out some references on hyperspherical harmonics, and Massimo Porrati for pointing out some references on theta functions. One of us (A.P.) thanks New York University for generous hospitality while much of this work was being carried out.

Two of the authors (A.C. and T.M.) thank the "McDonald's" on Broadway and Washington Place, New York, for satisfying and *time-saving* food consumed during the writing of this paper. Another of the authors (A.D.S.) thanks the waitresses at Caffè Pane Cioccolato for helpful and enjoyable conversations.

This work was supported in part by the Istituto Nazionale di Fisica Nucleare (A.P.) and by NSF grant DMS-9200719 (T.M. and A.D.S.).